\documentclass[pre,aps,twocolumn]{revtex4-2}
\usepackage{amsfonts}
\usepackage{times}
\usepackage{amsmath, amsthm}
\usepackage[final]{graphicx}
\usepackage{amssymb, algorithm, algpseudocode}
\algnewcommand\algorithmicforeach{\textbf{for each}}
\algdef{S}[FOR]{ForEach}[1]{\algorithmicforeach\ #1\ \algorithmicdo}
\usepackage[title]{appendix}
\usepackage{xcolor,comment}
\newtheorem{definition}{Definition}

\newcommand{\eq}[2]{\begin{equation}\begin{split}#1\end{split}\label{#2}\end{equation}}
\newcommand{\eqnn}[1]{\begin{equation}\begin{split}#1\end{split}\nonumber\end{equation}}

\newcommand{\co}[1]{{\color{orange}{#1}}}

\begin{document}

\title{PainleveBacklundCheck: A Sympy-powered Kivy app for the Painlev\'e property of nonlinear dispersive PDEs and auto-B\"acklund transformations}
\author{Shrohan Mohapatra}
\affiliation{Department of Mathematics and Statistics and Department of Physics, University
of Massachusetts, Amherst, Massachusetts 01003-4515, USA}

\author{P. G. Kevrekidis}
\affiliation{Department of Mathematics and Statistics, University
of Massachusetts, Amherst, Massachusetts 01003-4515, USA}

\author{Stephane Lafortune}
\affiliation{Department of Mathematics, College of Charleston, Charleston, SC}

\date{\today}

\begin{abstract}
In the present work we revisit the Painlev\'e property for partial
differential equations. We consider the PDE variant of the relevant
algorithm on the basis of the fundamental work of Weiss, Tabor
and Carnevale and explore a number of relevant examples. Subsequently,
we present an implementation of the relevant algorithm in an open-source
platform in Python and discuss the details of a Sympy-powered Kivy app
that enables checking of the property and the derivation of
associated auto-B{\"a}ck{u}nd transform when the property is present.
Examples of the relevant code and its implementation are also provided, as well
as details of its open access for interested potential users.
\end{abstract}

\maketitle





\section{Introduction}

The study of dispersive nonlinear partial differential equations
is a topic of widespread interest over a broad range of
fields, as motivated by a diverse array of applications.
For instance, the dynamics of the electric field
in optical fibers~\cite{hasegawa,agrawal}, nonlinear effects in
plasmas~\cite{plasmabook}, or the study of ultracold atomic systems
in the realm of Bose-Einstein condensates~\cite{pethick,stringari} 
lead to the emergence of variants of the so-called  
nonlinear Schr{\"o}dinger equation~\cite{ablowitz3,ablowitz2,ablowitz1,sulem,siambook}.
Similarly, the evolution of shallow water waves leads naturally
to the emergence of the Korteweg-de Vries (KdV)
equation~\cite{ablowitz3,ablowitz2,infeld,water,drazinjohnson}, while the study of superconducting Josephson junctions, as well as of mechanical systems 
of coupled torsion pendula lattices leads to the
emergence (in appropriate limits) of Klein-Gordon models, such
as the sine-Gordon equation~\cite{dauxois,floyd}. These are but a 
few examples of the diverse and wide range of models of interest.

When such a system emerges in applications, a natural question
is that of integrability~\cite{ablowitz3,ablowitz2,water}. The 
prototypical tool in that regard is the potential identification
of a Lax pair~\cite{ablowitz3,ablowitz2,ablowitz1,water,dauxois}, which allows 
one to use the machinery of the inverse scattering transform to
identify soliton and multi-soliton solutions, as well as, in principle,
to develop the evolution dynamics from arbitrary initial data. However,
there is no generally known approach for constructing such Lax pairs.
In light of that, the presence of ``integrability tests'' such as the
Painlev\'e test~~\cite{painleve1,painleve2,bountis,ablowitz}, i.e., 
that the only 'movable' singularities for the solution of differential equation
systems are simple poles, is of particular value. 
Ablowitz et. al. \cite{ablowitz} used an exact similarity reduction to develop the connection between the nonlinear partial differential equations (PDEs) solvable by the inverse scattering transform and nonlinear ordinary differential equations (ODEs) that admit 
this property. 

Importantly also, the work of Weiss, Tabor and Carnevale~\cite{weiss} (WTC; see also~\cite{weiss2})
offered a systematic way to consider the Painlev\'e property directly at
the level of PDEs, treating the integrable behavior in a unified manner. 
Indeed, this groundbreaking work allowed not only to retrieve the Cole-Hopf
transform for the Burgers' equation, but also to (re)discover B{\"a}cklund
transformations for the KdV equation and similarly for other problems such
as the sine-Gordon, the modified KdV, the Boussinesq model etc. 
Subsequently, numerous researchers have sought to provide an
algorithmic approach~\cite{fudingxie} and accordingly a symbolic software
package~\cite{hereman1,hereman2} (see also~\cite{honechapter}) for the
realization of the WTC approach for PDEs. 

Our aim herein is to provide an open-source package based on Python ---while earlier attempts such as those of~\cite{hereman1,hereman2} were based on proprietary software such as Mathematica--- and a corresponding app that allows for a direct identification of the Painlev\'e property {of the partial differential equation taken in as input. Additionally, in contrast to the earlier attempts in  ~\cite{weiss, weiss2, hereman1, hereman2, honechapter, rconte} where special emphasis has been placed on the simple execution of the Painlev\'e test by means of substituting the Painlev\'e expansion in the PDE after carrying out the appropriate leading-order power balancing, looking for resonance integers and the corresponding redundancies; our work also provides a comprehensive roadmap to the proof or disproof of the compatibility of the other additional conditions}. We believe that this will place such a formulation directly at the disposal of interested researchers that wish to check the potential for integrability of a given model via the WTC approach. Additionally, when the system does turn out to pass the relevant test, the approach can not only test for the compatibility of the additional constraints, but also, in the case of an integrable system, provide a suitable B{\"a}cklund transform that may allow to obtain interesting solutions of such models with a relatively minimal effort, as discussed in case examples below.

Our presentation below will be structured as follows. First, in section
II, we provide the background of the PDE formulation of the Painlev\'e 
approach, while in section III, we will provide a number of examples
for completeness. 
In section IV, we will provide the algorithm and implementation details,
while in section V, we will provide some conclusions and directions for
future study. 

\section{Painlev\'e property and B{\"a}cklund transformation} \label{painleve}

{Similar to the definition of a polynomial of $n$-variables of degree $N$,}
\begin{multline} \label{eqn1}
F(u_1, u_2, \cdots, u_n) =\\ \sum_{0\le k_1+k_2+\cdots+k_n\le N}
c_{k_1, k_2, \cdots k_n} u_1^{k_1} u_2^{k_2}\cdots u_n^{k_n}
\end{multline}
{(with $c_{k_1, k_2, \cdots k_n}, u_1, u_2, \cdots, u_n \in \mathbb{C}$, $k_1, k_2, \cdots, k_n \in \mathbb{N}$), we can extend the definition of a 
non-linear partial differential equation
involving powers of partial derivatives
to a polynomial involving relevant partial differential operators: for a function $u : \mathbb{C}^n\to \mathbb{C}$, we can define a generalized PDE of order $M$ and degree $D$ to be,}
\begin{multline}  \label{eqn2}
F\bigg(\bigg\{\frac{\partial^{\alpha_1+\alpha_2+\cdots+\alpha_n}u}{\partial x_1^{\alpha_1}\partial x_2^{\alpha_2}\cdots\partial x_n^{\alpha_n}}\bigg\}\bigg|_{0 \le \sum_{1\le k\le n} \alpha_k \le M}\bigg)= \\
\sum_{0 \le k_{\alpha_1,\alpha_2,\cdots,\alpha_n} \le D} c_{\{k_{\alpha_1,\alpha_2,\cdots,\alpha_n}\}} \times \prod_{\alpha_1, \alpha_2, \cdots, \alpha_n \in \mathbb{N}}^{ 0 \le \sum_{k=1}^{n} \alpha_k \le M}\\ \bigg(\frac{\partial^{\alpha_1+\alpha_2+\cdots+\alpha_n}u}{\partial x_1^{\alpha_1}\partial x_2^{\alpha_2}\cdots\partial x_n^{\alpha_n}}\bigg)^{k_{\alpha_1,\alpha_2,\cdots,\alpha_n}} = 0,
\end{multline}
where
\begin{equation*}
c_{\{k_{\alpha_1,\alpha_2,\cdots,\alpha_n}\}} = c_{\{k_{\alpha_1,\alpha_2,\cdots,\alpha_n}\}}(x_1, x_2, \cdots, x_n).
\end{equation*}
A partial differential equation is said to have the 
Painlev\'e property when the solutions of the PDE are ``single-valued" about the movable, singularity manifold $\phi = \phi(x_1, x_2, \cdots, x_n) = 0$ \cite{weiss}, i.e., the solutions $u = u(x_1, x_2, \cdots, x_n)$ are of the form:
\begin{equation} \label{eqn3}
u = \phi^{\alpha}\sum_{m=0}^{\infty}u_m \phi^{m} = \sum_{m=0}^{\infty}u_m \phi^{m-\beta}
\end{equation}
where $\forall m \in \mathbb{N},$ $ m\ge 0,$ $u_m = u_m(x_1, x_2, \cdots, x_n)$ and $\phi = \phi(x_1, x_2, \cdots, x_n)$ 
are analytic functions in the neighbourhood of the manifold; and $\alpha = -\beta, \beta > 0, \alpha, \beta \in \mathbb{Z}$. 
This formalism also encompasses the definitions for ODEs \cite{painleve1, painleve2, bountis}. For the prototypical PDEs of interest in this article, we restrict to constant coefficients $c_{\{k_{\alpha_1,\alpha_2,\cdots,\alpha_n}\}}(x_1, x_2, \cdots, x_n) = c_{\{k_{\alpha_1,\alpha_2,\cdots,\alpha_n}\}}$ in the equation \eqref{eqn2} to simplify the relevant equation which acquires the form:
\begin{multline} \label{eqn4}
F\bigg(\bigg\{\frac{\partial^{\alpha_1+\alpha_2+\cdots+\alpha_n}u}{\partial x_1^{\alpha_1}\partial x_2^{\alpha_2}\cdots\partial x_n^{\alpha_n}}\bigg\}\bigg)=\\ \sum_{k_{\alpha_1,\alpha_2,\cdots,\alpha_n}\in\mathbb{N}} c_{\{k_{\alpha_1,\alpha_2,\cdots,\alpha_n}\}}\prod_{\alpha_1, \alpha_2, \cdots, \alpha_n \in \mathbb{N}} \\
 \bigg(\frac{\partial^{\alpha_1+\alpha_2+\cdots+\alpha_n}u}{\partial x_1^{\alpha_1}\partial x_2^{\alpha_2}\cdots\partial x_n^{\alpha_n}}\bigg)^{k_{\alpha_1,\alpha_2,\cdots,\alpha_n}} = 0.
\end{multline}
Using the product rule for the partial derivatives, we can simplify,
\begin{multline} \label{eqn5}
\frac{\partial^{\alpha_1+\alpha_2+\cdots+\alpha_n}u}{\partial x_1^{\alpha_1}\partial x_2^{\alpha_2}\cdots\partial x_n^{\alpha_n}} = \sum_{m = 0}^{\infty}
\frac{\partial^{\alpha_1+\alpha_2+\cdots+\alpha_n} \left(u_m \phi^{m-\beta}\right)}{\partial x_1^{\alpha_1}\partial x_2^{\alpha_2}\cdots\partial x_n^{\alpha_n}} \\ = \sum_{m = 0}^{\infty}
 \sum_{0\le\beta_k\le\alpha_k} \prod_{k=1}^{n} \binom{\alpha_k}{\beta_k}
\frac{\partial^{\beta_1+\beta_2+\cdots+\beta_n}\left(\phi^{m-\beta}\right)}{\partial x_1^{\beta_1}\partial x_2^{\beta_2}\cdots\partial x_n^{\beta_n}} \times \\
\frac{\partial^{(\alpha_1-\beta_1)+(\alpha_2-\beta_2)+\cdots+(\alpha_n-\beta_n)}u_m}{\partial x_1^{\alpha_1-\beta_1}\partial x_2^{\alpha_2-\beta_2}\cdots\partial x_n^{\alpha_n-\beta_n}}.
\end{multline}

{The check of the Painlev\'e property of the PDE follows the consequently (tedious yet) straightforward simplification 
of equation \eqref{eqn5} and, in turn, substitution of the latter into the equation \eqref{eqn4}. After determining the value of $\alpha$ using the leading-order analysis in equation \eqref{eqn4} \cite{weiss, fudingxie}, one can show that the simplification leads to an equivalent expression of the PDE into a form equivalent to the one in \cite{fudingxie},
i.e., a power series in $\phi$ as:}
\begin{equation} \label{eqn6}
\frac{1}{\phi^{\mu_{min}}}\bigg(P_0 + P_1 \phi + P_2 \phi^2 + P_3 \phi^3 + \cdots\bigg) = 0,
\end{equation}
where $P_k = P_k\bigg(u_0, u_1, \cdots u_k, \bigg\{\frac{\partial^{\alpha_1+\alpha_2+\cdots+\alpha_n}\phi}{\partial x_1^{\alpha_1}\partial x_2^{\alpha_2}\cdots\partial x_n^{\alpha_n}}\bigg\}\bigg), 0 \le \sum_{1\le k\le n} \alpha_k \le M, \mu_{min} \in \mathbb{N}$.
The inherent recursion relations of $u_n, n \ge 1$ in terms of $\{u_0, u_1, \cdots, u_{n-1}\}$ and the natural occurrence of the resonance integers have been systematically addressed in~\cite{weiss, fudingxie, hereman1, hereman2} and will be 
used accordingly here as well. The work of~\cite{fudingxie} also highlights the direct Painlev\'e test in terms of the smallest positive resonance integer and the consequent implicit compatibility equations. But several examples in \cite{weiss} show an interesting pattern, namely  that the extraction of the B{\"a}cklund transform comes from the truncation of the Painlev\'e expansion in equations \eqref{eqn3} and \eqref{eqn4} at $m = -\alpha = \beta > 0$. The aforementioned truncation expansion up to $m = -\alpha = \beta$ which when substituted into the original PDE results in a similar expansion of the form;
\begin{equation} \label{eqn7}
\frac{1}{\phi^{\mu_{min}}}\bigg(P_0 + P_1 \phi + P_2 \phi^2 + P_3 \phi^3 + \cdots + P_{\mu_{min}} \phi^{\mu_{min}}\bigg) = 0
\end{equation}
where it can be observed that,
\begin{multline}
P_{\mu_{min}} = F\bigg(\bigg\{\frac{\partial^{\alpha_1+\alpha_2+\cdots+\alpha_n}u_{\beta}}{\partial x_1^{\alpha_1}\partial x_2^{\alpha_2}\cdots\partial x_n^{\alpha_n}}\bigg\}\bigg)=\\ \sum_{k_{\alpha_1,\alpha_2,\cdots,\alpha_n}\in\mathbb{N}} c_{\{k_{\alpha_1,\alpha_2,\cdots,\alpha_n}\}}\prod_{\alpha_1, \alpha_2, \cdots, \alpha_n \in \mathbb{N}} \\
 \bigg(\frac{\partial^{\alpha_1+\alpha_2+\cdots+\alpha_n}u_{\beta}}{\partial x_1^{\alpha_1}\partial x_2^{\alpha_2}\cdots\partial x_n^{\alpha_n}}\bigg)^{k_{\alpha_1,\alpha_2,\cdots,\alpha_n}} = 0.
\end{multline}
which generates back the same original PDE with the substitution of $u = u_{\beta}$. Thus the set of equations
\begin{equation}
P_i=0,\;\;i=0\dots \mu_{min},
\label{PB}
\end{equation}
will be an auto-B\"acklund transformation if ``consistent''. Below, in Definition \ref{BTD}, we define precisely what we mean by \eqref{PB} being consistent, and thus being a \textit{valid} B\"acklund transformation. Also, the more specific concept of auto-B\"{a}cklund transformation is  discussed at the start of the section \ref{backlundexamples}.
We first define the differential commutation operation as follows: let $\delta$ be a derivation and
let $R$ be a differential ring, i.e. a commutative ring with a unit element
and a derivation $\delta$ acting on it. $S = R\{X_1, \cdots , X_m\} = R[\delta^n X_i: i = 1, ... , m, n \in \mathbb{N}_0]$ is the differential ring of the differential polynomials in the differential indeterminates
${X_1, \cdots, X_m}$ (the functions which we want to solve for) with coefficients in the differential ring $R$, where $\delta(\delta^n X_i) = \delta^{n+1} X_i$, $\forall n \in \mathbb{N}_0, i$. 
The operation called ``differential commutation'' $D_c(f_1, f_2)$ where $f_1, f_2 \in S$ are two differential polynomials, $f_1$ of order $m$ and $f_2$ of order $n$, $m < n$ is defined as 
\begin{equation}
    D_c(f_1, f_2) = \delta^n f_1 - \delta^m f_2.
\end{equation}
The same definition can be extended to the case of partial differential rings: let $\delta_1, \delta_2, \cdots, \delta_k$ be mutually commuting derivations and $R$ be a differential ring equipped with these $k$ derivations.  $S_{1, 2, \cdots m} = R\{X_1, \cdots , X_m\} = R[\delta_1^{n_1} \delta_2^{n_2} \cdots \delta_k^{n_k} X_i: i = 1, ... , m, n_1, n_2, \cdots n_k \in \mathbb{N}_0]$ is the differential ring of the \textit{partial} differential polynomials in the differential indeterminates ${X_1, \cdots, X_m}$ with coefficients in the differential ring $R$. For a differential polynomial $f \in S_{1, 2, \cdots m}$, the order of $f$ is $ord(f) = \max\{n_1+n_2+\cdots+n_k: f$ contains a power product in $ X_1, ..., X_m, \delta_1 X_1, \cdots \delta_1^{n_1} \delta_2^{n_2} \cdots \delta_k^{n_k} X_m $ $\text{ with nonzero coefficient}\}$. Also the leading term in the differential polynomial $f \in S_{1, 2, \cdots m}$ is defined as $lt(f) = \delta_1^{m_1} \delta_2^{m_2} \cdots \delta_k^{m_k} X_l$, $m_1 + m_2 + \cdots + m_k = ord(f), 1\le l \le m$. 

The differential commutation $D_c(f_1, f_2)$ where $f_1, f_2 \in S$ are two differential polynomials, $f_1$ of order $m$ and $f_2$ of order $n$, $m < n$ is given by 
\begin{equation}
    D_c(f_1, f_2) = \delta_1^{n_1} \delta_2^{n_2} \cdots \delta_k^{n_k} f_1 - \delta_1^{m_1} \delta_2^{m_2} \cdots \delta_k^{m_k} f_2
\end{equation}
where $m_1 + m_2 + \cdots + m_k = m$, $n_1 + n_2 + \cdots + n_k = n$, $lt(f_1) = \delta_1^{m_1} \delta_2^{m_2} \cdots \delta_k^{m_k} X_l$ and $lt(f_2) = \delta_1^{n_1} \delta_2^{n_2} \cdots \delta_k^{n_k} X_l$. For a set of $m$ differential polynomials $f_1$, $f_2$, ..., $f_m$ $\in R\{X_1, \cdots , X_r\}$ in $r$ differential indeterminates $X_1, X_2, \cdots X_r$ such that $m > r$ and $f_1 = 0, f_2 = 0, \cdots, f_m = 0$ form an overdetermined system of PDEs; let us define a sequence $g_1, g_2, g_3, g_4 \cdots$ as follows:
\begin{enumerate}
    \item \begin{equation} \label{eqnRef11}
    g_1 = \langle f_1, f_2, \cdots, f_m\rangle
    \end{equation}
    \item \begin{multline} \label{eqnRef12}
    g_k = \bigg\langle D_c(g_{k-1, 1}, g_{k-1, 2}), D_c(g_{k-1, 2}, g_{k-1, 3}), \\
    \cdots, D_c(g_{k-1, m-1}, g_{k-1, m}), D_c(g_{k-1, m}, g_{k-1, 1})\bigg\rangle, k \ge 2.
    \end{multline}
\end{enumerate}
We call the set $\{f_1, f_2, \cdots, f_m\}$ to be ``context-free differential commutation compatible'' (or simply context-free compatible) if there exists a finite positive integer $M$ such that $0 \in g_M$. 

Let the following be a set $S_{DAE}$ of $\kappa+n$ differential algebraic equations in $\kappa+1$ differential indeterminates $[\phi, u_1, u_2, u_3, \cdots u_{\kappa}]$ with $n > 1$ (thus forming an overdetermined system of PDEs):
\begin{itemize}
\item \begin{multline} \label{eqnRef13}
\delta_x^{A_1} \delta_t^{B_1} \phi = f_1(\phi, \phi_x, \phi_t, \cdots, \delta_x^{A_1} \delta_t^{B_1-1} \phi,\\
\delta_x^{A_1-1} \delta_t^{B_1} \phi, u_1, u_2, \cdots, u_\kappa, \cdots, \delta_x^{C_{\kappa}}\delta_t^{D_{\kappa}-1} u_{\kappa})
\end{multline}
\item \begin{multline} \label{eqnRef14}
\delta_x^{A_2} \delta_t^{B_2} \phi = f_2(\phi, \phi_x, \phi_t, \cdots, \delta_x^{A_2} \delta_t^{B_2-1} \phi,\\
\delta_x^{A_2-1} \delta_t^{B_2} \phi, u_1, u_2, \cdots, u_\kappa, \cdots, \delta_x^{C_{\kappa}}\delta_t^{D_{\kappa}-1} u_{\kappa})
\end{multline}
\item $\cdots \cdots \cdots \cdots \cdots \cdots \cdots \cdots \cdots \cdots \cdots \cdots \cdots \cdots \cdots \cdots \cdots$
\item \begin{multline} \label{additionaleqnRef14}
\delta_x^{A_n} \delta_t^{B_n} \phi = f_n(\phi, \phi_x, \phi_t, \cdots, \delta_x^{A_n} \delta_t^{B_n-1} \phi,\\
\delta_x^{A_n-1} \delta_t^{B_n} \phi, u_1, u_2, \cdots, u_\kappa, \cdots, \delta_x^{C_{\kappa}}\delta_t^{D_{\kappa}-1} u_{\kappa})
\end{multline}
\item \begin{multline} \label{eqnRef15}
\delta_x^{C_1} \delta_t^{D_1} u_1 = g_1(\phi, \phi_x, \phi_t, \cdots, \delta_x^{A_{max}-1} \delta_t^{B_{max}} \phi,\\
\delta_x^{A_{max}} \delta_t^{B_{max}} \phi, u_1, u_2, \cdots, u_\kappa, \cdots, \delta_x^{C_{\kappa}}\delta_t^{D_{\kappa}-1} u_{\kappa})
\end{multline}
\item \begin{multline} \label{eqnRef16}
\delta_x^{C_2} \delta_t^{D_2} u_2 = g_2(\phi, \phi_x, \phi_t, \cdots, \delta_x^{A_{max}-1} \delta_t^{B_{max}} \phi,\\
\delta_x^{A_{max}} \delta_t^{B_{max}} \phi, u_1, u_2, \cdots, u_\kappa, \cdots, \delta_x^{C_{\kappa}}\delta_t^{D_{\kappa}-1} u_{\kappa})
\end{multline}
\item $\cdots \cdots \cdots \cdots \cdots \cdots \cdots \cdots \cdots \cdots \cdots \cdots \cdots \cdots \cdots \cdots$
\item \begin{multline} \label{eqnRef17}
\delta_x^{C_{\kappa}} \delta_t^{D_{\kappa}} u_{\kappa} = g_\kappa(\phi, \phi_x, \phi_t, \cdots, \delta_x^{A_{max}-1} \delta_t^{B_{max}} \phi,\\
\delta_x^{A_{max}} \delta_t^{B_{max}} \phi, u_1, u_2, \cdots, u_\kappa, \cdots, \delta_x^{C_{\kappa}}\delta_t^{D_{\kappa}-1} u_{\kappa})
\end{multline}
\end{itemize}
where,
\begin{multline*}
A_{max} = max\{A_1, A_2, \cdots, A_n\}, \\
B_{max} = max\{B_1, B_2, \cdots, B_n\}
\end{multline*}
If $W_k =$ $ \delta_x^{A_{max}-A_k}$ $\delta_t^{B_{max}-B_k} f_k$ $ - \delta_x^{A_{max}-A_{k+1}}$ $\delta_t^{B_{max}-B_{k+1}} f_{k+1}, $ $\forall$ $1 \le k \le n-1$ and $W_n =$ $\delta_x^{A_{max}-A_n}$ $\delta_t^{B_{max}-B_n} f_n $ $ - \delta_x^{A_{max}-A_1} $ $\delta_t^{B_{max}-B_1}$ $f_1$, and for all $k, 1 \le k \le n$, if the following holds true,
\begin{multline} \label{eqnRef18}
W_k\bigg(\delta_x^{C_1} \delta_t^{D_1} u_1 = g_1,
\cdots, \\
\delta_x^{C_{\kappa-1}} \delta_t^{D_{\kappa-1}} u_{\kappa-1} = g_{\kappa-1}, \delta_x^{C_\kappa} \delta_t^{D_\kappa} u_\kappa = g_\kappa\bigg) = 0
\end{multline}
we call the set $S_{DAE}$ to be ``context-sensitive differential commutation compatible'' (or simply context-sensitive  compatible).

We are now ready to introduce four concepts that we will use to define what we mean when we say that Equation \eqref{PB} is a valid B\"acklund transformation or not. For this paper, we restrict ourselves the partial differential ring $R_{1+1}$ in (1+1) dimensions $x$ and $t$ equipped with just the two mutually commuting partial derivations $\delta_x$ and $\delta_t$; and $S_{x, t}$ the differential ring of partial differential polynomials,
\begin{multline*}
S_{x, t} = R_{1+1}\{\phi, u_0, u_1, u_2, \cdots, u_{\beta}\} \\
= R_{1+1}[\delta_1^{n_1} \delta_2^{n_2} X, X \in \{\phi, u_0, u_1, u_2, \cdots, u_{\beta}\}, n_1, n_2 \in \mathbb{N}].
\end{multline*}
For any set $T \in S_{x, t}$, let us denote
\begin{enumerate}
    \item the truth value to $T$ being perfectly determinate or simply \textit{determinate} (i.e. the number of algebraic partial differential polynomials or equations in $T$ is exactly equal to the number of differential indeterminates or unknown functions) by the predicate $PerfectDeterminate(T)$;
    \item the truth value to $T$ being overdeterminate (i.e. the number of algebraic partial differential polynomials or equations in $T$ is strictly greater than the number of differential indeterminates or unknown functions) by the predicate $OverDeterminate(T)$;
    \item the truth value to $T$ being context-free compatible by the predicate $ContextFreeCompatible(T)$, and,
    \item the truth value to $T$ being context-sensitive compatible by the predicate $ContextSensitiveCompatible(T)$.
\end{enumerate}
For example, by definition, for any set of PDEs $T$, if $PerfectDeterminate(T)$ is true, then $OverDeterminate(T)$ is false and vice-versa. Also if $OverDeterminate(T)$ is true, then either of $ContextFreeCompatible(T)$ or $ContextSensitiveCompatible(T)$ is true. From the different examples of the PDEs considered as special cases in section \ref{backlundexamples}, it is evident that if $ContextFreeCompatible(T)$ {{is true}}, there exists at least one $f \in T$, non-integers integers $a_1$ and $a_2$, and a partial differential polynomial $M \in R_{1+1}\{T \backslash \{f\}\} \subset S_{x, t}$, such that,
\begin{multline}
\{g_1, g_2, g_3, \cdots g_{|T|-1}\} = T \backslash \{f\} \implies \\
M(g_1 = 0, g_2 = 0, g_3 = 0, \cdots g_{|T|-1} = 0) = 0
\end{multline}
and,
\begin{equation} \label{eqn21From12142023}
\delta_x^{a_1} \delta_t^{a_2} f = M(g_1, g_2, g_3, \cdots g_{|T|-1})
\end{equation}
so that whenever $g_k = 0, \forall$ $1 \le k < |T|$, $f = 0$, and thereby we define this partial differential polynomial $f \in T$ to be $f_{Rec}(T)$ and the equation \eqref{eqn21From12142023} to be a \textit{reconciliation relation}. As will be seen in many different examples in section \ref{backlundexamples} and also in the notion explained in the following paragraph, the test of a set of partial differential polynomials/equations to be context-free compatible primarily serves the role of determining the existence of a reconciliation relation amongst the PDEs owing to the equation $f_{Rec}(T){{=0}}$ corresponding to a resonance integer, as has been also seen in \cite{weiss, weiss2, hereman1, hereman2, honechapter, rconte}. However, if a set of PDEs $T$ is context-free compatible, it does not necessarily imply that the set $T\backslash\{f_{Rec}(T)\}$ is context-sensitive compatible; this also plays a key role in distinguishing between a valid compatible B\"{a}cklund transform from an invalid one.

Following the aforementioned construct which is also laid down in \cite{weiss}, the substitution of the finite truncated expansion in equation \eqref{eqn3} into the algebraic partial differential equation \eqref{eqn2} leads to the set of PDEs \eqref{PB} that can be read off from the respective coefficients of the powers of $\phi$ in equation \eqref{eqn7}. Consider the set 
\begin{equation}\label{Qdef}
Q = \{P_0, P_1, P_2, \cdots, P_{\mu_{min}}\}
\end{equation}
made of the LHS's of the PDEs given in \eqref{PB}.
By construction, we have that $Q \in S_{x, t}$. We define a sequence $R_{seq}(Q)$ of sets $S_1, S_2, S_3, S_4, \cdots$ in the following way:
\begin{equation}
S_1 = Q 
\end{equation}
and for $k \ge 1$,
\begin{equation}
S_{k+1} = \begin{cases}
        S_k \backslash \{f_{Rec}(S_k)\} & \text{if } ContextFreeCompatible(S_k) \\
        S_k & \text{otherwise}
    \end{cases}
    \label{seq}
\end{equation}
and $R_{seq}(Q)$ terminates at $S_{k_{stop}}$ where $k_{stop}$
is the smallest value of $k$  such that  $ContextFreeCompatible(S_{k_{stop}})$ is false. 
\begin{definition}
\label{BTD}
Consider the sequence $R_{seq}(Q)$ defined above and associated to the set $Q$ in \eqref{Qdef}. If 
\begin{multline} \notag\label{eqn24From12132023}
PerfectDeterminate(S_{k_{stop}}) \lor 
\\ ContextSensitiveCompatible(S_{k_{\rm stop}}) 
\end{multline}
is true, then we call the set $Q$ a \textit{valid B\"{a}cklund transformation}. Otherwise, we call it an \textit{invalid B\"{a}cklund transformation}.
\end{definition}
In effect, Definition \ref{BTD} is saying that the set of PDEs \eqref{PB} defines a B\"acklund transformation if it is equivalent to a determinate set, or to a set of equations that is context-sensitive.

We are going to explore a set of examples of integrable and non-integrable models in the section \ref{backlundexamples} with the test of the context-free compatibility of the consequent additional PDEs in the identification of the coefficients of the Painlev\'e expansion corresponding to resonance integers (if any) and the test of the context-sensitive compatibility to completely examine the compatibility of the remnant PDEs. {All the results in the section \ref{backlundexamples} demonstrate that the integrable PDEs  admit a valid and compatible (auto)-B\"{a}cklund transform whereas non-integrable PDEs do not admit a valid B\"{a}cklund transform.}


\section{Examples of the WTC And Compatibility Test for Select PDEs} \label{backlundexamples}

Before we embark on the examples, let us define what B{\"a}cklund transformations are. Following~\cite{drazinjohnson}, let $P(u) = 0$ and $Q(v) = 0$ be two uncoupled partial differential equations involving two functions $u$ and $v$, where $P$ and $Q$ are two non-linear operators. Let $R_i = 0$ be a pair of relations between the two functions $u$ and $v$:
\begin{equation*}
    R_i(u, v, \partial_{\alpha}u, \partial_{\beta}v, \cdots) = 0, \, i = 1, 2.
\end{equation*}
Then $R_i = 0$ constitute a \textit{B\"acklund transformation} if it is integrable for $v$ when $P(u) = 0$ and if the resulting $v$ satisfies $Q(v) = 0$, and vice-versa. A special case is that of $P = Q$, in which case it is known as an \textit{auto-B\"acklund transformation}. The practical relevance  of B\"acklund transformations is that they allow one to generate another solution to a nonlinear PDE when one to another PDE is already known. {Similarly, an auto-B\"{a}cklund transformation to generate another solution to the same nonlinear PDE from an already known solution; this is often considered as an anchoring property that is known to decide the integrability of the PDE \cite{ablowitz, weiss, weiss2, hereman1, hereman2, honechapter, rconte}}. The notations in the rest of this section resemble the ones used in \cite{weiss}. Section \ref{wtcpainleveintegrablePDE} explores the existence of an auto-B\"{a}cklund transform and the compatibility of the associated equations that comes as a consequence of the WTC test for the case of some known integrable PDEs (Burgers' equation, KdV equation and the modified KdV equation in sections \ref{burgersEquationWTC}, \ref{KdVEquationWTC} and \ref{mKdVEquationWTC} respectively). On the contrary, section \ref{wtcpainlevenonintegrablePDE} discusses the non-existence of an auto-B\"{a}cklund transform that comes as a consequence of the incompatibility of the associated equations for the case of some expected-to-be non-integrable PDEs (BBM equation, KdV-Burgers equation and the quartic-interactions Klein-Gordon equation in sections \ref{BBMEquationWTC}, \ref{KdVBurgersEquationWTC} and \ref{phi4TheoryWTC} respectively).

\subsection{WTC And Compatibility Tests for Integrable PDEs} \label{wtcpainleveintegrablePDE}

\subsubsection{Burgers' equation} \label{burgersEquationWTC}
Let us take the example of the procedure sketched in section \ref{painleve} for the Burgers' equation for $u : \mathbb{R}^2 \to \mathbb{R}, x, t, \sigma \in \mathbb{R}$
\begin{equation} \label{eqn8}
u_t + u u_x - \sigma u_{xx} = 0.
\end{equation}
Substituting the equation \eqref{eqn3} in equation \eqref{eqn8}, by leading-order analysis one can obtain $\beta = 1$ $(\alpha = -1)$~\cite{weiss, fudingxie} and the expansion reminiscent of equation \eqref{eqn7}: $\frac{1}{\phi^3}(P_0 + P_1 \phi + P_2 \phi^2 + P_3 \phi^3 + \cdots) = 0$, where,
\begin{equation}
P_0 = \left(-2\sigma\phi_x-u_{0}\right)u_{0}\phi_x,
\end{equation}
\begin{equation}
P_1 = \sigma (u_{0} \phi_{xx} + 2 u_{0, x} \phi_x) - u_{0} u_{1} \phi_x + u_{0} u_{0, x} - u_{0}\phi_t,
\end{equation}
\begin{equation}
P_2 = - \sigma u_{0, xx} + u_{0} u_{1, x} + u_{1} u_{0, x} + u_{0, t},
\end{equation}
\begin{multline}
P_3 = u_2 \phi_t + u_{1, t} + (u_1 u_2 + u_0 u_3) \phi_x \\ - 2 \sigma u_3 \phi_x^2 + u_2 u_{0, x} + u_1 u_{1, x} + u_0 u_{2, x} \\ - 2 \sigma \phi_x u_{2, x} - \sigma u_2 \phi_{xx} - \sigma u_{1, xx},
\end{multline}
and similar relations for higher order coefficients. The particular independence of the coefficient $P_2$ of the function $u_2$ is a direct implication of $m = 2$ being the only positive resonance integer \cite{weiss}. Let us truncate the expansion by setting $u_j = 0, j \ge 2$. For $u_0 \ne 0, \phi_x \ne 0$, $P_0 = 0$ implies $u_0 = -2 \sigma \phi_x$ which when substituted in the equation $P_1 = 0$ implies
\begin{equation} \label{eqn13}
\phi_t + u_1 \phi_x - \sigma \phi_{xx} = L_1[u_1, \phi] = 0 \implies u_1 = \frac{\sigma \phi_{xx} - \phi_t}{\phi_x},
\end{equation}
and when substituted in the equation $P_2 = 0$ implies
\begin{equation} \label{eqn14}
\frac{\partial}{\partial x}\left(\phi_t + u_1 \phi_x - \sigma \phi_{xx} \right) = \frac{\partial}{\partial x}L_1[u_1, \phi] = 0,
\end{equation}
and in $P_3 = 0$ implies
\begin{equation} \label{eqn15}
u_{1, t} + u_1 u_{1, x} - \sigma u_{1, xx} = 0.
\end{equation}
{One can also check that all the other coefficients $P_k = 0, k \ge 4$.} For {$Q_{1, \text{Burgers}} = \phi_t + u_1 \phi_x - \sigma \phi_{xx}$, $Q_{2, \text{Burgers}}$ = $ \frac{\partial}{\partial x}\left(\phi_t + u_1 \phi_x - \sigma \phi_{xx} \right)$, the set $\{Q_{1, \text{Burgers}}, Q_{2, \text{Burgers}}\}$ is context-free compatible since $D_c(Q_{1, \text{Burgers}}, Q_{2, \text{Burgers}}) = 0$}, a consequence of the fact that equation \eqref{eqn13} implies \eqref{eqn14}. Also, apart from \eqref{eqn14}, the pair of equations \eqref{eqn13} and \eqref{eqn15}, {i.e., the set $\{P_1, P_3\}$} forms a perfectly determined system of PDEs to adhere to the layout of the definition \ref{BTD} of the B\"{a}cklund transform discussed in section \ref{painleve}. {Thus, with the transformation $u = \frac{u_0}{\phi} + u_1$, $u_0 = -2 \sigma \phi_x$, $u_1 = \frac{\sigma \phi_{xx} - \phi_t}{\phi_x}$, we end up with the \textit{valid} auto-B{\"a}cklund transformation associated with equation \eqref{eqn15}; this suggests
the complete integrability of the viscous Burgers' equation~\cite{weiss} (a feature that is well known to be associated with the Cole-Hopf transformation---see also below).}\\

We now truncate the expansion one coefficient earlier by setting $u_j = 0, j \ge 1$; then all the coefficients $P_n, n \ge 3$ vanish, provided,
\begin{equation}
\label{Dif}
u = -\frac{2\sigma\phi_x}{\phi}, \phi_t = \sigma \phi_{xx}
\end{equation}
which comes from $P_0 = 0$ and $P_1 = 0$ respectively; as an added compatibility condition, we obtain
\begin{equation}
\label{DifDer}
\frac{\partial}{\partial x}\left(\phi_{t} - \sigma \phi_{xx}\right) = 0
\end{equation}
that comes from $P_2 = 0$. {Also, for $Q_{1, \text{Burgers}, -} = \phi_{t} - \sigma \phi_{xx}$, $Q_{2, \text{Burgers}, -}$ = $ \frac{\partial}{\partial x}\left(\phi_{t} - \sigma \phi_{xx}\right)$, the set $\{Q_{1, \text{Burgers}, -}, Q_{2, \text{Burgers}, -}\}$ is context-free compatible since $D_c(Q_{1, \text{Burgers}, -}, Q_{2, \text{Burgers}, -}) = 0$.} Indeed, \eqref{DifDer} is automatically satisfied if the second equation of \eqref{Dif} holds. Here, we retrieve the remarkable Cole-Hopf transformation that helps transform from a solution to the 1D Burgers' equation to one of the standard diffusion equation (and thus
offers a path to obtaining its solution). Thus the transform, $u = -\frac{2\sigma\phi_x}{\phi}$ defines a B\"{a}cklund transform from the Burgers' equation $u_t + u u_x - \sigma u_{xx} = 0$ to the diffusion equation $\phi_t = \sigma \phi_{xx}$, since the system \eqref{Dif} is determinate. However, it does not constitute an auto-B{\"a}cklund transformation that would take a solution to the Burgers' equation to another solution of the same equation~\cite{weiss}.

{Finally upon the truncation of the expansion one coefficient ahead by setting $u_j = 0, j \ge 3$, such that $u = \frac{u_0}{\phi} + u_1 + u_2 \phi$, we obtain
\begin{equation} \label{eqn36From03032024}
    P_0 = 0 \implies u_0 = -2\sigma\phi_x,
\end{equation}
\begin{equation} \label{eqn37From03032024}
    P_1 = 0 \implies u_1 = \frac{\sigma \phi_{xx}-\phi_t}{\phi_x}.
\end{equation}
The substitution of $u_0$ and $u_1$ from the equations \eqref{eqn36From03032024} and \eqref{eqn37From03032024} leads to the coefficient $P_2$ to be identically zero as $j = 2$ happens to be the resonance integer of this system \cite{weiss}. The system of equations $P_3 = P_4 = P_5 = 0$ upon the substitution of $u_0$ and $u_1$ from the equations \eqref{eqn36From03032024} and \eqref{eqn37From03032024} (in which case, $P_k = 0, k \ge 6$) happens to be an overdetermined system of PDEs in terms of the functions $u_2$ and $\phi$; {which necessitates the tests of context-free and context-sensitive compatibilities towards the check of the validity of the B\"{a}cklund transformation $u = \frac{\sigma \phi_{xx}-\phi_t}{\phi_x} - \frac{2\sigma\phi_x}{\phi} + u_2 \phi$ cf. the definition \ref{BTD}.
These tests of compatibility of} the set $\{ Q_{1, \text{Burgers}, +},  Q_{2, \text{Burgers}, +}, Q_{3, \text{Burgers}, +}\}$ (where $P_3 = Q_{1, \text{Burgers}, +}$, $P_4 = Q_{2, \text{Burgers}, +}$ and $P_5 = Q_{3, \text{Burgers}, +}$) have been rigorously verified by means of symbolic computation which is available in the associated ``sandbox'' programs in the open-source repository in \cite{painleveapprepo}. These computations show that the set $\{ Q_{1, \text{Burgers}, +},  Q_{2, \text{Burgers}, +}, Q_{3, \text{Burgers}, +}\}$ is \textit{context-free compatible} with the appropriate reconciliation relation, but the set $\{ Q_{1, \text{Burgers}, +},  Q_{2, \text{Burgers}, +}\}$ is neither \textit{context-free} nor \textit{context-sensitive} compatible. Conclusively, this computation leads to a contradiction to the assumption that the expansion $u = \frac{\sigma \phi_{xx}-\phi_t}{\phi_x} - \frac{2\sigma\phi_x}{\phi} + u_2 \phi$ is a valid B\"{a}cklund transformation. Thus, there does not exist a B\"{a}cklund transformation in the functions $\langle u_2, \phi\rangle$ upon the truncation of the expansion one coefficient ahead of the case where we got a B{\"a}cklund transformation.}

\subsubsection{KdV equation} \label{KdVEquationWTC}
The next example of an integrable PDE we consider is the case of Korteweg-De Vries (KdV) equation,
in the form:
\begin{equation} \label{eqn20}
u_t + u u_x + \sigma u_{xxx} = 0.
\end{equation}
By leading-order analysis, one can obtain $\beta = 2$ $(\alpha = -2)$ \cite{weiss, fudingxie} and the expansion reminiscent of equation \eqref{eqn7} $\frac{1}{\phi^5}(P_0 + P_1 \phi + P_2 \phi^2 + P_3 \phi^3 + P_4 \phi^4 + P_5 \phi^5 \cdots) = 0$, where,
\begin{equation}
P_0 = -2 u_0 \phi_x \left(u_0 + 12\sigma\phi_x^2\right),
\end{equation}
\begin{multline}
P_1 = u_0 u_{0, x} - 3 u_0 u_1 \phi_x - 6 \sigma u_1 \phi_x^3 \\
+ 18 \sigma \phi_x^2 u_{0, x} + 18 \sigma u_0 \phi_x \phi_{xx},
\end{multline}
\begin{multline}
P_2 = \left(u_0 u_1\right)_x - 2 u_0 \phi_t - u_1^2 \phi_x \\ - 2 u_0 u_2 \phi_x
+ 6 \sigma (\phi_x^2 u_{1, x} + u_1 \phi_x \phi_{xx} \\ - u_{0, x}\phi_{xx} - \phi_x u_{0, xx}) - 2\sigma u_0 \phi_{xxx},
\end{multline}
\begin{multline}
P_3 = - u_1 \phi_t + u_{0, t} - (u_1 u_2 + u_0 u_3) \phi_x \\ + \left(u_0 u_2\right)_x
+ u_1 u_{1, x} - 3 \sigma u_{1, x}\phi_{xx} \\ - 3\sigma \phi_x u_{1, xx}
- \sigma u_1 \phi_{xxx} + \sigma u_{0, xxx},
\end{multline}
\begin{equation}
P_4 = u_{1, t} + (u_0 u_3 + u_1 u_2 + \sigma u_{1, xx})_x,
\end{equation}
\begin{multline} \label{eqn26}
P_5 = u_3 \phi_t + u_{2, t} + (u_1 u_4 u_2 u_3)\phi_x \\
+ (u_0 u_4 + u_1 u_3)_x + u_2 u_{2, x} + 6 \sigma \phi_x^2 u_{4, x} \\
+ 6 \sigma u_4 \phi_x \phi_{xx} + 3 \sigma u_{3, x}\phi_{xx} + 3\sigma \phi_x u_{3, xx} \\
+ \sigma u_3 \phi_{xxx} + \sigma u_{2, xxx},
\end{multline}
and similar such relations for higher order coefficients. The particular independence of the coefficient $P_4$ of the function $u_4$ is a direct implication of $m = 4$ being the least positive resonance integer~\cite{weiss} (the only other larger resonance integer being $m=6$). If we truncate the expansion by setting $u_j = 0, j \ge 3$, then all the coefficients $P_n, n \ge 5$ vanish, provided,
\begin{equation} \label{eqn27}
u = \frac{12 \sigma \phi_{xx}}{\phi} - \frac{12 \sigma \phi_x^2}{\phi^2} + u_2,
\end{equation}
with the following compatibility conditions that have also been obtained in~\cite{weiss} by setting $P_2$, $P_3$ and $P_4$ respectively to zero.
\begin{equation} \label{equationRef7}
\phi_x \phi_t + \phi_x^2 u_2 + 4\sigma \phi_x \phi_{xxx} - 3\sigma\phi_{xx}^2 = 0,
\end{equation}
\begin{multline} \label{equationRef8}
u_{2,x} \phi_x^2 + 2 \phi_x \phi_{xt} + \phi_t \phi_{xx} + 3 u_2 \phi_x \phi_{xx} \\
- 2 \sigma \phi_{xx} \phi_{xxx} + 5 \sigma \phi_x \phi_{xxxx} = 0,
\end{multline}
\begin{equation} \label{equationRef9}
\frac{\partial}{\partial x}\left(\phi_{xt}+\phi_{xx}u_2+\sigma \phi_{xxxx}\right) = 0,
\end{equation}
and from the equation \eqref{eqn26} the auto-B{\"a}cklund transformation (for this case) reveals itself automatically
\begin{equation} \label{equationRef10}
u_{2, t} + u_2 u_{2, x} + \sigma u_{2, xxx} = 0.
\end{equation}
One can check and verify that
\begin{multline} \label{equationRef12}
    \frac{\partial}{\partial x}\bigg(\phi_x \phi_t + \phi_x^2 u_2 + 4\sigma \phi_x \phi_{xxx} - 3\sigma\phi_{xx}^2\bigg) \\
    + \phi_x(\phi_{xt}+\phi_{xx}u_2+\sigma \phi_{xxxx}) = u_{2,x} \phi_x^2 + 2 \phi_x \phi_{xt} + \phi_t \phi_{xx} \\
    + 3 u_2 \phi_x \phi_{xx} - 2 \sigma \phi_{xx} \phi_{xxx} + 5 \sigma \phi_x \phi_{xxxx}.
\end{multline}
On the choice of $u_2 = \frac{3 \sigma \phi_{xx}^2 - 4 \sigma \phi_x \phi_{xxx} - \phi_x \phi_t}{\phi_x^2}$, we get that equation \eqref{equationRef8} implies 
\begin{equation} \label{equationRef11}
\phi_{xt}+\phi_{xx}u_2+\sigma \phi_{xxxx} = 0.
\end{equation}
This reconciliation is reflected in the appearance of equation \eqref{equationRef9} which is also stated as an associated compatibility condition in \cite{weiss}. If
\begin{multline*}
F_{\text{KdV}} = {\{Q_{1, \text{KdV}}, Q_{2, \text{KdV}}, Q_{3, \text{KdV}}, Q_{4, \text{KdV}}\}} = \{\phi_x \phi_t \\
+ \phi_x^2 u_2 + 4\sigma \phi_x \phi_{xxx} - 3\sigma\phi_{xx}^2, u_{2,x} \phi_x^2 + 2 \phi_x \phi_{xt} + \phi_t \phi_{xx} \\
+ 3 u_2 \phi_x \phi_{xx} - 2 \sigma \phi_{xx} \phi_{xxx} + 5 \sigma \phi_x \phi_{xxxx}, \\
\frac{\partial}{\partial x}\left(\phi_{xt}+\phi_{xx}u_2+\sigma \phi_{xxxx}\right) , u_{2, t} + u_2 u_{2, x} + \sigma u_{2, xxx}\}
\end{multline*}
then following the 
definition in the equations \eqref{eqnRef11} and \eqref{eqnRef12}, one can verify that $F_{\text{KdV}}$ is a context-free compatible set as $0 \in g_{12}$, and this happens as a consequence of the reconciliation equation \eqref{equationRef12}. \cite{weiss} also observes the equation \eqref{equationRef9} as a "compatibility condition" with $j = 4$ as the resonance integer, and one can also infer that $f_{Rec}(F_{\text{KdV}}) = \phi_{xt}+\phi_{xx}u_2+\sigma \phi_{xxxx}$. But 
the reduced set $F_{\text{KdV}}\backslash \{Q_{3, \text{KdV}}\} = \{Q_{1, \text{KdV}}, Q_{2, \text{KdV}}, Q_{4, \text{KdV}}\}$ 
is not context-free compatible (as the sequence of sets of algebraic partial differential equations following the definition in the equations \eqref{eqnRef11} and \eqref{eqnRef12} keep increasing in the order without the reduction of number of terms), which necessitates the check of context-sensitive compatibility.

From equations \eqref{equationRef7}, \eqref{equationRef10} and \eqref{equationRef11}, following the differential-algebraic setting of equations \eqref{eqnRef13}, \eqref{eqnRef14}, \eqref{eqnRef15}, \eqref{eqnRef16}, \eqref{eqnRef17} and \eqref{eqnRef18}, one can refer to the following higher derivatives,
\begin{equation} \label{equation46}
\phi_t = \frac{-1}{\phi_x}(\phi_x^2 u_2 + 4\sigma \phi_x \phi_{xxx} - 3\sigma\phi_{xx}^2),
\end{equation}
\begin{multline} \label{equation47}
\phi_{xxxx} = \frac{-1}{\sigma} \bigg(\phi_{xt}+\phi_{xx}u_2\bigg) = \frac{-1}{\sigma} \bigg(\phi_{xx} u_2 + \frac{\partial}{\partial x}\bigg(
\\
\frac{-1}{\phi_x}\bigg(\phi_x^2 u_2 + 4\sigma \phi_x \phi_{xxx} - 3\sigma\phi_{xx}^2\bigg)\bigg)\bigg) \implies \phi_{xxxx} =\\
\frac{6\sigma\phi_ x\phi_{xx}\phi_{xxx}-3\sigma\phi_{xx}^3+u_2\phi_x^2\phi_{xx}-u_{2, x}\phi_x^3}{3\sigma\phi_x^2},
\end{multline}
\begin{equation} \label{equation48}
u_{2, t} = - u_2 u_{2, x} - \sigma u_{2, xxx}.
\end{equation}
{For the context-sensitive compatibility, we need to show that the difference $\frac{\partial^4}{\partial x^4} \phi_t - \frac{\partial}{\partial t} \phi_{xxxx}$ vanishes. The PDEs \eqref{equation46} and \eqref{equation47} have the derivative terms $\phi_x$, $\phi_{xx}$, $\phi_{xxx}$, $u_2$ and $u_{2, x}$. Therefore, to get to the higher derivatives $\phi_{xxxxt}$ and $\phi_{txxxx}$ we would need the following higher derivatives wherein one needs careful substitution from the equations \eqref{equation46}, \eqref{equation47} and \eqref{equation48}:
}
\begin{equation} \label{equationRef49For03042024}
\phi_{x, 5} = \frac{6 \phi_{xxx}^2 \phi_{x} \sigma - 3 \phi_{xxx} \phi_{xx}^2 \sigma - 3 \phi_{xx} \phi_{x}^2 u_{2 ,x} - \phi_{x}^3 u_{2,xx}}{3 \phi_{x}^2 \sigma},  
\end{equation}
\begin{multline} \label{equationRef50For03042024}
\phi_{x, 6} = \frac{1}{3 \phi_{x}^4 \sigma} (12 \phi_{xxx}^2 \phi_{xx} \phi_{x}^2 \sigma - 12 \phi_{xxx} \phi_{xx}^3 \phi_{x} \sigma \\
- 7 \phi_{xxx} \phi_{x}^4 u_{2,x} + 3 \phi_{xx}^5 \sigma + \phi_{xx}^2 \phi_{x}^3 u_{2,x} \\
- 4 \phi_{xx} \phi_{x}^4 u_{2,xx} - \phi_{x}^5 u_{2,xxx}),
\end{multline}
\begin{multline} \label{equationRef51For03042024}
\phi_{x, 7} = \frac{1}{9 \phi_{x}^4 \sigma^2} (36 \phi_{xxx}^3 \phi_{x}^2 \sigma^2 - 36 \phi_{xxx}^2 \phi_{xx}^2 \phi_{x} \sigma^2 \\
+ 9 \phi_{xxx} \phi_{xx}^4 \sigma^2 - 60 \phi_{xxx} \phi_{xx} \phi_{x}^3 \sigma u_{2,x} - 33 \phi_{xxx} \phi_{x}^4 \sigma u_{2,xx} \\
+ 30 \phi_{xx}^3 \phi_{x}^2 \sigma u_{2,x} + 3 \phi_{xx}^2 \phi_{x}^3 \sigma u_{2,xx} - 15 \phi_{xx} \phi_{x}^4 \sigma u_{2,xxx} \\
- 3 \phi_{x}^5 \sigma u_{2,xxxx} + 7 \phi_{x}^5 u_{2,x}^2),
\end{multline}
\begin{equation} \label{equationRef52For03042024}
u_{2, xt} = -\sigma u_{2,xxxx} - u_2 u_{2,xx} - u_{2,x}^2,
\end{equation}
\begin{equation} \label{equationRef53For03042024}
\phi_{xt} = -\frac{2 \phi_{xxx} \phi_{xx} \sigma}{\phi_{x}} + \frac{\phi_{xx}^3 \sigma}{\phi_{x}^2} - \phi_{xx} u_2 + \frac{\phi_{x} u_{2,x}}{3},
\end{equation}
\begin{equation} \label{equationRef54For03042024}
\phi_{xxt} = -\frac{2 \phi_{xxx}^2 \sigma}{\phi_{x}} + \frac{\phi_{xxx} \phi_{xx}^2 \sigma}{\phi_{x}^2} - \phi_{xxx} u_2 + \frac{\phi_{x} u_{2,xx}}{3},
\end{equation}
\begin{multline} \label{equationRef55For03042024}
\phi_{xxxt} = -\frac{4 \phi_{xxx}^2 \phi_{xx} \sigma}{\phi_{x}^2} + \frac{4 \phi_{xxx} \phi_{xx}^3 \sigma}{\phi_{x}^3} - \frac{2 \phi_{xxx} \phi_{xx} u_2}{\phi_{x}} \\
+ \frac{\phi_{xxx} u_{2,x}}{3} - \frac{\phi_{xx}^5 \sigma}{\phi_{x}^4} + \frac{\phi_{xx}^3 u_2}{\phi_{x}^2} - \frac{\phi_{xx}^2 u_{2,x}}{3 \phi_{x}} \\
+ \frac{\phi_{xx} u_{2,xx}}{3} + \frac{\phi_{x} u_{2,xxx}}{3} + \frac{\phi_{x} u_2 u_{2,x}}{3 \sigma}.
\end{multline}
The algorithm \ref{ContextSensitiveDifferentialCommutationCompatibility} elucidates the sequential substitution and the consequent computation of these requisite higher derivatives in section \ref{algorithm}. Using the above equations \eqref{equationRef49For03042024}, \eqref{equationRef50For03042024}, \eqref{equationRef51For03042024}, \eqref{equationRef52For03042024}, \eqref{equationRef53For03042024}, \eqref{equationRef54For03042024} and \eqref{equationRef55For03042024} for the higher derivatives, one can show that $\phi_{xxxxt} = \phi_{txxxx}$. This renders the set of differential polynomials {$\{Q_{1, \text{KdV}},$ $ Q_{2, \text{KdV}},$ $ Q_{4, \text{KdV}}\}$} context-sensitive compatible, thereby establishing, by definition \ref{BTD}, the validity of the auto-B\"{a}cklund transform made out of equation \eqref{eqn27}, and the two equations \eqref{equationRef7} and \eqref{equationRef8}.


The contradictions with the truncations which go one order behind $\bigg($i.e. $u = \frac{12 \sigma \phi_{xx}}{\phi}-\frac{12 \sigma \phi_{x}^2}{\phi^2}\bigg)$ and one order ahead $\bigg($i.e. $u = \frac{12 \sigma \phi_{xx}}{\phi}-\frac{12 \sigma \phi_{x}^2}{\phi^2} + u_2 + u_3 \phi\bigg)$ {and the computational details in terms of the test of the existence of the Painlev\'e expansion and the context-free and context-sensitive compatibilities of the associated PDEs to test the affirmation of the non-existence of the B\"{a}cklund transforms for both the truncations have been extensively presented in appendix \ref{appendixA}.}
\subsubsection{The modified KdV equation} \label{mKdVEquationWTC}
A final example of an integrable PDE we consider is the case of the modified KdV (mKdV) equation:
\begin{equation} \label{addseteqn61At02162024}
u_t + 2 \sigma^2 u_{xxx} - 3 u^2 u_{x} = 0.
\end{equation}
Upon a Miura transformation of the form,
\begin{equation*}
v = u_x - \frac{1}{2 \sigma}u^2
\end{equation*}
one can obtain the following form of the KdV equation,
\begin{equation*}
v_t + 6 \sigma v v_x + 2 \sigma^2 v_{xxx} = 0
\end{equation*}
wherein with the rescaled coordinates $x = \xi \sqrt{\frac{\sigma}{3}}$, $\tau = 6 \sqrt{3 \sigma} t$, one can show that
\begin{equation}
v_\tau + v v_\xi + \sigma v_{\xi\xi\xi} = 0
\end{equation}
which is the original KdV equation \eqref{eqn20} in $\tau$ and $\xi$. The value of $\alpha$ from the leading-order analysis of substitution of the equation \eqref{eqn3} in the equation \eqref{addseteqn61At02162024} is $-1$ (i.e. $u = \frac{u_0}{\phi} + u_1 + u_2 \phi + u_3 \phi^2 + \cdots$) and the resultant expansion upon the substitution into equation \eqref{addseteqn61At02162024} gives the following form of equation \eqref{eqn7} $\frac{1}{\phi^4}(P_0 + P_1 \phi + P_2 \phi^2 + P_3 \phi^3 + P_4 \phi^4 + P_5 \phi^5 + \cdots) = 0$, where,
\begin{equation}
\label{mkdvu_0}
P_0 = 3 \phi_{x} u_0 (-4 \phi_{x}^2 \sigma^2 + u_0^2)  = 0.
\end{equation}
Since we are looking for an expansion with leading exponent $\alpha = -1$, $u_0\neq 0$ and the equation \eqref{mkdvu_0} has two solutions $u_0 =  \pm 2 \sigma \phi_x$. 
{Furthermore, since equation \eqref{addseteqn61At02162024} is invariant under the transformation $u \rightarrow -u$, this simply changes the sign of all the coefficients in the Painlev\'e expansion and thereby just multiplies every associated equation (which need to be used during the test of compatibility) by $-1$ on both sides. Therefore, we choose the branch $u_0 = 2 \sigma \phi_x$}, for which one can obtain the Painlev\'e expansion truncated up to $u_k, k = \beta = -\alpha = 1$,
\begin{equation} \label{addseteqn63At02162024}
    u = \frac{2 \sigma \phi_x}{\phi} + u_1
\end{equation}
with $u_k = 0, k \ge 2$ and the equations satisfied by $(u_1, \phi)$ that come from the following coefficients set to zero,
\begin{equation} \label{addseteqn64At02162024}
{P_1 = 0 \implies Q_{1, \text{mKdV}} = \phi_{xx} \sigma + \phi_{x} u_1  = 0},
\end{equation}
\begin{multline} \label{addseteqn65At02162024}
{P_2 = 0 \implies Q_{2, \text{mKdV}} = -\phi_{t} \phi_{x} - 8 \phi_{xxx} \phi_{x} \sigma^2 - 6 \phi_{xx}^2 \sigma^2} \\
- 12 \phi_{xx} \phi_{x} \sigma u_1 - 6 \phi_{x}^2 \sigma u_{1,x} + 3 \phi_{x}^2 u_1^2  = 0,
\end{multline}
\begin{multline} \label{addseteqn66At02162024}
P_3 = 0 \implies Q_{3, \text{mKdV}} = \phi_{tx} + 2 \phi_{xxxx} \sigma^2 \\
- 3 \phi_{xx} u_1^2 - 6 \phi_{x} u_1 u_{1,x}  =  \frac{\partial}{\partial x} (\phi_t + 2 \sigma^2 \phi_{xxx} - 3 \phi_x u_1^2) = 0,
\end{multline}
\begin{equation} \label{addseteqn67At02162024}
P_4 = 0 \implies Q_{4, \text{mKdV}} = u_{1,t} + 2 \sigma^2 u_{1,xxx} - 3 u_1^2 u_{1,x} = 0.
\end{equation}
One can reconcile the above four equations \eqref{addseteqn64At02162024}, \eqref{addseteqn65At02162024}, \eqref{addseteqn66At02162024} and \eqref{addseteqn67At02162024} using the following equation, 

\begin{multline} \label{addseteqn68At02162024}
\phi_x (\phi_t + 2 \sigma^2 \phi_{xxx} - 3 \phi_x u_1^2) + \frac{\partial}{\partial x} \left(6 \phi_x \left(\phi_{xx} \sigma + \phi_{x} u_1\right)\right) \\
= \phi_{t} \phi_{x} + 8 \phi_{xxx} \phi_{x} \sigma^2 + 6 \phi_{xx}^2 \sigma^2 \\
+ 12 \phi_{xx} \phi_{x} \sigma u_1 + 6 \phi_{x}^2 \sigma u_{1,x} - 3 \phi_{x}^2 u_1^2.
\end{multline}


The above reconciliation relation \eqref{addseteqn68At02162024} can be worked out upon the comparison of the coefficients of terms $- 6 \phi_{xx}^2 \sigma^2$ and $- 6 \phi_{x}^2 \sigma u_{1,x}$ in $Q_{2, \text{mKdV}}$ with their differential-algebraic counterparts in $Q_{1, \text{mKdV}}$; and then comparison of the remaining terms with the integral of $Q_{3, \text{mKdV}}$ with respect to $x$.
From the solution to the equation \eqref{addseteqn64At02162024} $u_1 = -\sigma \frac{\phi_{xx}}{\phi_x}$, we get that equation \eqref{addseteqn65At02162024} implies 
\begin{equation} \label{addseteqn69At02162024}
\phi_t + 2 \sigma^2 \phi_{xxx} - 3 \phi_x u_1^2 = 0.
\end{equation}
Equation \eqref{addseteqn68At02162024} shows that the equation \eqref{addseteqn69At02162024} {is automatically satisfied} (and thus consequently the equation \eqref{addseteqn66At02162024} as it involves the partial $x$ derivative of the RHS of the former) if equations \eqref{addseteqn64At02162024} and \eqref{addseteqn65At02162024} are. This is a consequence that the resonance condition in the Painlev\'e analysis of the mKdV appears at power 3 \cite{weiss}.
One can show that the set of partial differential polynomials {$\{Q_{1, \text{mKdV}}, Q_{2, \text{mKdV}}, Q_{3, \text{mKdV}}, Q_{4, \text{mKdV}}\}$} is context-free compatible as $0 \in g_9$ which is a consequence of the reconciliation relation \eqref{addseteqn68At02162024}.
But the set {$\{Q_{1, \text{mKdV}}, Q_{2, \text{mKdV}}, Q_{4, \text{mKdV}}\}$} is not context-free compatible (checked up to 150 recursions), which necessitates the check of context-sensitive compatibility.

From equations \eqref{addseteqn64At02162024}, \eqref{addseteqn65At02162024} and \eqref{addseteqn67At02162024}, we obtain the following expressions for the higher derivatives, following the prescribed procedure given in equations \eqref{eqnRef13}, \eqref{eqnRef14}, \eqref{eqnRef15}, \eqref{eqnRef16}, \eqref{eqnRef17} and \eqref{eqnRef18}
\begin{equation} \label{addseteqn70At02162024}
\phi_{xx} = -\frac{\phi_{x} u_1}{\sigma},
\end{equation}
\eq{ 
\phi_t& = -8 \phi_{xxx} \sigma^2 - \frac{6 \phi_{xx}^2 \sigma^2}{\phi_{x}} - 12 \phi_{xx} \sigma u_1 \\
&- 6 \phi_{x} \sigma u_{1,x} + 3 \phi_{x} u_1^2,
}{addseteqn71At02162024}
\begin{equation} \label{addseteqn72At02162024}
u_{1, t} = -2 \sigma^2 u_{1,xxx} + 3 u_1^2 u_{1,x}.
\end{equation}
We use \eqref{addseteqn70At02162024} to simplify \eqref{addseteqn71At02162024}, then differentiate the simplified equation to obtain $\phi_{xt}$; we also differentiate \eqref{addseteqn70At02162024} to obtain $\phi_{xxx}$, which leads to the expressions
\begin{equation} \label{addseteqn73At02162024}
\phi_t = 2 \phi_{x} \sigma u_{1,x} + \phi_{x} u_1^2,
\end{equation}
\begin{equation} \label{addseteqn74At02162024}
\phi_{xt} = 2 \phi_{x} \sigma u_{1,xx} - \frac{\phi_{x} u_1^3}{\sigma},
\end{equation}
\begin{equation} \label{addseteqn75At02162024}
\phi_{xxx} = - \frac{\phi_{x} u_{1,x}}{\sigma} + \frac{\phi_{x} u_1^2}{\sigma^2}.
\end{equation}
Finally with the help of equations \eqref{addseteqn72At02162024}, \eqref{addseteqn73At02162024}, \eqref{addseteqn74At02162024} and \eqref{addseteqn75At02162024}, one can prove that the difference $\frac{\partial}{\partial t} \phi_{xx} - \frac{\partial}{\partial x} \phi_{xt}$ vanishes $\bigg($ with $\phi_{txx} = \phi_{xxt} = 2 \phi_{x} \sigma u_{1,xxx} - 2 \phi_{x} u_1 u_{1,xx} - \frac{3 \phi_{x} u_1^2 u_{1,x}}{\sigma} + \frac{\phi_{x} u_1^4}{\sigma^2}\bigg)$. This renders the set of differential polynomials {$\{Q_{1, \text{mKdV}}, Q_{2, \text{mKdV}}, Q_{4, \text{mKdV}}\}$} context-sensitive compatible, thereby establishing, by definition \ref{BTD}, the validity of the auto-B\"{a}cklund transform from the equations \eqref{addseteqn63At02162024}, \eqref{addseteqn64At02162024} and \eqref{addseteqn65At02162024}.

The contradictions with the truncations which go one order behind $\bigg($i.e. $u = \frac{2 \sigma \phi_{x}}{\phi}\bigg)$ and one order ahead $\bigg($ i.e. $u = \frac{2 \sigma \phi_{x}}{\phi} - \frac{\sigma \phi_{xx}}{\phi_x} + u_2 \phi \bigg)$ and the computational details in terms of the test of the existence of the Painlev\'e expansion and the context-free and context-sensitive compatibilities of the associated PDEs to affirm the non-existence of the B\"{a}cklund transforms for both the truncations have been extensively verified by means of symbolic computation which is available in the associated ``sandbox'' programs in the open-source repository in \cite{painleveapprepo}.

\subsection{WTC And Compatibility Tests for Non-integrable PDEs} \label{wtcpainlevenonintegrablePDE}
\subsubsection{Benjamin-Bona-Mahony equation} \label{BBMEquationWTC}
The aforementioned cases were those where Painlev\'e expansion tests for known integrable PDEs give rise to valid B\"{a}cklund transforms. Now we will focus on the equation(s) for which {similar B\"{a}cklund transforms are not known to exist}. As an example, let us consider the Benjamin-Bona-Mahony (BBM) equation \cite{honechapter,bbm}:
\begin{equation} \label{NewEqnRef57}
    u_t + u u_x + u_x - u_{xxt} = 0.
\end{equation}
The exponent $\alpha$ for this equation is $-2$ ($\beta = 2$) which leads to the Painlev\'e expansion $\frac{1}{\phi^5}\left(P_0 + P_1 \phi + P_2 \phi^2 + P_3 \phi^3 + P_4 \phi^4\cdots\right)$, wherewithal we can obtain $u_k$, ($k \ge 0$ and  $\phi_x \ne 0$),
\begin{multline} \label{refeqn46}
   P_0 = -2 u_0 \phi_x (u_0 - 12 \phi_x \phi_x) = 0 \implies u_0 = 12 \phi_x \phi_t,
\end{multline}
\begin{multline} \label{refeqn47}
    P_1 = -6\phi_x(5 u_1 \phi_x \phi_t + 12 \phi_x^2 \phi_{tt} + 36 \phi_x \phi_t \phi_{xt} + 12 \phi_t^2 \phi_{xx}) \\ = 0
    \implies u_1 = \frac{-12}{5 \phi_x \phi_t}\bigg(\frac{\partial}{\partial t}(\phi_x^2 \phi_t) + \phi_t \frac{\partial}{\partial x} \phi_x \phi_t \bigg),
\end{multline}
and substituting the equations \eqref{refeqn46} and \eqref{refeqn47} for $u_0$ and $u_1$ in the equation $P_2 = 0$ gives
\begin{multline} \label{refeqn48}
    u_2 = \frac{1}{25 \phi_x^3 \phi_t^3} \bigg[14 \phi_t^2 \phi_x^3 \frac{\partial^2}{\partial t^2}(\log{\phi_x}) + 31 \phi_t^3 \phi_x^3 \frac{\partial^2}{\partial x \partial t}\\\log{(\phi_x \phi_t)} + 14 \phi_t^2 \phi_x^2 \frac{\partial}{\partial t}(\phi_t \phi_{xx}) + 11 \phi_t^4 \phi_x^2 \frac{\partial^2}{\partial x^2}(\log{\phi_x})\\ + 5 \phi_x^4 \phi_t^2 \frac{\partial^2}{\partial t^2}(\log{\phi_t}) + \frac{\partial}{\partial t}(\phi_t^3)\frac{\partial}{\partial x}(\phi_x^3) - 25 \phi_t^3 \phi_x^2 (\phi_x + \phi_t) \\
    - 6 \phi_x (\phi_{tt}^2 \phi_x^3 + \phi_t^4 \phi_{xxx}) + 14 \phi_x^3 \phi_t^2 (\phi_x-\phi_t) \frac{\partial^2}{\partial t^2}\log{\phi_x}
    \bigg],
\end{multline}
and similar {such relations for $u_k, k \ge 3$}. If we truncate the expansion by setting $u_j = 0, j \ge 3$, then all the {coefficients $P_n, n \ge 6$} vanish, provided
\begin{equation}
\label{uBBM}
u = \frac{u_0}{\phi^2} + \frac{u_1}{\phi} + u_2.
\end{equation}
For this truncation, substituting $u_0, u_1$ and $u_2$ from the equations \eqref{refeqn46}, \eqref{refeqn47} and \eqref{refeqn48} into the equation $P_3 = 0$ with some simplification implies:
\begin{multline} \label{NewRefEqn62}
    u_{0, t} + (1+u_2)u_{0, x} + u_0 u_{2, x} + 2 \phi_x u_{1, xt} + 2 u_{1,x} \phi_{xt} \\
    + \phi_t u_{1, xx} + u_{1,t}\phi_{xx} - u_1(\phi_t - u_{1,x} + (1+u_2) \phi_x - \phi_{xt}) = 0,
\end{multline}
and setting $P_4 = 0$ and $P_5 = 0$ respectively imply,
\begin{equation} \label{NewRefEqn63}
    u_{1, t} + \frac{\partial}{\partial x}(u_1(1+u_2) - u_{1, xt}) = 0,
\end{equation}
\begin{equation} \label{NewRefEqn64}
    u_{2, t} + u_{2, x} + u_2 u_{2, x} - u_{2, xxt} = 0.
\end{equation}
Notably, the equation \eqref{NewRefEqn64} does reveal an auto-B\"{a}cklund transformation but obscures within itself a certain contradiction to the generic existence of the above Painlev\'e expansion: as an observation, equations \eqref{refeqn46}, \eqref{refeqn47}, \eqref{refeqn48}, \eqref{NewRefEqn62}, \eqref{NewRefEqn63} and \eqref{NewRefEqn64} form an over-determined system of 6 equations in 4 variables $\langle u_0, u_1, u_2, \phi \rangle$. Expressing the same system only in terms of $\langle u_1, u_2, \phi\rangle$ yields,
\begin{multline} \label{addseteqn22}
P_1 = Q_{1, \text{BBM}} = 12\phi_{tt}\phi_{x}^2 + 36\phi_{tx}\phi_{t}\phi_{x} \\
+ 12\phi_{t}^2\phi_{xx} + 5\phi_{t}\phi_{x}u_1 = 0,
\end{multline}
\begin{multline} \label{addseteqn23}
P_2 = Q_{2, \text{BBM}} = 48\phi_{ttx}\phi_{x}^2 + 72\phi_{tt}\phi_{xx}\phi_{x} + 96\phi_{txx}\phi_{t}\phi_{x} \\
+ 96\phi_{tx}^2\phi_{x} + 120\phi_{tx}\phi_{t}\phi_{xx} + 8\phi_{tx}\phi_{x}u_1 \\
 + 24\phi_{t}^2\phi_{xxx} - 24\phi_{t}^2\phi_{x} + 10\phi_{t}\phi_{xx}u_1 \\
- 24\phi_{t}\phi_{x}^2u_2 - 24\phi_{t}\phi_{x}^2 + 8\phi_{t}\phi_{x} u_{1, x} - 2\phi_{x}^2 u_{1, t} - \phi_{x}u_1^2 = 0,
\end{multline}
\begin{multline} \label{addseteqn24}
P_3 = Q_{3, \text{BBM}} = 12\phi_{ttxx}\phi_{x} + 24\phi_{ttx}\phi_{xx} + 12\phi_{tt}\phi_{xxx} \\
- 12\phi_{tt}\phi_{x} + 12\phi_{txxx}\phi_{t} + 36\phi_{txx}\phi_{tx} - \phi_{txx}u_1
\\
- 12\phi_{tx}\phi_{t} - 12\phi_{tx}\phi_{x}u_2 - 12\phi_{tx}\phi_{x} \\
- 2\phi_{tx} u_{1, x} - 12\phi_{t}\phi_{xx}u_2 - 12\phi_{t}\phi_{xx} \\
- 12\phi_{t}\phi_{x}u_{2,x} + \phi_{t}u_1 - \phi_{t} u_{1, xx} - \phi_{xx} u_{1, t} \\
+ \phi_{x}u_1u_2 + \phi_{x}u_1 - 2\phi_{x} u_{1, xt} - u_1 u_{1, x} = 0,
\end{multline}
\begin{equation}  \label{addseteqn25}
P_4 = Q_{4, \text{BBM}} = u_1 u_{2,x} -  u_{1, txx} +  u_{1, t} +  u_{1, x} u_2 +  u_{1, x} = 0,
\end{equation}
\begin{equation}  \label{addseteqn26}
P_5 = Q_{5, \text{BBM}} = u_2 u_{2,x} - u_{2,txx} + u_{2,t} + u_{2,x} = 0.
\end{equation}
With systematic computation, one can verify that the set of differential polynomials 
{$\{Q_{1, \text{BBM}},$ $Q_{2, \text{BBM}},$ $Q_{3, \text{BBM}},$ $Q_{4, \text{BBM}},$ $Q_{5, \text{BBM}}\}$} is \textit{not} a context-free compatible set (checked up to 150 recursions with the ever-increasing order of the differential polynomials in the consequent differential ideal). Using equation \eqref{addseteqn22} and \eqref{addseteqn24}, we can solve $\phi_{tt}$ and $\phi_{txxx}$ and {obtain the following relations,}
\begin{equation} \label{addseteqn27}
\phi_{tt} = -3\frac{\phi_{tx}\phi_{t}}{\phi_{x}} - \frac{\phi_{t}^2\phi_{xx}}{\phi_{x}^2} - \frac{5\phi_{t}u_1}{12\phi_{x}},
\end{equation}
\begin{multline} \label{addseteqn28}
\phi_{txxx} = -\frac{\phi_{ttxx}\phi_{x}}{\phi_{t}} - \frac{2\phi_{ttx}\phi_{xx}}{\phi_{t}} - \frac{3\phi_{txx}\phi_{tx}}{\phi_{t}} \\
+ \frac{\phi_{txx}u_1}{12\phi_{t}} + \frac{3\phi_{tx}\phi_{xxx}}{\phi_{x}} - 2\phi_{tx}\\
+ \frac{\phi_{tx}\phi_{x}u_2}{\phi_{t}} + \frac{\phi_{tx}\phi_{x}}{\phi_{t}} + \frac{\phi_{tx} u_{1, x}}{6\phi_{t}} + \frac{\phi_{t}\phi_{xxx}\phi_{xx}}{\phi_{x}^2} \\
- \frac{\phi_{t}\phi_{xx}}{\phi_{x}} + \frac{5\phi_{xxx}u_1}{12\phi_{x}} + \phi_{xx}u_2 + \phi_{xx} + \frac{\phi_{x}u_{2,x}} {\phi_{t}} - \frac{u_1}{2} \\
+ \frac{ u_{1, xx}}{12} + \frac{\phi_{xx} u_{1, t}}{12\phi_{t}} - \frac{\phi_{x}u_1u_2}{12\phi_{t}} - \frac{\phi_{x}u_1}{12\phi_{t}} + \frac{\phi_{x} u_{1, xt}}{6\phi_{t}} + \frac{u_1 u_{1, x}}{12\phi_{t}}.
\end{multline}
Using equations \eqref{addseteqn25} and \eqref{addseteqn26}, we can obtain the following higher derivatives,
\begin{equation} \label{addseteqn29}
u_{1,xxt} = u_1 u_{2,x} +  u_{1, t} +  u_{1, x}u_2 +  u_{1, x},
\end{equation}
\begin{equation} \label{addseteqn30}
u_{2,xxt} = u_2 u_{2,x} + u_{2,t} + u_{2,x}.
\end{equation}
{If one observes the equation \eqref{addseteqn28}, the PDE in terms of the partial derivative term $\phi_{txxx}$ has terms $\phi_{ttx} = \frac{\partial}{\partial x}\phi_{tt}$ and $\phi_{ttxx} = \frac{\partial^2}{\partial x^2}\phi_{tt}$ which are higher derivatives of $\phi_{tt}$ that can be} {computed from the corresponding higher orders of differentiation of both sides of } {equation \eqref{addseteqn27}}. 
Using the four equations \eqref{addseteqn27}, \eqref{addseteqn28}, \eqref{addseteqn29} and \eqref{addseteqn30}, one can simplify to obtain,
\begin{multline} \label{addseteqn31}
\phi_{txxx} = - \frac{3 \phi_{txx} \phi_{tx}}{\phi_{t}} - \frac{\phi_{txx} \phi_{xx}}{\phi_{x}} - \frac{\phi_{txx} u_1}{4 \phi_{t}} - \frac{\phi_{tx}^2 \phi_{xx}}{\phi_{t} \phi_{x}}\\
- \frac{2 \phi_{tx} \phi_{xxx}}{\phi_{x}} + \frac{2 \phi_{tx} \phi_{xx}^2}{\phi_{x}^2} + \phi_{tx} - \frac{\phi_{tx} \phi_{x} u_2}{2 \phi_{t}} - \frac{\phi_{tx} \phi_{x}}{2 \phi_{t}} \\
- \frac{\phi_{tx} u_{1,x}}{2 \phi_{t}} - \frac{\phi_{t} \phi_{xxxx}}{2 \phi_{x}} + \frac{3 \phi_{t} \phi_{xxx} \phi_{xx}}{2 \phi_{x}^2} - \frac{\phi_{t} \phi_{xx}^3}{\phi_{x}^3} + \frac{\phi_{t} \phi_{xx}}{2 \phi_{x}} \\
- \frac{\phi_{xx} u_2}{2} - \frac{\phi_{xx}}{2} - \frac{\phi_{x} u_{2,x}}{2} + \frac{u_1}{4} - \frac{u_{1,xx}}{4} \\
- \frac{\phi_{xx} u_{1,t}}{24 \phi_{t}} + \frac{\phi_{x} u_1 u_2}{24 \phi_{t}} + \frac{\phi_{x} u_1}{24 \phi_{t}} - \frac{\phi_{x} u_{1,tx}}{12 \phi_{t}} - \frac{u_1 u_{1,x}}{24 \phi_{t}},
\end{multline}
\begin{multline} \label{addseteqn32}
\phi_{ttx} = \frac{-3 \phi_{txx} \phi_{t}}{\phi_{x}} - \frac{3  \phi_{tx}^2}{\phi_{x}} + \frac{\phi_{tx} \phi_{t} \phi_{xx}}{\phi_{x}^2} - \frac{5 \phi_{tx} u_1}{12 \phi_{x}} \\
- \frac{\phi_{t}^2 \phi_{xxx}}{\phi_{x}^2} + \frac{2 \phi_{t}^2 \phi_{xx}^2}{\phi_{x}^3} + \frac{5 \phi_{t} \phi_{xx} u_1}{12 \phi_{x}^2} - \frac{5 \phi_{t} u_{1,x}}{12 \phi_{x}},
\end{multline}
\begin{multline} \label{addseteqn33}
\phi_{ttxx}  = \frac{7 \phi_{txx} \phi_{t} \phi_{xx}}{\phi_{x}^2} + \frac{\phi_{txx} u_1}{3 \phi_{x}} + \frac{7 \phi_{tx}^2 \phi_{xx}}{\phi_{x}^2} \\
+ \frac{5 \phi_{tx} \phi_{t} \phi_{xxx}}{\phi_{x}^2} - \frac{4 \phi_{tx} \phi_{t} \phi_{xx}^2}{\phi_{x}^3} - \frac{3 \phi_{tx} \phi_{t}}{\phi_{x}} + \frac{5 \phi_{tx} \phi_{xx} u_1}{6 \phi_{x}^2} \\
+ \frac{3 \phi_{tx} u_2}{2} + \frac{3 \phi_{tx}}{2} + \frac{2 \phi_{tx} u_{1,x}}{3 \phi_{x}} + \frac{\phi_{t}^2 \phi_{xxxx}}{2 \phi_{x}^2} + \frac{3 \phi_{t}^2 \phi_{xxx} \phi_{xx}}{2 \phi_{x}^3} \\
- \frac{3 \phi_{t}^2 \phi_{xx}^3}{\phi_{x}^4} - \frac{3 \phi_{t}^2 \phi_{xx}}{2 \phi_{x}^2} + \frac{5 \phi_{t} \phi_{xxx} u_1}{12 \phi_{x}^2} - \frac{5 \phi_{t} \phi_{xx}^2 u_1}{6 \phi_{x}^3} + \frac{3 \phi_{t} \phi_{xx} u_2}{2 \phi_{x}}\\
+ \frac{3 \phi_{t} \phi_{xx}}{2 \phi_{x}} + \frac{5 \phi_{t} \phi_{xx} u_{1,x}}{6 \phi_{x}^2} + \frac{3 \phi_{t} u_{2,x}}{2} - \frac{3 \phi_{t} u_1}{4 \phi_{x}} \\
+ \frac{\phi_{t} u_{1,xx}}{3 \phi_{x}} + \frac{\phi_{xx} u_{1,t}}{8 \phi_{x}} - \frac{u_1 u_2}{8} - \frac{u_1}{8} + \frac{u_{1,tx}}{4} + \frac{u_1 u_{1,x}}{8 \phi_{x}}.
\end{multline}
Using the three equations \eqref{addseteqn31}, \eqref{addseteqn32} and \eqref{addseteqn33}, one can verify that the difference $\frac{\partial}{\partial x}\phi_{ttxx} - \frac{\partial}{\partial t}\phi_{txxx}$ does not identically vanish. 

This confirms that the set of differential polynomials {$\{Q_{1, \text{BBM}},$ $Q_{2, \text{BBM}},$ $Q_{3, \text{BBM}},$ $Q_{4, \text{BBM}},$ $Q_{5, \text{BBM}}\}$} is \textit{neither} context-free compatible \textit{nor} context-sensitive compatible, thus precluding the {existence} of the B\"{a}cklund transform and hence suggesting the {non-integrability} of the BBM equation. In fact, the BBM equation was shown not to be of the Painlev\'e type in \cite[Section 3.4]{PainlBBM}.

The contradictions with the truncations which go one order behind $\bigg($i.e. $u = \frac{u_0}{\phi^2}+\frac{u_1}{\phi}\bigg)$ and one order ahead $\bigg($i.e. $u = \frac{u_0}{\phi^2}+\frac{u_1}{\phi}+u_2+u_3 \phi\bigg)$ and the computational details in terms of the test of the existence of the Painlev\'e expansion and the context-free and context-sensitive compatibilities of the associated PDEs to {affirm the non-existence} of the B\"{a}cklund transforms have been extensively presented in appendix \ref{appendixA}.


\subsubsection{KdV-Burgers equation} \label{KdVBurgersEquationWTC}
As the next example of a non-integrable PDE, let us consider the case of the KdV-Burgers equation for $u : \mathbb{R}^2 \to \mathbb{R}, x, t \in \mathbb{R}$, $\sigma, \kappa \ge 0$
\begin{equation} \label{addseteqn1}
u_t + u u_x - \sigma u_{xx} + \kappa u_{xxx} = 0.
\end{equation}
The trivial cases of $\kappa = 0, \sigma \ne 0$ and $\kappa \ne 0, \sigma = 0$ lead to Burgers and KdV equations respectively (each of which have been proven to be integrable in the past). The value of $\alpha$ (for the Painlev\'e equations) for this equation is $-2$. It must be noted that the limit $\sigma \ne 0, \kappa \to 0$ gives rise to the Burgers equations where the value of $\alpha$ steps up to $-1$ (as seen in section \ref{burgersEquationWTC}). 
Importantly, this changes the complete nature of the consequent Painlev\'e expansion, 
thereby rendering this limit as singular in this context, where the steps performed below cannot
be completed. Indeed, this shall also be reflected in the compatibility results analyzed below. 
One can obtain the relevant Painlev\'e expansion in the form of,
\begin{equation*}
u = -\frac{12 \kappa \phi_x^2}{\phi^2} + \frac{12 \kappa \phi_{xx} - 12 \sigma \phi_x}{5 \phi} + u_2
\end{equation*}
such that $(u_2, \phi)$ satisfy the following equations:


\begin{multline} \label{equationRef69}
{P_2 = 0 \implies Q_{1, \text{KdVBurgers}} = -100\kappa^2\phi_{xxx}\phi_{x} + 75\kappa^2\phi_{xx}^2} \\
{- 25\kappa\phi_{t}\phi_{x} + 30\kappa\phi_{xx}\phi_{x}\sigma - 25\kappa\phi_{x}^2 u_2 + \phi_{x}^2\sigma^2  = 0},
\end{multline}
\begin{multline} \label{equationRef70}
{P_3 = 0 \implies Q_{2, \text{KdVBurgers}} = - 5 \kappa^2 \phi_{xxxx} \phi_{x} }\\
{ + 2 \kappa^2 \phi_{xxx} \phi_{xx} - 2 \kappa \phi_{tx} \phi_{x} - \kappa \phi_{t} \phi_{xx} + \frac{12 \kappa \phi_{xxx} \phi_{x} \sigma}{5}  }\\
{+ \frac{6 \kappa \phi_{xx}^2 \sigma}{5} - 3 \kappa \phi_{xx} \phi_{x} u_2 - \kappa \phi_{x}^2 u_{2, x} } \\
{+ \frac{\phi_{t} \phi_{x} \sigma}{5} 
- \frac{3\phi_{xx} \phi_{x} \sigma^2}{25} + \frac{\phi_{x}^2 \sigma u_2}{5} = 0},
\end{multline}
\begin{multline} \label{equationRef71}
{P_4 = 0 \implies Q_{3, \text{KdVBurgers}} = \kappa^2 \phi_{xxxxx} + \kappa \phi_{txx}}\\
{ - \frac{6 \kappa \phi_{xxxx} \sigma}{5} + \kappa \phi_{xxx} u_2 + \kappa \phi_{xx} u_{2,x} - \frac{\phi_{tx} \sigma}{5} + \frac{\phi_{xxx} \sigma^2}{5} }\\
{ - \frac{\phi_{xx} \sigma u_2}{5} - \frac{\phi_{x} \sigma u_{2,x}}{5} = 0},
\end{multline}
\begin{multline} \label{equationRef72}
{P_5 = 0 \implies Q_{4, \text{KdVBurgers}} = \kappa u_{2,xxx} }\\
{- \sigma u_{2,xx} + u_2 u_{2,x} + u_{2,t} = 0}.
\end{multline}
One can verify that the set {$\{Q_{1, \text{KdVBurgers}},$ $Q_{2, \text{KdVBurgers}},$ $Q_{3, \text{KdVBurgers}},$ $Q_{4, \text{KdVBurgers}}\}$ is \textit{context-free compatible} } ($0 \in g_6$ in the sense of the definition reflected in the equations \eqref{eqnRef11} and \eqref{eqnRef12}), but {the set $\{Q_{1, \text{KdVBurgers}},$ $Q_{2, \text{KdVBurgers}},$ $Q_{4, \text{KdVBurgers}}\}$ is} \textit{not} context-free compatible (checked up to 150 recursions with the ever increasing order of the differential polynomials in the consequent differential ideal). This happens as one can reconcile the four equations \eqref{equationRef69}, \eqref{equationRef70}, \eqref{equationRef71} and \eqref{equationRef72} using the following correlation, which shows that the antiderivative of equation \eqref{equationRef71} can be obtained by {the following} linear combination of equation \eqref{equationRef72} and the derivative of \eqref{equationRef70}:


\begin{multline} \label{equationRef96For04032024}
\bigg(25 \kappa u_{2, x} \phi_x^2 + 50 \kappa \phi_x \phi_{xt} + 3 \sigma^2 \phi_x \phi_{xx} - 30 \sigma \kappa \phi_{xx}^2 \\
- 5 u_2 \phi_x (\sigma \phi_x - \kappa \phi_{xx}) - 5 \phi_t (\sigma \phi_x - 5 \kappa \phi_{xx})
- 60 \sigma \kappa \phi_x \phi_{xxx} \\
- 50 \kappa^2 \phi_{xx} \phi_{xxx} + 125 \kappa^2 \phi_x \phi_{xxxx} \bigg) - \frac{\partial}{\partial x} \bigg(
75 \kappa^2 \phi_{xx}^2 + 10 \kappa (3 \sigma \phi_{xx} \\
- 10 \kappa \phi_{xxx}) + \sigma^2 \phi_x^2 - 25 \kappa u_2 \phi_x^2 - 25 \kappa \phi_x \phi_t \bigg) = 5 \phi_x (5 \kappa \phi_{xt} \\
+ \sigma^2 \phi_{xx} + 5 \kappa^2 \phi_{xxxx} + 5 \kappa u_2 \phi_{xx} - 6 \sigma \kappa \phi_{xxx} - \sigma u_2 \phi_x - \sigma \phi_x).
\end{multline}
The above reconciliation relation \eqref{equationRef96For04032024} can be worked out upon the comparison of the coefficients of terms $25 \kappa u_{2, x} \phi_x^2$ and $25 \kappa \phi_t \phi_{xx}$ in $Q_{2, \text{KdVBurgers}}$ with their differential-algebraic counterparts in $\frac{\partial}{\partial x} Q_{1, \text{KdVBurgers}}$; and then comparison of the remaining terms after subtraction with $Q_{3, \text{KdVBurgers}}$. 
From the equation \eqref{equationRef69}, one can obtain,
\begin{multline} \label{equationRef124}
\phi_t = - 4 \kappa \phi_{xxx} + \frac{3 \kappa \left(\phi_{xx}\right)^{2}}{\phi_x} + \frac{6 \sigma \phi_{xx}}{5} - u_2 \phi_x + \frac{\sigma^{2} \phi_x}{25 \kappa},
\end{multline}
\begin{multline} \label{equationRef125}
\implies \phi_{xt} = - \frac{2 \kappa \phi_{xx}\phi_{xxx}}{\phi_x} + 
\frac{\kappa \left(\phi_{xx}\right)^{3}}{\left(\phi_x\right)^{2}} + 
\frac{2 \sigma \phi_{xxx}}{15} + \frac{4 \sigma \left(\phi_{xx}\right)^{2}}{5 \phi_x} \\
- u_2 \phi_{xx} + \frac{\phi_x u_{2,x}}{3} + \frac{\sigma^{2} \phi_{xx}}{25 \kappa} + \frac{4 \sigma^{3} \phi_x}{375 \kappa^{2}}.
\end{multline}
From the equation \eqref{equationRef70}, one can obtain,
\begin{multline} \label{equationRef126}
\phi_{xxxx} = \frac{2 \phi_{xx}\phi_{xxx}}{\phi_x} - \frac{\left(\phi_{xx}\right)^{3}}{\left(\phi_x\right)^{2}} + \frac{4 \sigma \phi_{xxx}}{15 \kappa} \\
- \frac{\sigma \left(\phi_{xx}\right)^{2}}{5 \kappa \phi_x} - \frac{\phi_x u_{2,x}}{3 \kappa} - \frac{\sigma^{3} \phi_x}{375 \kappa^{3}}.
\end{multline}
From the equation \eqref{equationRef72}, one can obtain,
\begin{equation} \label{equationRef127}
u_{2, t} = - \kappa u_{2, xxx} + \sigma u_{2,xx} - u_2 u_{2,x}.
\end{equation}
Using the equations \eqref{equationRef124}, \eqref{equationRef126} and \eqref{equationRef127}, one can obtain equations for the higher derivatives $u_{2, xt}$, $\phi_{x, 5}$, $\phi_{x, 6}$ and $\phi_{x, 7}$ which help in obtaining the following time derivatives,
\eqnn{\phi_{xxt} = - \frac{2 \kappa \left(\phi_{xxx}\right)^{2}}{\phi_x} + \frac{\kappa \left(\phi_{xx}\right)^{2} \phi_{xxx}}{\left(\phi_x\right)^{2}} + \frac{4 \sigma \phi_{xx}\phi_{xxx}}{3 \phi_x} \\
- \frac{8 \sigma \left(\phi_{xx}\right)^{3}}{15 \left(\phi_x\right)^{2}} 
- u_2 \phi_{xxx} + \frac{\phi_x u_{2,xx}}{3} + \frac{17 \sigma^{2} \phi_{xxx}}{225 \kappa} \\- \frac{2 \sigma^{2} \left(\phi_{xx}\right)^{2}}{75 \kappa \phi_x} - \frac{2 \sigma \phi_x u_{2,x}}{45 \kappa} + \frac{2 \sigma^{3} \phi_{xx}}{125 \kappa^{2}} - \frac{2 \sigma^{4} \phi_x}{5625 \kappa^{3}},
}
\begin{multline}
\phi_{xxxt} = - \frac{4 \kappa \phi_{xx}\left(\phi_{xxx}\right)^{2}}{\left(\phi_x\right)^{2}} 
+ \frac{4 \kappa \left(\phi_{xx}\right)^{3} \phi_{xxx}}{\left(\phi_x\right)^{3}} \\
- \frac{\kappa \left(\phi_{xx}\right)^{5}}{\left(\phi_x\right)^{4}} + \frac{\phi_{xx}u_{2,xx}}{3} + \frac{\phi_{xxx} u_{2,x}}{3} + \cdots\\
 - \frac{17 \sigma^{2} \phi_x u_{2,x}}{675 \kappa^{2}} - \frac{22 \sigma^{4} \phi_{xx}}{5625 \kappa^{3}} + \frac{\sigma^{3} u_2 \phi_x}{375 \kappa^{3}} - \frac{17 \sigma^{5} \phi_x}{84375 \kappa^{4}}.
\end{multline}
Finally using all of the aforementioned higher derivatives, one can show that the difference of the following two {mixed} partial derivatives is given by
\begin{multline*}
\frac{\partial}{\partial t} \phi_{xxxx} - \frac{\partial}{\partial x} \phi_{xxxt} = \frac{12 \sigma \phi_{xx}\left(\phi_{xxx}\right)^{2}}{5 \left(\phi_x\right)^{2}} \\
- \frac{24 \sigma \left(\phi_{xx}\right)^{3} \phi_{xxx}}{5 \left(\phi_x\right)^{3}} + \frac{12 \sigma \left(\phi_{xx}\right)^{5}}{5 \left(\phi_x\right)^{4}} - \frac{28 \sigma^{2} \left(\phi_{xxx}\right)^{2}}{75 \kappa \phi_x} \\
+ \frac{16 \sigma^{2} \left(\phi_{xx}\right)^{2} \phi_{xxx}}{25 \kappa \left(\phi_x\right)^{2}} - \frac{6 \sigma^{2} \left(\phi_{xx}\right)^{4}}{25 \kappa \left(\phi_x\right)^{3}} - \frac{\sigma \phi_x u_{2, xxx}}{5 \kappa} \\
+ \frac{2 \sigma \phi_{xx}u_{2,xx}}{5 \kappa} + \frac{4 \sigma \phi_{xxx} u_{2,x}}{5 \kappa} - \frac{\sigma \left(\phi_{xx}\right)^{2} u_{2,x}}{\kappa \phi_x} \\
+ \frac{\sigma^{2} \phi_x u_{2,xx}}{75 \kappa^{2}} + \frac{8 \sigma^{4} \phi_{xxx}}{1875 \kappa^{3}} - \frac{4 \sigma^{4} \left(\phi_{xx}\right)^{2}}{625 \kappa^{3} \phi_x} \\
+ \frac{\sigma^{3} \phi_x u_{2,x}}{375 \kappa^{3}} + \frac{2 \sigma^{6} \phi_x}{46875 \kappa^{5}}.
\end{multline*}
So the difference does not identically vanish, which implies that the equations \eqref{equationRef69}, \eqref{equationRef70} and \eqref{equationRef72} are \textit{neither} context-free compatible \textit{nor} context-sensitive compatible; thereby arriving at a contradiction to the assumption that the B\"{a}cklund transform $u = -\frac{12 \kappa \phi_x^2}{\phi^2} + \frac{12 \kappa \phi_{xx} - 12 \sigma \phi_x}{5 \phi} + u_2$ for the KdV-Burgers equation exists. Therefore the B\"{a}cklund transform of the suggested form does not exist, thus suggesting the non-integrability of the KdV-Burgers equation. In fact, the KdV-Burgers equation is shown to not pass the Painlev\'e test in \cite[Section 2]{PainlKdVB}. But interestingly, the difference $\frac{\partial}{\partial t} \phi_{xxxx} - \frac{\partial}{\partial x} \phi_{xxxt}$ vanishes for $\sigma = 0$ which reduces the original KdV-Burgers equation to the KdV equation, which reaffirms the agreement really well. On the contrary, this difference of the two equivalent mixed partial derivatives has a pole of order 5 at $\kappa = 0$, which reflects back on the singularity in this limit as the KdV-Burgers equation reduces to the KdV equation, as has been highlighted in the beginning of this section. 



The contradictions with the truncations which go one order behind $\bigg($i.e. $u = -\frac{12 \kappa \phi_x^2}{\phi^2} + \frac{12 \kappa \phi_{xx} - 12 \sigma \phi_x}{5 \phi}\bigg)$ and one order ahead $\bigg($i.e. $u = -\frac{12 \kappa \phi_x^2}{\phi^2} + \frac{12 \kappa \phi_{xx} - 12 \sigma \phi_x}{5 \phi} + u_2 + u_3 \phi\bigg)$ and the computational details in terms of the test of the existence of the Painlev\'e expansion and the context-free and context-sensitive compatibilities of the associated PDEs to {affirm the non-existence of the B\"{a}cklund transform in each truncation have been extensively verified by means of symbolic computation which is available in the associated ``sandbox'' programs in the open-source repository in \cite{painleveapprepo}.}


\subsubsection{Quartic-interactions Klein-Gordon equation ($\phi^4$ theory)} \label{phi4TheoryWTC}
A final example of a non-integrable model is the quartic-interactions Klein-Gordon theory equation for $u, \phi, u_0, u_1: \mathbb{R}^2 \to \mathbb{C}$
\begin{equation} \label{eqnRef123From12222023}
u_{tt} = u_{xx} + 2 u - 2 u^3.
\end{equation}
The value of $\alpha$ (for the Painlev\'e equations) for this is $-1$. One can obtain the relevant Painlev\'e expansion in the form of,
\begin{equation*}
u = \frac{u_0}{\phi} + u_1
\end{equation*}
such that the functions $u_0, u_1, \phi$ satisfy the following equations:

\begin{equation} \label{addseteqn114}
{P_0 = 0 \implies Q_{1, \phi^4} = \phi_{t}^2 - \phi_{x}^2 - u_0^2  = 0},
\end{equation}
\begin{multline} \label{addseteqn115}
{P_1 = 0 \implies Q_{2, \phi^4} = \phi_{tt} u_0 + 2 \phi_{t} u_{0,t} }\\
{- \phi_{xx} u_0 - 2 \phi_{x} u_{0,x} + 6 u_0^2 u_1  = 0},
\end{multline}
\begin{equation} \label{addseteqn116}
{P_2 = 0 \implies Q_{3, \phi^4} = -6 u_0 u_1^2 + 2 u_0 + u_{0,tt} - u_{0,xx}  = 0},
\end{equation}
\begin{equation} \label{addseteqn117}
{P_3 = 0 \implies Q_{4, \phi^4} = -2 u_1^3 + 2 u_1 + u_{1,tt} - u_{1,xx} = 0}.
\end{equation}
By computation, one can verify that {the set $\{Q_{1, \phi^4},$ $Q_{2, \phi^4},$ $Q_{3, \phi^4},$ $Q_{4, \phi^4}\}$ is \textit{not} context-free compatible} (as the sequence of sets of algebraic partial differential equations following the definition in the equations \eqref{eqnRef11} and \eqref{eqnRef12} keep increasing in the order without the reduction of number of terms).
From equation \eqref{addseteqn114}, we choose, without loss of generality, the following one of the two solution branches
\begin{equation} \label{addseteqn118}
\phi_x = \sqrt{\phi_{t}^2 - u_0^2}.
\end{equation}
{We recognize that this is a limitation of our considerations herein, but believe that the essence of the relevant argument would not be modified under variable monotonicity as a function of $x$; i.e. the choice of the other branch of the solution, i.e. in this case $\phi_x = - \sqrt{\phi_{t}^2 - u_0^2}$ does not change the final outcome from the consequent context-sensitive compatibility test for that branch.} Then, differentiating equation \eqref{addseteqn118}
 with respect to $x$ we get:
\begin{equation} \label{addseteqn119}
\phi_{xx} = \frac{\phi_{tx} \phi_{t} - u_0 u_{0, x}}{\sqrt{\phi_{t}^2 - u_0^2}}.
\end{equation}
Finally, differentiating equation {\eqref{addseteqn118}} 
with respect to $t$, we get,
\begin{equation} \label{addseteqn120}
\phi_{xt} = \frac{\phi_{tt} \phi_{t} - u_0 u_{0, t}}{\sqrt{\phi_{t}^2 - u_0^2}}.
\end{equation}
Solving equation \eqref{addseteqn115} and substituting \eqref{addseteqn118} and \eqref{addseteqn119} , we obtain:
\begin{equation} \label{addseteqn121}
\phi_{tt} = \frac{\phi_{tx} \phi_{t} - u_0 u_{0, x}}{\sqrt{\phi_{t}^2 - u_0^2}} - \frac{2 \phi_{t} u_{0,t}}{u_0} - 6 u_0 u_1 + \frac{2 u_{0,x}\sqrt{\phi_{t}^2 - u_0^2}}{u_0}.
\end{equation}
Solving equations \eqref{addseteqn120} and \eqref{addseteqn121} for the variables $\phi_{xt}$ and $\phi_{tt}$ we have:
\begin{multline*}
\phi_{xt} = \frac{2 \phi_{t}^4 u_{0,t}}{u_0^3 \sqrt{\phi_{t}^2 - u_0^2}} + \frac{6 \phi_{t}^3 u_1}{u_0 \sqrt{\phi_{t}^2 - u_0^2}} - \frac{2 \phi_{t}^3 u_{0,x}}{u_0^3}
\end{multline*}
\begin{equation} \label{addseteqn122}
 - \frac{\phi_{t}^2 u_{0,t}}{u_0 \sqrt{\phi_{t}^2 - u_0^2}} - \frac{6 \phi_{t} u_0 u_1}{\sqrt{\phi_{t}^2 - u_0^2}} + \frac{3 \phi_{t} u_{0,x}}{u_0} - \frac{u_0 u_{0,t}}{\sqrt{\phi_{t}^2 - u_0^2}},
\end{equation}
\begin{multline} \label{addseteqn123}
\phi_{tt} = \frac{-2 \phi_{t}^4 u_{0,x}}{u_0^3 \sqrt{\phi_{t}^2 - u_0^2}} + \frac{2 \phi_{t}^3 u_{0,t}}{u_0^3} + \frac{5 \phi_{t}^2 u_{0,x}}{u_0 \sqrt{\phi_{t}^2 - u_0^2}} \\
+ \frac{6 \phi_{t}^2 u_1}{u_0} - \frac{\phi_{t} u_{0,t}}{u_0}  - \frac{3 u_0 u_{0,x}}{\sqrt{\phi_{t}^2 - u_0^2}} - 6 u_0 u_1.
\end{multline}
Then, substituting equation \eqref{addseteqn122} in the equation \eqref{addseteqn119},
\begin{multline*}
\phi_{xx} = \frac{2 \phi_{t}^5 u_{0,t}}{\phi_{t}^2 u_0^3 - u_0^5} + \frac{6 \phi_{t}^4 u_1}{\phi_{t}^2 u_0 - u_0^3} - \frac{2 \phi_{t}^4 u_{0,x}}{u_0^3 \sqrt{\phi_{t}^2 - u_0^2}} - \frac{\phi_{t}^3 u_{0,t}}{\phi_{t}^2 u_0 - u_0^3}
\end{multline*}
\begin{multline} \label{addseteqn124}
- \frac{6 \phi_{t}^2 u_0 u_1}{\phi_{t}^2 - u_0^2}+ \frac{3 \phi_{t}^2 u_{0,x}}{u_0 \sqrt{\phi_{t}^2 - u_0^2}} - \frac{\phi_{t} u_0 u_{0,t}}{\phi_{t}^2 - u_0^2} - \frac{u_0 u_{0,x}}{\sqrt{\phi_{t}^2 - u_0^2}}.
\end{multline}
Also from the equations \eqref{addseteqn116} and \eqref{addseteqn117}, we obtain $u_{0, tt}$ and $u_{1, tt}$ as follows,
\begin{equation} \label{addseteqn125}
u_{0, tt} = 6 u_0 u_1^2 - 2 u_0 + u_{0,xx},
\end{equation}
\begin{equation} \label{addseteqn126}
u_{1, tt} = 2 u_1^3 - 2 u_1 + u_{1,xx}.
\end{equation}


One can verify by systematic computation the differences $\frac{\partial}{\partial t} \phi_{xt} - \frac{\partial}{\partial x} \phi_{tt}$ and $\frac{\partial}{\partial x} \phi_{xt} - \frac{\partial}{\partial t} \phi_{xx}$ do not identically vanish, which implies that {the set $\{Q_{1, \phi^4},$ $Q_{2, \phi^4},$ $Q_{3, \phi^4},$ $Q_{4, \phi^4}\}$ is  \textit{not} context-sensitive compatible. Therefore we arrive at a contradiction to the  assumption that the expansion $u = \frac{u_0}{\phi} + u_1$ is a valid B\"{a}cklund transformation. Thus, there does not exist a B\"{a}cklund transformation in the functions $\langle u_0, u_1, \phi \rangle$ for this level of truncation of the Painlev\'e expansion, thereby suggesting the non-integrability of the $\phi^4$ or quartic-interactions Klein-Gordon theory equation. In fact, as a consequence of  \cite[Theorem 4.4 and paragraph below]{PainlBBM}, the quartic-interactions Klein-Gordon theory equation is not Painlev\'e-integrable.




{If we truncate one order earlier with $u_j = 0, j \ge 1$, one obtains,
\begin{equation} \label{eqnRef123From08032024}
u = \frac{u_0}{\phi}
\end{equation}
such that the functions $\langle u_0, \phi \rangle$ satisfy the following equations:
\begin{equation} \label{eqnRef124From08032024}
P_0 = 0 \implies Q_{1, \phi^4, +} = \phi_{t}^2 - \phi_{x}^2 - u_0^2  = 0,
\end{equation}
\begin{multline} \label{eqnRef125From08032024}
P_1 = 0 \implies Q_{2, \phi^4, +} = \phi_{tt} u_0 + 2 \phi_{t} u_{0,t} \\
- \phi_{xx} u_0 - 2 \phi_{x} u_{0,x} = 0,
\end{multline}
\begin{equation} \label{eqnRef126From08032024}
P_2 = 0 \implies Q_{3, \phi^4, +} =  2 u_0 + u_{0,tt} - u_{0,xx}  = 0.
\end{equation}
By computation, one can verify that the set $\{Q_{1, \phi^4, +},$ $Q_{2, \phi^4, +},$ $Q_{3, \phi^4, +}\}$ is \textit{not} context-free compatible (as the sequence of sets of algebraic partial differential equations following the definition in the equations \eqref{eqnRef11} and \eqref{eqnRef12} keep increasing in the order without the reduction of number of terms).
From equation 
\eqref{eqnRef124From08032024}, we choose, once again, the following one of the two solution branches
\begin{equation} \label{eqnRef127From08032024}
\phi_x = \sqrt{\phi_{t}^2 - u_0^2}.
\end{equation}
{It must be noted that, similar to the earlier case of truncation, the choice of the other branch of the solution $\phi_x = - \sqrt{\phi_{t}^2 - u_0^2}$ does not change the final outcome from the consequent context-sensitive compatibility test for that branch.} Solving equations \eqref{eqnRef125From08032024} and \eqref{eqnRef126From08032024} for the terms $\phi_{tt}$ and $u_{0, tt}$ respectively we get,
\begin{equation} \label{eqnRef128From09032024}
    \phi_{tt} = \frac{-2 \phi_{t} u_{0,t}}{u_0} + \phi_{xx} + \frac{2 \phi_{x} u_{0,x}}{u_0},
\end{equation}
\begin{equation} \label{eqnRef129From09032024}
    u_{0,tt} = -2 u_0 + u_{0,xx}.
\end{equation}
The equation \eqref{eqnRef128From09032024} for $\phi_{tt}$ has the term $\phi_{xx}$ which can be borrowed from the differentiation of both sides of the equation \eqref{eqnRef127From08032024} with respect to $x$ as follows,
\begin{equation} \label{eqnRef130From09032024}
\phi_{xx} = \frac{\phi_{tx} \phi_{t}}{\sqrt{\phi_{t}^2 - u_0^2}} - \frac{u_0 u_{0,x}}{\sqrt{\phi_{t}^2 - u_0^2}}.
\end{equation}
The equation \eqref{eqnRef130From09032024} for $\phi_{xx}$ has the term $\phi_{tx}$ which can be borrowed from the differentiation of both sides of the equation \eqref{eqnRef127From08032024} with respect to $t$ as follows,
\begin{equation} \label{eqnRef131From09032024}
\phi_{tx} = \frac{\phi_{tt} \phi_{t}}{\sqrt{\phi_{t}^2 - u_0^2}} - \frac{u_0 u_{0,t}}{\sqrt{\phi_{t}^2 - u_0^2}}.
\end{equation}
Additionally the equation \eqref{eqnRef128From09032024} contains a term with $\phi_x$ which can be substituted from the equation \eqref{eqnRef127From08032024} to get
\begin{equation} \label{eqnRef132From09032024}
    \phi_{tt} = -\frac{2 \phi_{t} u_{0,t}}{u_0} + \phi_{xx} + \frac{2 u_{0,x} \sqrt{\phi_{t}^2 - u_0^2}}{u_0},
\end{equation}
Visibly so, the equations \eqref{eqnRef130From09032024}, \eqref{eqnRef131From09032024} and \eqref{eqnRef132From09032024} form a linear system in the terms $\phi_{xx}$, $\phi_{tx}$ and $\phi_{tt}$ which can be solved to obtain the following solutions,
\begin{multline} \label{eqnRef133From09032024}
    \phi_{xx} = \frac{2 \phi_{t}^3 u_{0,t}}{u_0^3} - \frac{2 \phi_{t}^2 u_{0,x} \sqrt{\phi_{t}^2 - u_0^2}}{u_0^3} \\ + \frac{\phi_{t} u_{0,t}}{u_0} + \frac{u_{0,x} \sqrt{\phi_{t}^2 - u_0^2}}{u_0}
\end{multline}
\begin{multline} \label{eqnRef134From09032024}
    \phi_{xt} = \frac{2 \phi_{t}^4 u_{0,t}}{u_0^3 \sqrt{\phi_{t}^2 - u_0^2}} - \frac{2 \phi_{t}^3 u_{0,x}}{u_0^3} - \frac{\phi_{t}^2 u_{0,t}}{u_0 \sqrt{\phi_{t}^2 - u_0^2}} \\ + \frac{3 \phi_{t} u_{0,x}}{u_0} - \frac{u_0 u_{0,t}}{\sqrt{\phi_{t}^2 - u_0^2}}
\end{multline}
\begin{multline} \label{eqnRef135From09032024}
    \phi_{tt} = \frac{2 \phi_{t}^3 u_{0,t}}{u_0^3} - \frac{2 \phi_{t}^2 u_{0,x} \sqrt{\phi_{t}^2 - u_0^2}}{u_0^3} \\
    - \frac{\phi_{t} u_{0,t}}{u_0} + \frac{3 u_{0,x} \sqrt{\phi_{t}^2 - u_0^2}}{u_0}.
\end{multline}
One can verify by systematic computation the differences $\frac{\partial}{\partial t} \phi_{xt} - \frac{\partial}{\partial x} \phi_{tt}$ and $\frac{\partial}{\partial x} \phi_{xt} - \frac{\partial}{\partial t} \phi_{xx}$ do not identically vanish, which implies that the set $\{Q_{1, \phi^4, +},$ $Q_{2, \phi^4, +},$ $Q_{3, \phi^4, +}\}$ is \textit{not} context-sensitive compatible. Therefore we arrive at a contradiction to the  assumption that the expansion $u = \frac{u_0}{\phi}$ is a valid B\"{a}cklund transformation. Thus, there does not exist a B\"{a}cklund transformation in the functions $\langle u_0, \phi \rangle$ for this level of truncation of the Painlev\'e expansion.
}

{Therefore as a final continuation of the proof-of-concept, let us take one order ahead of the original case, i.e. we now set $u_j = 0, j \ge 3$ such that the relevant Painlev\'e expansion reads,
\begin{equation} \label{eqnRef116From03032024}
u = \frac{u_0}{\phi} + u_1 + u_2 \phi.
\end{equation}
}
By order-by-order comparison and reading-off of the coefficients with the truncation up to $u_2$ as mentioned above, one can verify that $P_6 = 0 \implies u_2 = 0$ wherein we affinely fall back to the case with the expansion $u = \frac{u_0}{\phi} + u_1$ where we noticed the expansion to lead to an overdetermined yet context-sensitive \textit{incompatible} system of PDEs {that leads to a contradiction in connection to the existence of a B\"{a}cklund transform \eqref{eqnRef116From03032024}.}

There is a host of symbolically verified proofs for the Painlev{\'e} test for the various PDEs explored in this section proven in {separate ``sandbox'' programs} with the algorithm explained in section \ref{algorithm} \cite{painleveapprepo}. With the above cases, one can observe that for a PDE or a system of PDEs that satisfy the Painlev\'e property with the exponent $\alpha$, we can obtain an auto-B{\"a}cklund transformation for a Painlev\'e expansion truncated until the term with the exponent $m = -\alpha$. In section \ref{algorithm}, an algorithm based on this truncation is described and explained (along with the procedural and implementational details for the Sympy-based Kivy app that performs the WTC test \cite{weiss} and obtains the consequent resulting B{\"a}cklund transforms).



\section{The algorithm and implementational details} \label{algorithm}
Motivated by the above examples, we now consider the development of an algorithm with the following modular strategy: since our cases of interest pertain to only integer values of $\beta = 1, 2, 3, \cdots $, one can check for a given $\beta$ for the PDE if an appropriate B\"acklund transform with the Painlev\'e expansion $u = \sum_{m=0}^{\beta-1} \frac{u_m \phi^m}{\phi^{\beta}} + u_{\beta}$ works {(where $\beta = -\alpha$ cf. equations \eqref{eqn6} and \eqref{eqn7})}. If the expansion is unsuccessful, one can check for the same with the successor of $\beta$; and this step is repeated until the actual value of $\beta$ for the PDE is reached and the corresponding B\"acklund transformation emerges as a byproduct. The details of obtaining the B\"acklund transformation involve the substitution of the Painlev\'e expansion $u = \sum_{m=0}^{\beta} u_m \phi^{m-\beta}$ into the equation \eqref{eqn4}, reading off the coefficients $P_k$, and setting them equal to zero to obtain the transformation; the part of the problem reformulation is taken care by algorithm \ref{precursorAlgorithm}. The iterative part is taken care of by the algorithm \ref{formatorAlgorithm}, which uses both the context-free and context-sensitive forms of differential commutation compatibility described by algorithms \ref{ContextFreeDifferentialCommutationCompatibility} and \ref{ContextSensitiveDifferentialCommutationCompatibility} respectively. The overall complexity of this scheme of construction can be shown to be $O\bigg(|\mu_{min}| 
 N_{\text{context-sensitive}}(N_{\text{solution-loop}}+|\mu_{min}|)D^2 + N_{\text{context-free}}|\mu_{min}|^2\bigg)$ {(c.f. equation \eqref{eqn7})} where $D$ is the order of the input PDE, 
$N_{\text{solution-loop}}$ is the number of loops taken in algorithm \ref{ContextSensitiveDifferentialCommutationCompatibility} from the lines 60-87 to compute the required and independent solutions to all the higher derivatives for context-sensitive compatibility, and $N_{\text{context-free}}$ and $N_{\text{context-sensitive}}$ are the number of recursions in the context-free and context-sensitive algorithms (algorithms \ref{ContextFreeDifferentialCommutationCompatibility} and \ref{ContextSensitiveDifferentialCommutationCompatibility}) respectively; and under the assumption that the power-balancing leads to a relation such that $|\mu_{min}| = O(|\alpha|)$, then the overall complexity is $O\bigg(N_{\text{context-sensitive}}N_{\text{solution-loop}}|\alpha|D^2 + N_{\text{context-sensitive}} |\alpha|^2 D^2 + N_{\text{context-free}}|\alpha|^2\bigg)$. 
{It should be noted by the reader that the numbers of recursions $N_{\text{context-free}}$, $N_{\text{context-sensitive}}$ and $N_{\text{solution-loop}}$ may significantly 
grow and, in principle, diverge particularly in the case of non-integrability equations such as the BBM, KdV-Burgers, and quartic-interaction Klein-Gordon equations explored in sections \ref{BBMEquationWTC}, \ref{KdVBurgersEquationWTC} and \ref{phi4TheoryWTC} respectively {\cite{PainlBBM,PainlKdVB}}.


 



 
\begin{algorithm}[H]
\caption{TransformExistence: The test of the existence of a B\"acklund transformation for a given $\alpha$}\label{precursorAlgorithm}
\begin{algorithmic}[1]
\Require \text{The input PDE } $F(\partial^{\vec{\gamma}} u) = 0$, \text{ and the Painlev\'e exponent } $\beta$
\Ensure $S_{\text{backlundTransform}}, S_{\text{conditions}}$ \Comment{$S_{\text{backlundTransform}}$ is the set of equations behind the B\"acklund transforms and $S_{\text{conditions}}$ is the set of compatibility conditions associated with the same.}
    \State $\beta \gets -\alpha$
    \State $u \gets \sum_{m=0}^{\beta-1} \frac{u_m \phi^m}{\phi^{\beta}} + u_{\beta}$
    \State $ P_0 + P_1 \phi + P_2 \phi^2 + P_3 \phi^3 + \cdots P_{\mu_{min}} \phi^{\mu_{min}} \gets \phi^{\mu_{min}} F(\partial^{\vec{\gamma}} u)$
    \State $k \gets 0$
    \While{$k < \mu_{min}$}
        \If{$k = 0$}
            \State $U_k \gets Solve(P_k = 0, U_k)$
        \ElsIf{$k \le \beta$}
            \State $U_k \gets Solve(P_k.\text{subs}(U_0, U_1, \cdots, U_{k-1}) = 0, U_k)$
        \Else
            \State $U_k \gets 0$
        \EndIf
        \State $k \gets k + 1$
    \EndWhile
    \State $k \gets 0$
    \While{$k < \mu_{min}$}
        \If{$k < \beta$}
            \State $S_{\text{backlundTransform}} \gets S_{\text{backlundTransform}} \cup \{u_k = U_k\}$
        \ElsIf{ $\beta < k < \mu_{min}$ \text{and $P_k$
        is not identically 0} }
            \State $S_{\text{conditions}} \gets S_{\text{conditions}} \cup \{P_k = 0\}$
        \EndIf
    \EndWhile
\end{algorithmic}
\end{algorithm}

\begin{algorithm}[H]
\caption{ContextFreeCompatibility: To solve the decision problem with a set of PDEs being context-free compatible}\label{ContextFreeDifferentialCommutationCompatibility}
\begin{algorithmic}[1]
\Require \text{A set of} $m$ \text{differential polynomials} $f_1$, $f_2$, ..., $f_m$ $\in R\{X_1, \cdots , X_r\}$, $R[\delta_x^{n_1} \delta_t^{n_2} X_i:$ $i = 1, ... , r$, $n_1, n_2, \in \mathbb{N}]$ \text{is the differential ring of the partial differential} \text{polynomials in the differential indeterminates} $\{X_1, \cdots, X_r\}$ \text{with coefficients in the differential ring} $R$, $\phi \in \{X_1, \cdots, X_r\}$
\Ensure $\text{ContextFreeCompatible}(\{f_1, f_2, \cdots, f_m\})$
    \State $k \gets 1$
    \While{$ k \le m$}
        \State \text{Express } $f_k = 0$ \text{as } 
        \begin{equation*}
            f_k = 0 \implies \delta_x^{\alpha_k} \delta_t^{\beta_k} \phi = F_k(\phi, \cdots, \delta_x^{\alpha_k} \delta_t^{\beta_k-1} \phi, \delta_x^{\alpha_k-1} \delta_t^{\beta_k} \phi)
        \end{equation*}
        \text{where $\alpha_k+\beta_k = ord(f_k)$}
        \State $k \gets k + 1$
    \EndWhile
    \State $T \gets \{\}$
    \State $k \gets 1$
    \While{$ k \le m-1$}
        \State $T \gets T \cup \{\delta_x^{\alpha_{k+1}} \delta_t^{\beta_{k+1}} F_k - \delta_x^{\alpha_k} \delta_t^{\beta_k} F_{k+1}\}$
        \State $k \gets k + 1$
    \EndWhile
    \State $T \gets T \cup \{\delta_x^{\alpha_{1}} \delta_t^{\beta_{1}} F_m - \delta_x^{\alpha_m} \delta_t^{\beta_m} F_{1}\}$
    \If{$0 \in T$}
        \State \text{return True}
    \Else
        \State \text{return} ContextFreeCompatibility(T, $\phi$)
    \EndIf
\end{algorithmic}
\end{algorithm}

\begin{algorithm}[H]
\caption{ContextSensitiveCompatibility: To solve the decision problem with a set of PDEs being context-sensitive compatible}\label{ContextSensitiveDifferentialCommutationCompatibility}
\begin{algorithmic}[1]
\Require \text{A set of} $\kappa+n$ \text{differential polynomials} $p_1$, $p_2$, $\cdots$, $p_n$ $q_1$, $q_2$, $\cdots$, $q_{\kappa}$ $\in R\{\phi, u_1, u_2, \cdots , u_\kappa\}$, $R[\delta_x^{n_1} \delta_t^{n_2} X_i, X_i \in \{\phi, u_1, u_2, \cdots , u_\kappa\}, n_1, n_2, \in \mathbb{N}]$ \text{is the differential ring of the partial differential polynomials in the} \text{differential indeterminates $\{\phi, u_1, u_2, \cdots , u_\kappa\}$ with coefficients} \text{ in the differential ring} $R$
\Ensure $\text{ContextSensitiveCompatible}(\{p_1, p_2, p_3, \cdots, p_n, $
$q_1, q_2, q_3, \cdots, q_{\kappa}\})$
    \State $S_{equations}, S_{variables} \gets \{\}, \{\}$
    \State $k \gets 1$
    \While{$k \le \kappa$}
        \State
        \begin{multline*}
        q_k = 0 \implies \delta_x^{C_k} \delta_t^{D_k} u_k = g_k(\phi, \phi_x, \phi_t, \cdots, \delta_x^{A_{max}-1} \delta_t^{B_{max}} \phi,\\
        \delta_x^{A_{max}} \delta_t^{B_{max}} \phi, u_1, u_2, \cdots, u_\kappa, \cdots, \delta_x^{C_{\kappa}}\delta_t^{D_{\kappa}-1} u_{\kappa})
        \end{multline*}
        \State $S_{\text{equations}} \gets S_{\text{equations}} \cup \{\delta_x^{C_k} \delta_t^{D_k} u_k = g_k\}$
        \State $S_{\text{variables}} \gets S_{\text{variables}} \cup \{\delta_x^{C_k} \delta_t^{D_k} u_k\}$
        \State $k \gets k + 1$
    \EndWhile

    \State $k \gets 1$
    \While{$k \le n$}
        \State
        \begin{multline*}
        p_k = 0 \implies \delta_x^{A_k} \delta_t^{B_k} \phi = f_k(\phi, \phi_x, \phi_t, \cdots, \delta_x^{A_k} \delta_t^{B_k-1} \phi,\\
        \delta_x^{A_k-1} \delta_t^{B_k} \phi, u_1, u_2, \cdots, u_\kappa, \cdots, \delta_x^{C_{\kappa}}\delta_t^{D_{\kappa}-1} u_{\kappa})
        \end{multline*}
        \State $S_{\text{equations}} \gets S_{\text{equations}} \cup \{\delta_x^{A_k} \delta_t^{B_k} \phi = f_k\}$
        \State $S_{\text{variables}} \gets S_{\text{variables}} \cup \{\delta_x^{A_k} \delta_t^{B_k} \phi\}$
        \State $k \gets k + 1$
    \EndWhile
    \If{\begin{multline*}
        \exists M, N, u_1, v_1, \delta_x^{A_M}\delta_t^{B_M} \phi = f_M, f_M \in R[\delta_x^{A_M+u_1} \delta_t^{B_N+v_1} \phi]
    \end{multline*}}
        \State $T_{\text{equations}} \gets \{\delta_x^{A_M} \delta_t^{B_M} \phi = f_M, \delta_x^{A_N} \delta_t^{B_N} \phi = f_N\}$
        \State $T_{\text{equations}} \gets T_{\text{equations}} \cup \{\delta_x^{A_N + u_1} \delta_t^{B_N + v_1} \phi = \delta_x^{u_1} \delta_t^{v_1}f_N\}$
        \State $T_{\text{variables}} \gets \{\delta_x^{A_M} \delta_t^{B_M} \phi, \delta_x^{A_N} \delta_t^{B_N} \phi, \delta_x^{A_N + u_1} \delta_t^{B_N + v_1} \phi\}$
        \State $T_{\text{solutions}} \gets \text{Solve}(T_{\text{equations}}, T_{\text{variables}})$
        \State $S_{\text{equations}} \gets S_{\text{equations}} \backslash \{\delta_x^{A_M}\delta_t^{B_M} \phi = f_M, \delta_x^{A_N} \delta_t^{B_N} \phi = f_N\}$
        \State $S_{\text{equations}} \gets S_{\text{equations}} \cup \{\delta_x^{A_M}\delta_t^{B_M} \phi = T_{\text{solutions}}[\delta_x^{A_M}\delta_t^{B_M}\phi]\}$
        \State $S_{\text{equations}} \gets S_{\text{equations}} \cup \{\delta_x^{A_N}\delta_t^{B_N} \phi = T_{\text{solutions}}[\delta_x^{A_N}\delta_t^{B_N}\phi]\}$
        \State $S_{\text{equations}} \gets S_{\text{equations}} \cup \{\delta_x^{A_N+u_1}\delta_t^{B_N+v_1} \phi = T_{\text{solutions}}[\delta_x^{A_N+u_1}\delta_t^{B_N+v_1}\phi]\}$
        \State \text{return ContextSensitiveCompatibility}$(S_{\text{equations}} \cup$ $\{q_1, q_2, q_3, \cdots, q_{\kappa}\}, \{\phi, u_1, u_2, \cdots , u_\kappa\})$
    \EndIf
    \State $A_{min} \gets min\{A_1, A_2, \cdots, A_n\}$
    \State $B_{min} \gets min\{B_1, B_2, \cdots, B_n\}$
    \State $A_{max} \gets max\{A_1, A_2, \cdots, A_n\}$
    \State $B_{max} \gets max\{B_1, B_2, \cdots, B_n\}$

    \State $v \gets 1$
    \While{$v \le \kappa$}
        \State $i \gets 1$
        \While{$i \le A_{max}-A_{min}$}
            \State $j \gets 1$
\algstore{ContextSensitiveDifferentialCommutationCompatibilityFirst}
\end{algorithmic}
\end{algorithm}

\begin{algorithm}[H]                   
\begin{algorithmic}
\algrestore{ContextSensitiveDifferentialCommutationCompatibilityFirst}
            \While{$j \le B_{max}-B_{min}$}
                \State $M_{i, j} \gets \{\delta_x^{i+C_v} \delta_t^{j+D_v} u_v  = \delta_x^{i} \delta_t^{j} g_v\}$
                \State $S_{\text{equations}} \gets S_{\text{equations}} \cup M_{i, j}$
                \State $S_{\text{variables}} \gets S_{\text{variables}} \cup \{\delta_x^{i+C_v} \delta_t^{j+D_v} u_v\}$
                \State $j \gets j + 1$
            \EndWhile
            \State $i \gets i + 1$
        \EndWhile
        \State $v \gets v + 1$
    \EndWhile

    \State $w \gets 1$
    \While{$w \le n$}
        \State $\text{derivsInX} \gets 1$
        \While{$\text{derivsInX} \le A_{max}-A_w+1$}
            \State $\text{derivsInT} \gets 1$
            \While{$\text{derivsInT} \le B_{max}-B_w+1$}
                \State \begin{multline*}
                    S_{\text{equations}} \gets S_{\text{equations}} \cup \{\delta_x^{A_w+\text{derivsInX}} \delta_t^{B_w+\text{derivsInT}} \phi \\
                    = \delta_x^{\text{derivsInX}} \delta_t^{\text{derivsInT}} f_w\}
                \end{multline*}
                \State $S_{\text{variables}}$ $\gets$ $S_{\text{variables}}$ $\cup$ $\{\delta_x^{A_w+\text{derivsInX}} \delta_t^{B_w+\text{derivsInT}} \phi\}$
                \State $\text{derivsInT} \gets \text{derivsInT} + 1$
            \EndWhile
            \State $\text{derivsInX} \gets \text{derivsInX} + 1$
        \EndWhile
        \State $w \gets w + 1$
    \EndWhile
    \Repeat
        \State $flag_1 \gets \text{True}$
        \Repeat
        \ForEach{$v \in S_{\text{variables}}$}
            \State \text{Find } $f_v(\phi, u_1, u_2, \cdots, u_\kappa)$
            \State \text{such that } $\{v = f_v\} \subset S_{\text{equations}}$
            \State $f_{v, \text{old}} \gets f_v$
            \Repeat
            \ForEach{$w \in S_{\text{variables}}$}
                \State \text{Find } $f_w(\phi, u_1, u_2, \cdots, u_\kappa)$
                \State \text{such that } $\{w = f_w\} \subset S_{\text{equations}}$
                \If{$v \ne w$}
                    \State $f_v \gets f_v.\text{substitute}(w, f_w)$
                \EndIf
            \EndFor
            \State $flag_1 \gets flag_1 \land f_{v, \text{old}} - f_v == 0$
            \If{$f_{v, \text{old}} - f_v \ne 0$}
                \State $S_{\text{equations}} \gets S_{\text{equations}} \backslash\{v=f_{v, \text{old}}\}$
                \State $S_{\text{equations}} \gets S_{\text{equations}} \cup \{v = f_v\}$
            \EndIf
            \Until{$w \in S_{\text{variables}}$}
        \EndFor
        \Until{$v \in S_{\text{variables}}$}
        \If{$flag_1$}
            \State $\textbf{break}$
        \EndIf
    \Until{True}

    \State $flag_2 \gets \text{True}$
    \State $k \gets 1$
    \While{$k \le n-1$}
        \State \begin{multline*}
            W_k \gets \delta_x^{A_{max}-A_k}\delta_t^{B_{max}-B_k}f_k - \\
            \delta_x^{A_{max}-A_{k+1}}\delta_t^{B_{max}-B_{k+1}}f_{k+1}
        \end{multline*}
        \State $W_k \gets W_k.\text{substitute}(S_{\text{variables}}, S_{\text{solutions}})$
\algstore{ContextSensitiveDifferentialCommutationCompatibilitySecond}
\end{algorithmic}
\end{algorithm}

\begin{algorithm}[H]            
\begin{algorithmic}
\algrestore{ContextSensitiveDifferentialCommutationCompatibilitySecond}
        \State $flag_2 \gets flag_2 \land W_k == 0$
        \State $k \gets k + 1$
    \EndWhile
    \State $W_n \gets \delta_x^{A_{max}-A_n}\delta_t^{B_{max}-B_n}f_n - \delta_x^{A_{max}-A_{1}}\delta_t^{B_{max}-B_{1}}f_{1}$
    \State $W_n \gets W_n.\text{substitute}(S_{\text{variables}}, S_{\text{solutions}})$
    \State $k \gets 1$
    \While{$k \le n$}
        \State $flag_2 \gets flag_2 \land W_k == 0$
    \EndWhile
    \State \text{return } $flag_2$
\end{algorithmic}
\end{algorithm}

\begin{algorithm}[H]
\caption{PainleveBacklundTransformCheck: The iterative test of the existence of a Painlev\'e expansion and a B\"acklund transformation}\label{formatorAlgorithm}
\begin{algorithmic}[1]
\Require \text{Only the input PDE } $F(\partial^{\vec{\gamma}} u) = 0$
\Ensure $S_{\text{backlundTransform}}, S_{\text{additionalConditions}}$
    \State $\alpha, \text{flag } \gets -1, \text{True}$
    \While{flag is True}
         \State $S_{\text{backlundTransform}}, S_{\text{conditions}} \gets \text{TransformExistence}(F, \alpha)$
        \If{$P_{-\alpha}.\text{subs}(u_{-\alpha} = u) = F(\partial^{\vec{\gamma}} u)$}
            \State $\text{flag} \gets \text{False}$
            \State \textbf{break}
        \Else
            \State $\alpha \gets \alpha - 1$
            \State \textbf{continue}
        \EndIf
    \EndWhile
    \If{$\exists N \in \mathbb{N}, \forall -\alpha > N, \forall k < -\alpha, U_k = 0$}
        \State \text{Return that the equation IS NOT a good candidate} 
        \State \text{for integrability as it fails the WTC test.}
    \EndIf
    \Repeat
    \ForEach{ $f \in S_{\text{additionalConditions}}$}
        \If{$\frac{\delta f}{\delta u_{-\alpha}} = 0$}
            \State \text{Return that the equation IS NOT a good candidate for} 
            \State \text{integrability as it fails the WTC test.}
        \EndIf
    \EndFor
    \Until{$\frac{\mathrm{\delta} f}{\mathrm{\delta} u_{-\alpha}} = 0$}
    \State $k \gets 0$
    \While{$k \le -\alpha - 1$}
        \State $m \gets 0$
        \While{$m \le k$}
            \Repeat
            \ForEach{ $f_1 \in S_{\text{additionalConditions}}$}
                 \If{$\mathrm{\delta} f_1/\mathrm{\delta} u_{m} = 0$}
                    \State $u_{m, solution} \gets \text{Solve}(f_1, u_m)$
                    \State \textbf{break}
                \EndIf
            \EndFor
            \Until{$\mathrm{\delta} f_1/\mathrm{\delta} u_{m} = 0$}
            \State $S_{\text{extract}} \gets (S_{\text{additionalConditions}} \backslash \{f_1\}) \cup 
            S_{\text{backlundTransform}}$ 
            \Repeat
            \ForEach{ $f_2 \in S_{\text{extract}}$}
                \State $f_2 \gets f_2.\text{substitute}(u_m, u_{m,solution})$
            \EndFor
            \Until{$f_2 \in S_{\text{extract}}$}
            \State $m \gets m + 1$
        \EndWhile
        \State $S_{\text{pull}} \gets (S_{\text{additionalConditions}} \backslash \{f_1\}) \cup S_{\text{backlundTransform}}$
\algstore{PainlevePDEwithBacklundTransformFirst}
\end{algorithmic}
\end{algorithm}

\begin{algorithm}[H]                   
\begin{algorithmic}
\algrestore{PainlevePDEwithBacklundTransformFirst}
        \If{$\text{ContextFreeCompatibility}(S_{\text{pull}}, \phi)$}
            \State $S_{\text{newEqn}} \gets S_{\text{pull}}$
            \State $S_{\text{newEqn}} \gets S_{\text{newEqn}}\backslash\{S_{\text{newEqn}}[\text{size}(S_{\text{newEqn}})-1]\}$
            \State $\text{flag} \gets \text{ContextSensitiveCompatibility}(S_{\text{newEqn}}, \{\phi, u_{-\alpha}\})$
            \If{$\text{flag}$}
                \State $S_{\text{additional}} \gets S_{\text{newEqn}} \backslash S_{\text{backlundTransform}}$
                \State \text{Return that the equation IS}
                \State \text{a good candidate for integrability}
                \State \text{with the B\"acklund transform }$S_{\text{backlundTransform}}$
                \State \text{with the additional conditions } $S_{\text{additional}}$
            \Else
                \State \text{Return that the equation IS NOT}
                \State \text{a good candidate for integrability}
            \EndIf
        \Else
            \If{$\text{ContextSensitiveCompatibility}(S_{\text{pull}}, \{\phi, u_{-\alpha}\})$}
                \State \text{Return that the equation IS}
                \State \text{a good candidate for integrability}
                \State \text{with the B\"acklund transform }$S_{\text{backlundTransform}}$
                \State \text{with the additional conditions } $S_{\text{pull}} \backslash S_{\text{backlundTransform}}$
            \Else
                \State \text{Return that the equation IS NOT}
                \State \text{a good candidate for integrability}
            \EndIf
        \EndIf
        \State $k \gets k + 1$
    \EndWhile
\end{algorithmic}
\end{algorithm}

\begin{figure*}
    \centering
    \includegraphics[scale=0.24]{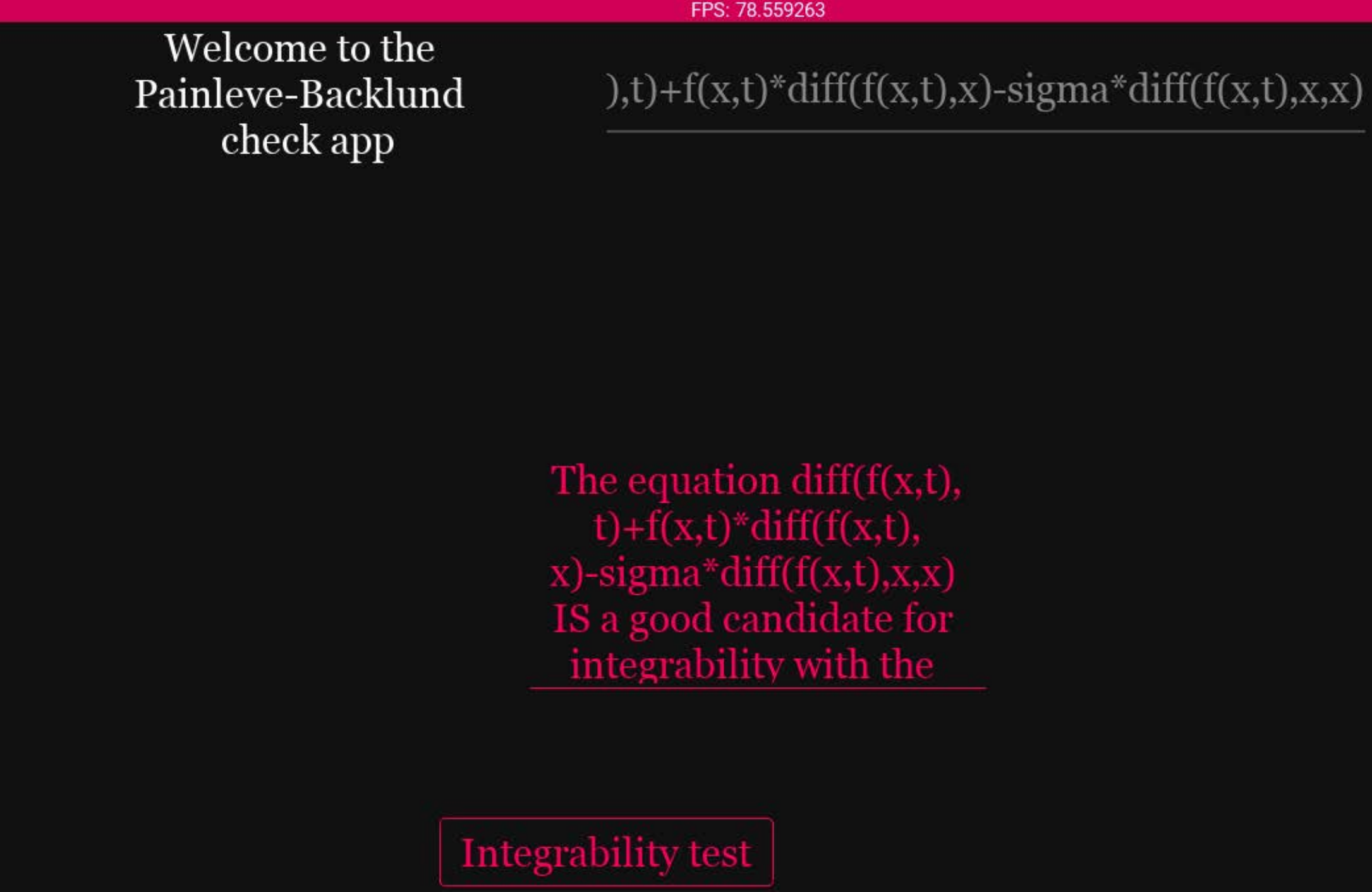}
    \includegraphics[scale=0.30]{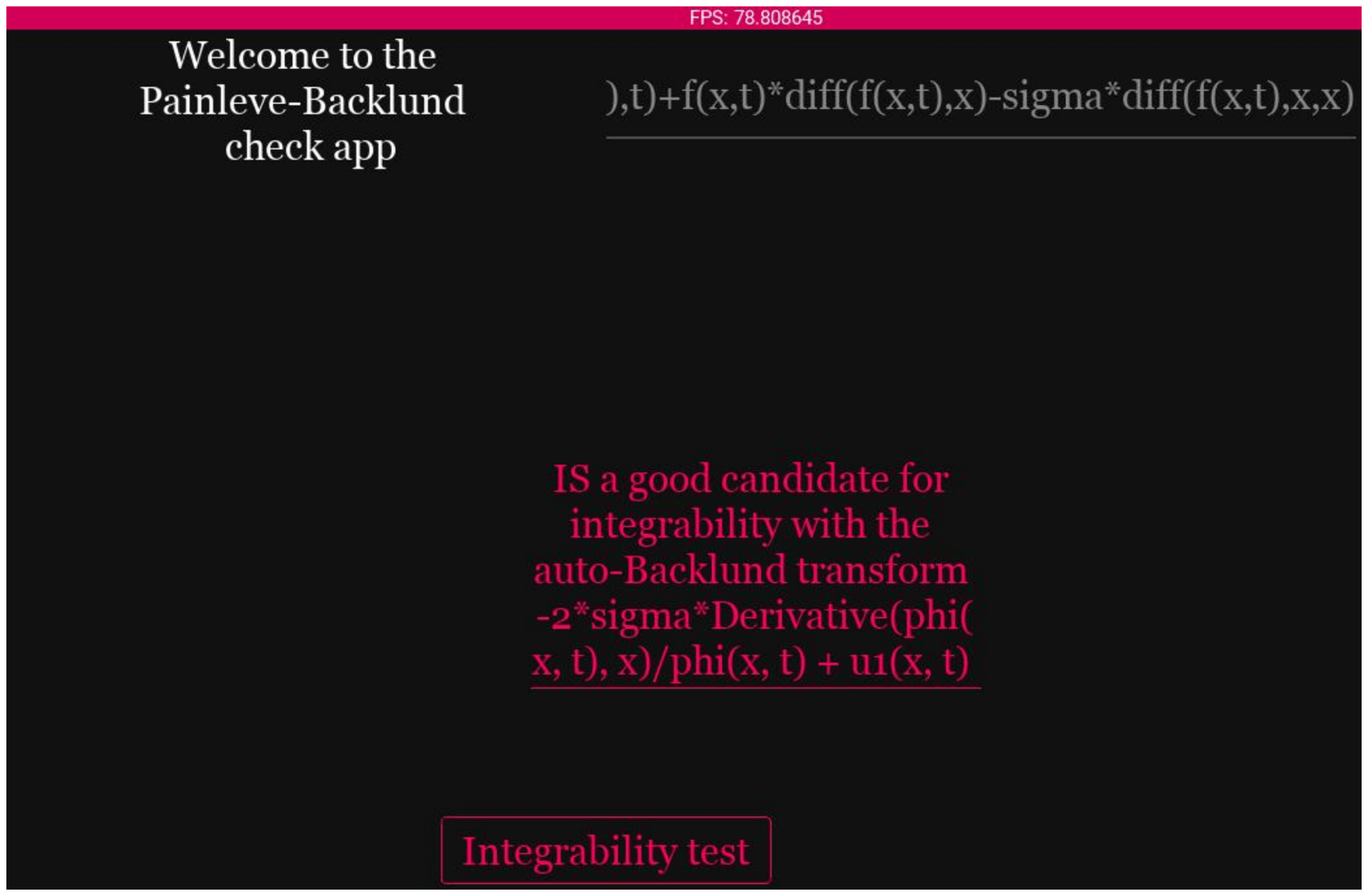}
    \caption{ \label{BurgerEquationScreenshot} The output of the PainleveBacklundCheck app with the Burgers' equation \eqref{eqn8} as the input. The B\"acklund transform obtained is the same as that in equation \eqref{eqn13} and the corresponding compatibility equations. 
    After pressing the `Integrability test' button, the context-free compatibility test happens too fast to capture from the execution screen (which has been shown more explicitly in figures \ref{KdVEquationScreenshot}, \ref{mKdVEquationScreenshot}, \ref{BBMEquationScreenshot}, \ref{KdVBurgersEquationScreenshot} and \ref{Phi4TheoryEquationScreenshot}). After the success of the test of the context-free compatibility and at the arrival of the perfectly determinate system of two equations \eqref{eqn13} and \eqref{eqn15}, the final result from the app comes as a statement that reads `The equation $u_{t} + u u_{x} - \sigma u_{xx} = 0$ IS a good candidate for integrability with the auto-B\"{a}cklund transform $u = \frac{- 2 \sigma \phi_x}{\phi} + u_1$ where $\sigma \phi_{xx} - u_1 \phi_x - \phi_t =  0$'. Both the input and the output of the app (which might have been truncated due to the sizing of the window in the image) are sympy-compatible equations.
    }
\end{figure*}
\begin{figure*}
    \centering
    \includegraphics[scale=0.29]{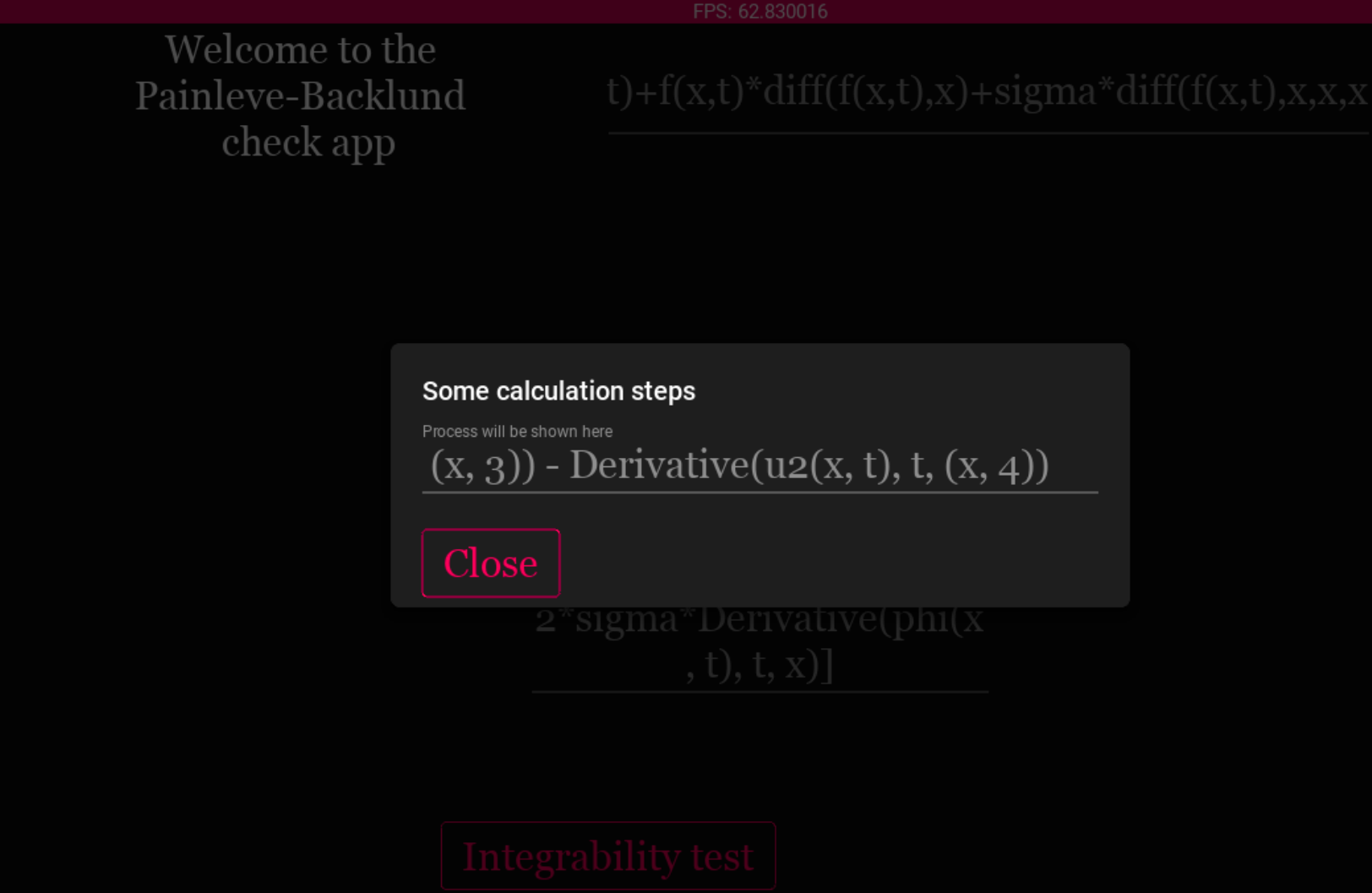}
    \includegraphics[scale=0.29]{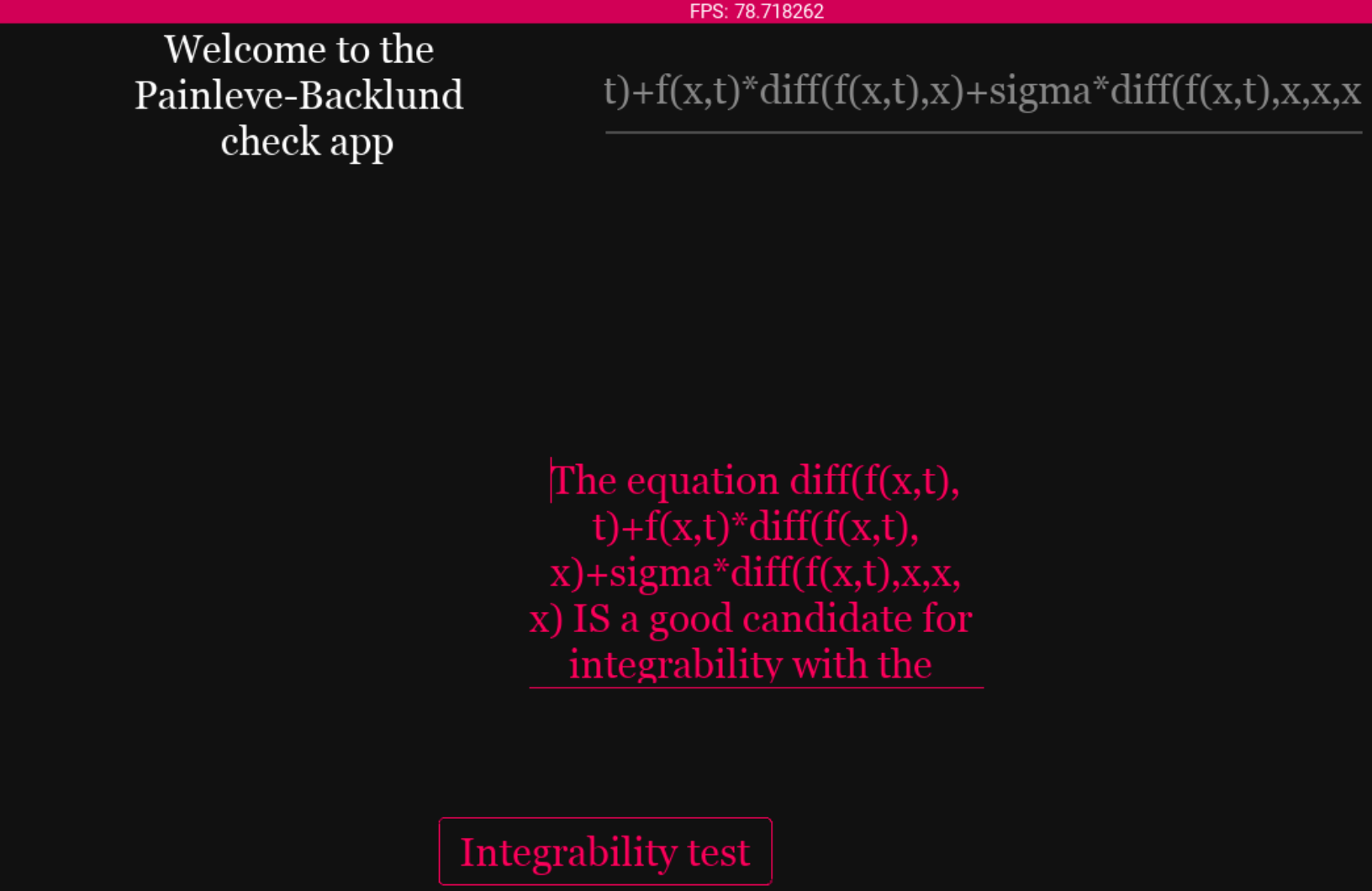}
    \includegraphics[scale=0.29]{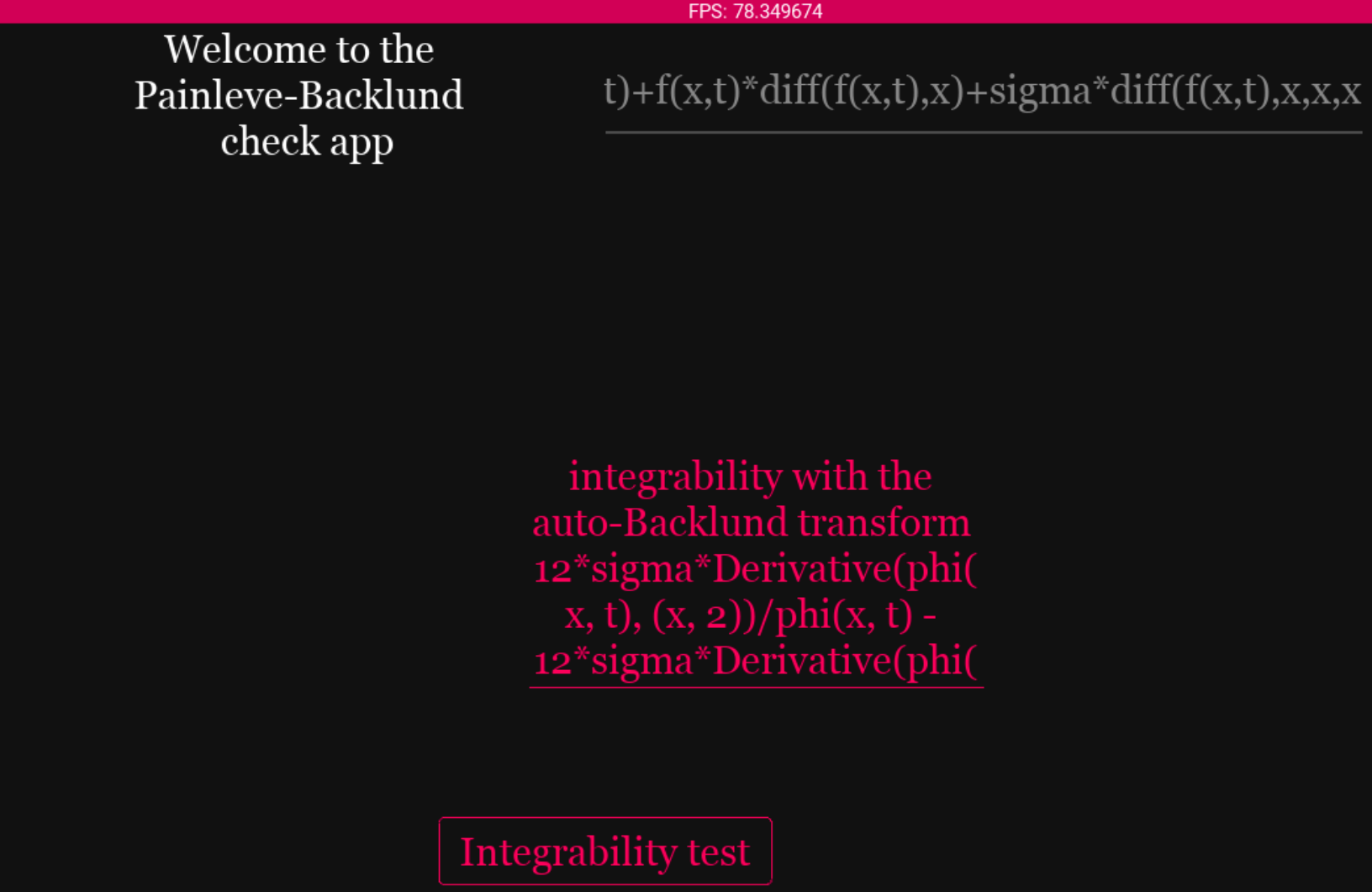}
    \caption{\label{KdVEquationScreenshot} The various parts of the output of the PainleveBacklundCheck app with the KdV equation \eqref{eqn20} as the input. The first (top left-corner) image shows a snapshot of the execution of algorithms \ref{ContextFreeDifferentialCommutationCompatibility} and \ref{ContextSensitiveDifferentialCommutationCompatibility}, and each step of the execution therewith after pressing the `Integrability test' button can be observed to be same as seen in section \ref{backlundexamples}. The B\"acklund transform obtained, as shown in all of the remnant images with the final result saying `The equation $u_t + u u_x + \sigma u_{xxx} = 0$ IS a good candidate for integrability ....', is the same in all as that in equation \eqref{eqn27} and the corresponding compatibility equations. Both the input and the output of the app (which might have been truncated due to the sizing of the window in the image) are sympy-compatible equations.}
\end{figure*}
\begin{figure*}
    \centering
    \includegraphics[scale=0.23]{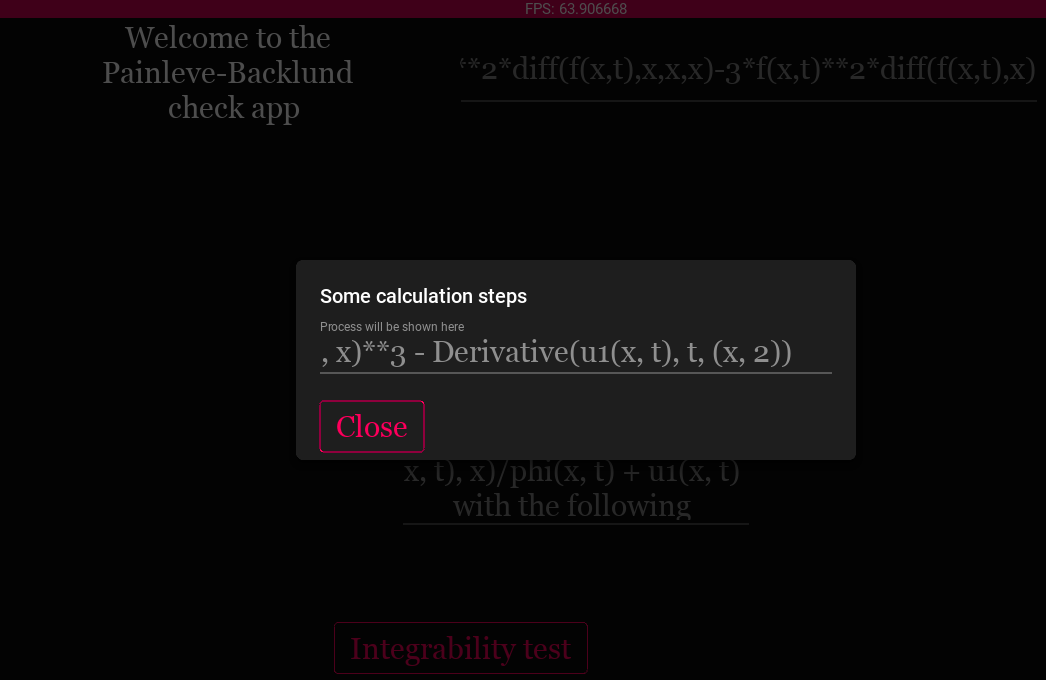}
    \includegraphics[scale=0.23]{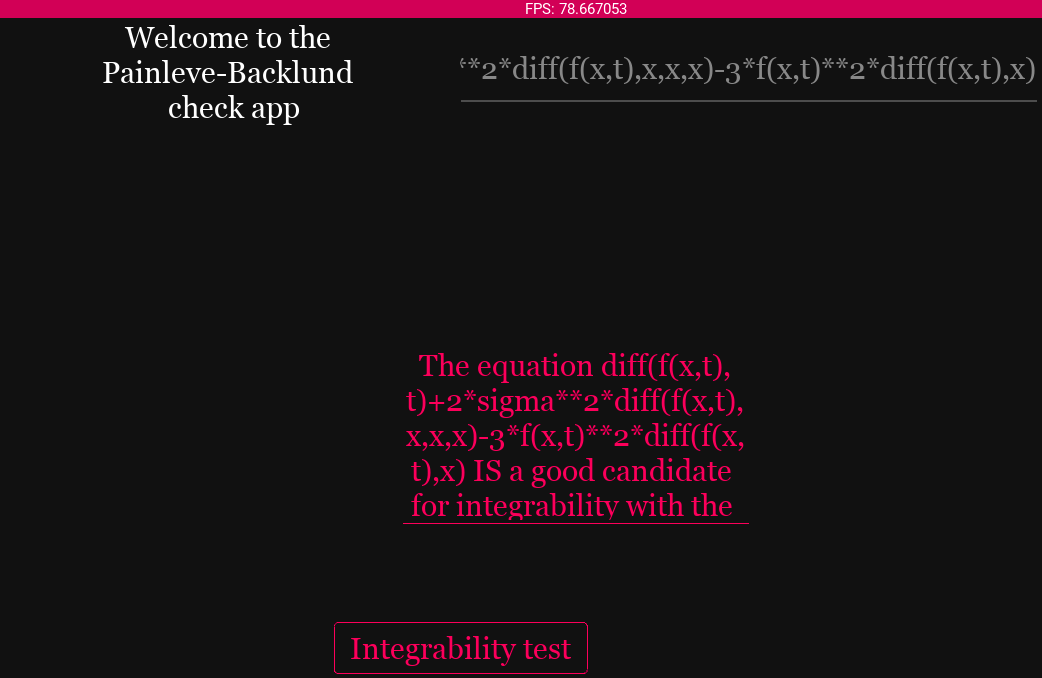}
    \includegraphics[scale=0.23]{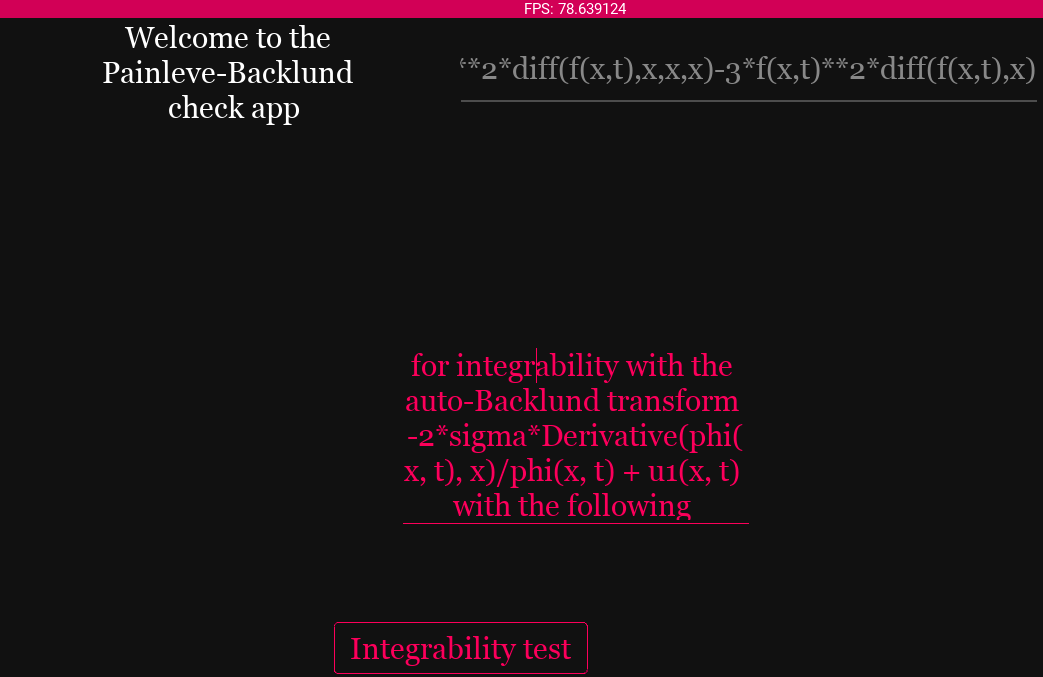}
    \caption{\label{mKdVEquationScreenshot} The various parts of the output of the PainleveBacklundCheck app with the modified KdV equation \eqref{addseteqn61At02162024} as the input. The first (top left-corner) image shows a snapshot of the execution of algorithms \ref{ContextFreeDifferentialCommutationCompatibility} and \ref{ContextSensitiveDifferentialCommutationCompatibility}, and each step of the execution therewith after pressing the `Integrability test' button can be observed to be same as seen in section \ref{backlundexamples}. The B\"acklund transform obtained, as shown in all of the remnant images with the final result saying `The equation $u_t + 2 \sigma^2 u_{xxx} - 3 u^2 u_x = 0$ IS a good candidate for integrability ....', is the same in all as that in equation \eqref{addseteqn63At02162024} and the corresponding compatibility equations. Both the input and the output of the app (which might have been truncated due to the sizing of the window in the image) are sympy-compatible equations.}
\end{figure*}

\begin{figure*}
    \centering
    \includegraphics[scale=0.29]{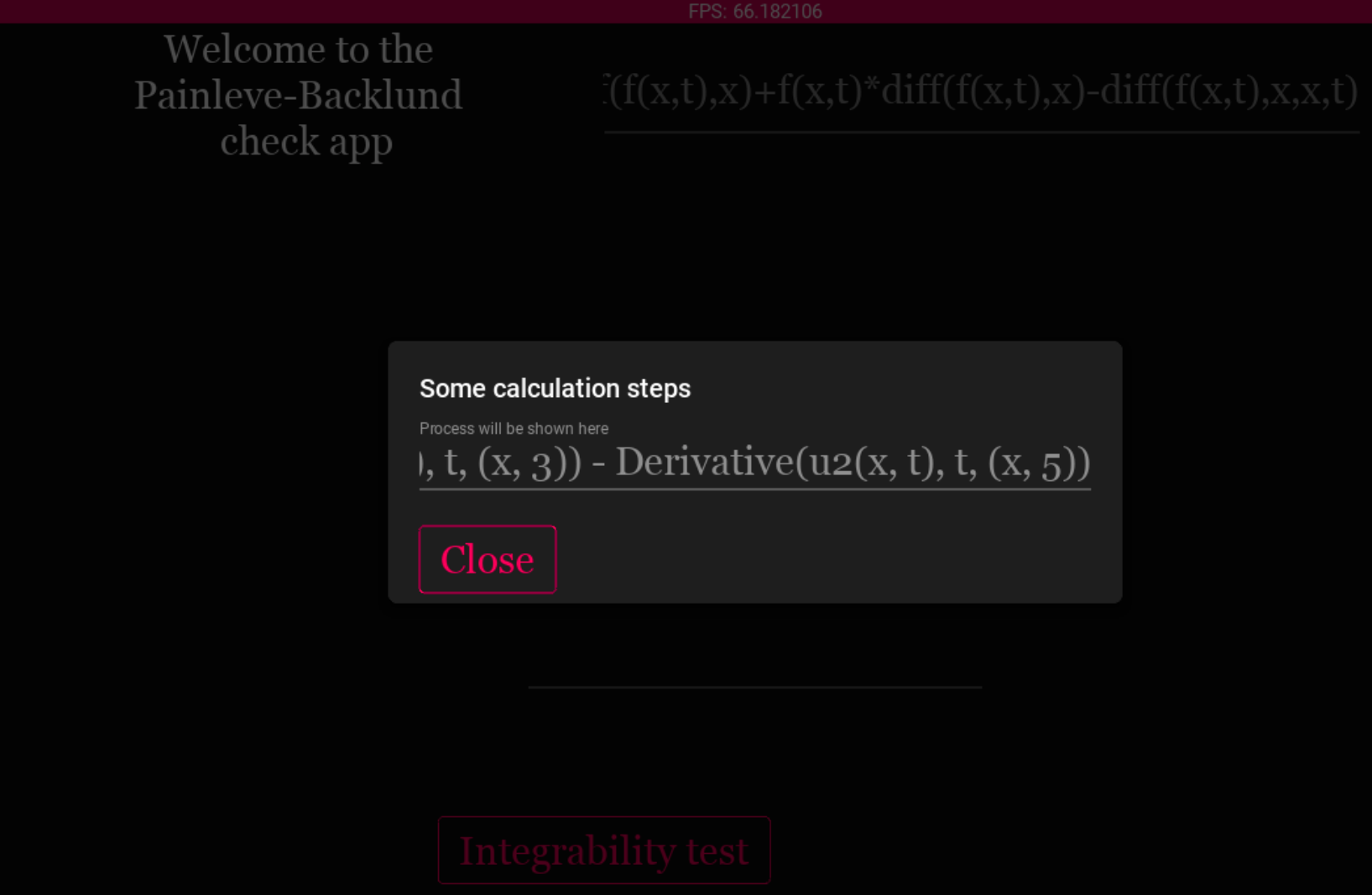}
    \includegraphics[scale=0.30]{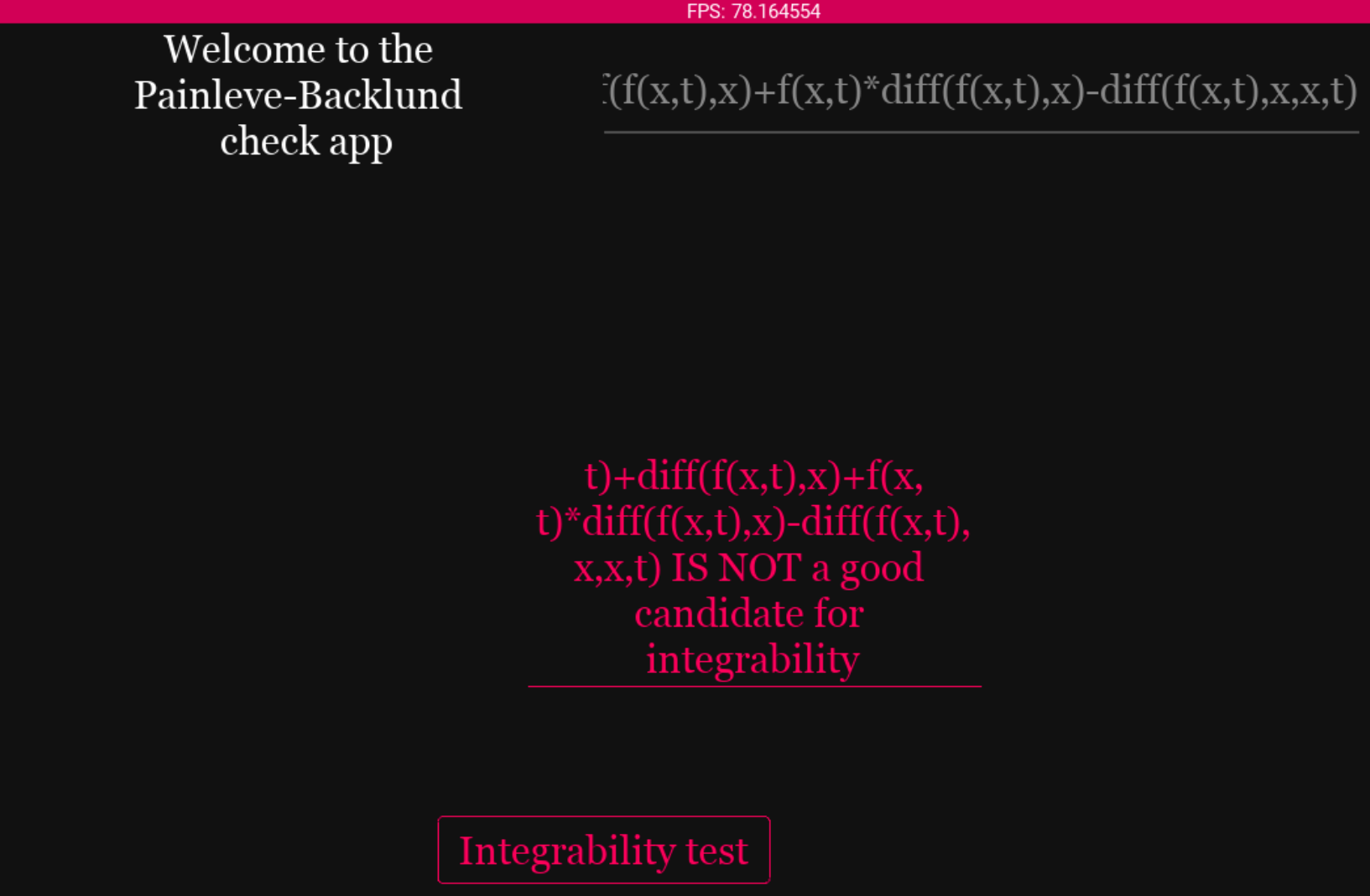}
    \caption{\label{BBMEquationScreenshot} The various parts of the output of the PainleveBacklundCheck app with the BBM equation \eqref{NewEqnRef57} as the input. The first image shows a snapshot of the execution of algorithms \ref{ContextFreeDifferentialCommutationCompatibility} and \ref{ContextSensitiveDifferentialCommutationCompatibility}, and each step of the execution therewith after pressing the `Integrability test' button  can be observed to be same as seen in section \ref{backlundexamples}. Both the input and the output of the app (which might have been truncated due to the sizing of the window in the image) are sympy-compatible equations. After the failure of the test of the context sensitivity, the final result from the app comes as a statement that reads `The equation $u_t + u_x + u u_x - u_{xxt} = 0$ IS NOT a good candidate for integrability'.}
\end{figure*}

\begin{figure*}
    \centering
    \includegraphics[scale=0.22]{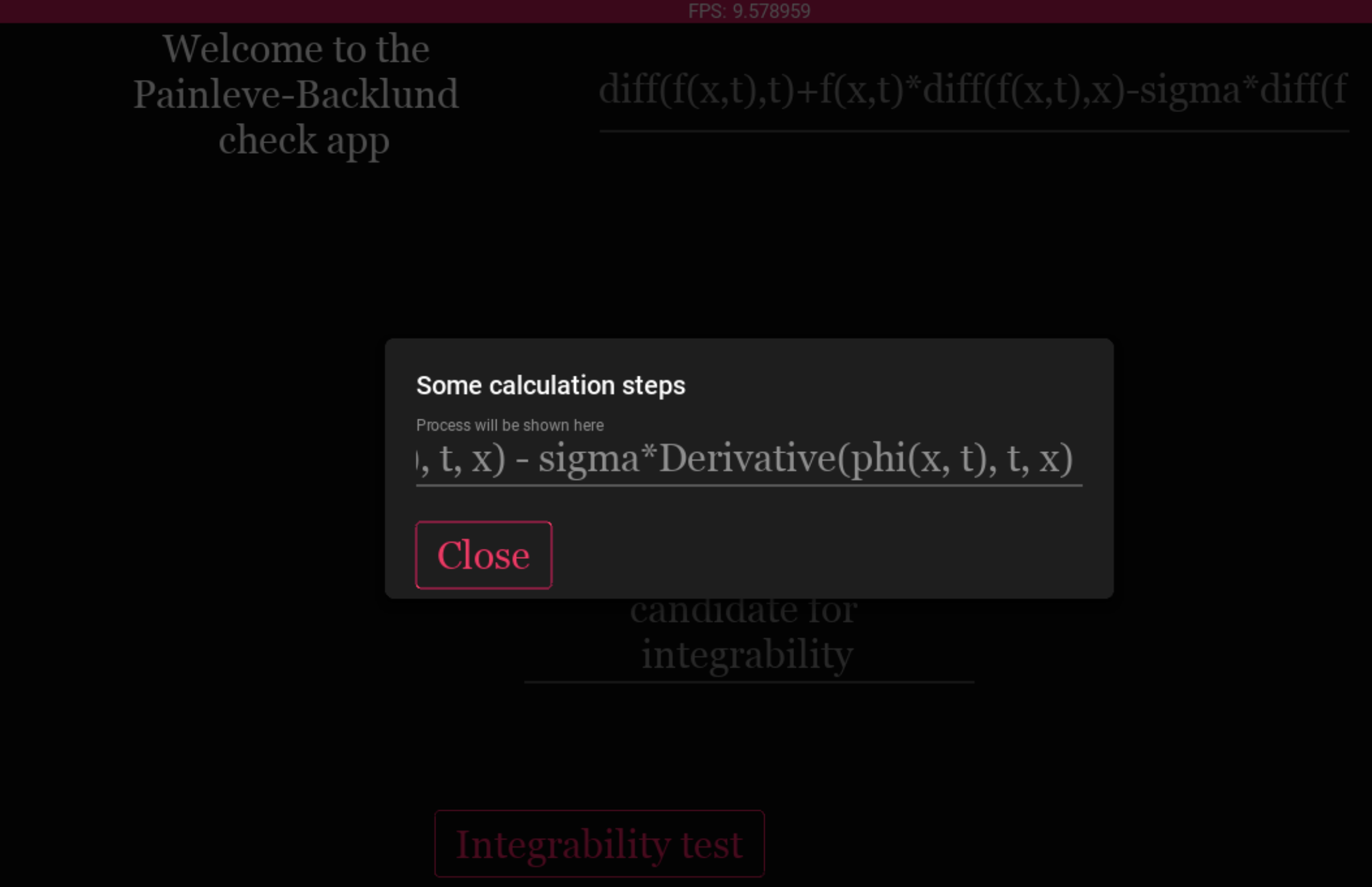}
    \includegraphics[scale=0.30]{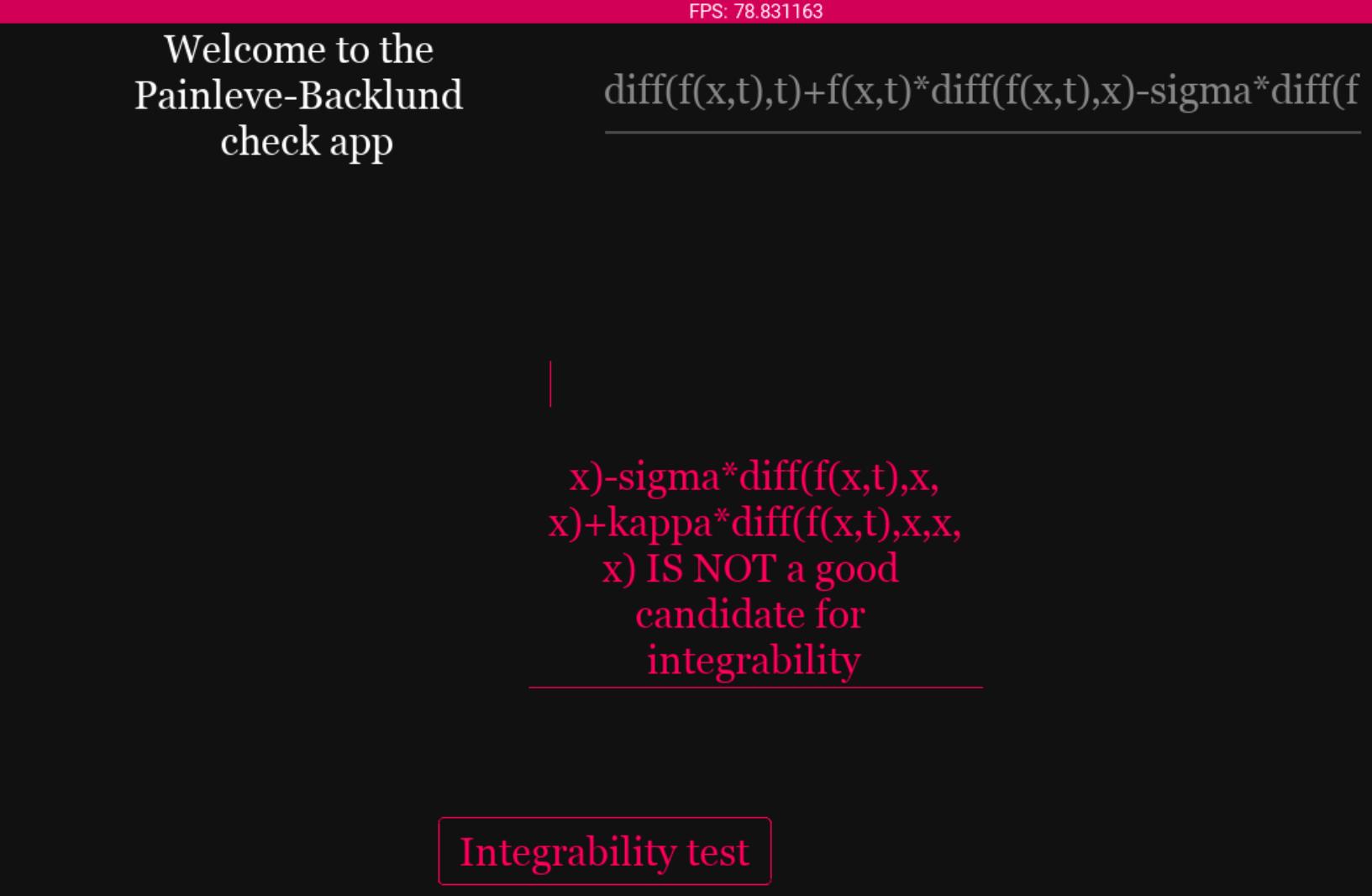}
    \caption{\label{KdVBurgersEquationScreenshot} The various parts of the output of the PainleveBacklundCheck app with the KdV-Burgers equation \eqref{addseteqn1} as the input. The first image shows a snapshot of the execution of algorithms \ref{ContextFreeDifferentialCommutationCompatibility} and \ref{ContextSensitiveDifferentialCommutationCompatibility}, and each step of the execution therewith after pressing the `Integrability test' button can be observed to be same as seen in section \ref{backlundexamples}. Both the input and the output of the app (which might have been truncated due to the sizing of the window in the image) are sympy-compatible equations. After the failure of the test of the context sensitivity, the final result from the app comes as a statement that reads `The equation $u_t + u u_x - \sigma u_{xx} + \kappa u_{xxx} = 0$ IS NOT a good candidate for integrability'.}
\end{figure*}

\begin{figure*}
    \centering
    \includegraphics[scale=0.29]{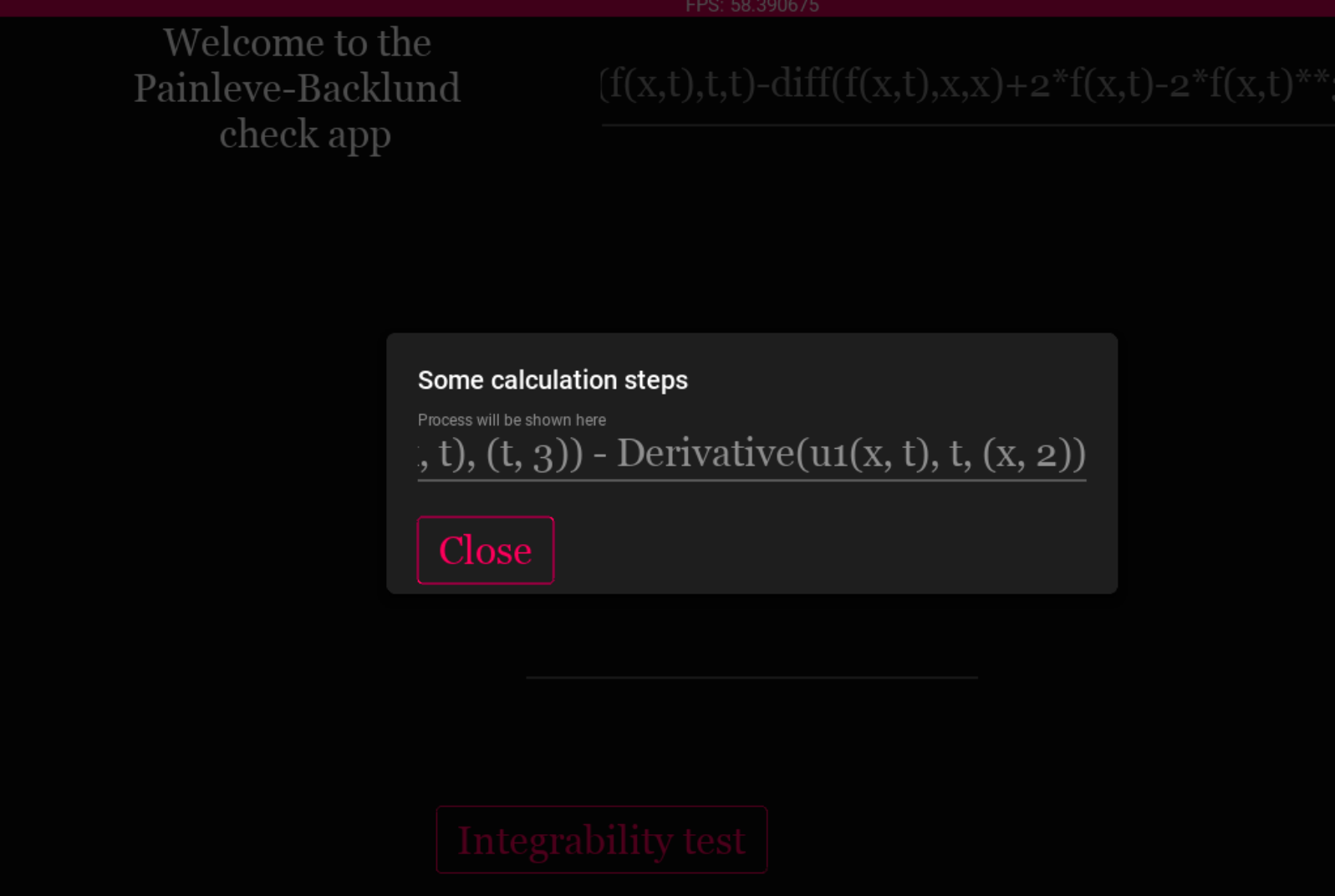}
    \includegraphics[scale=0.30]{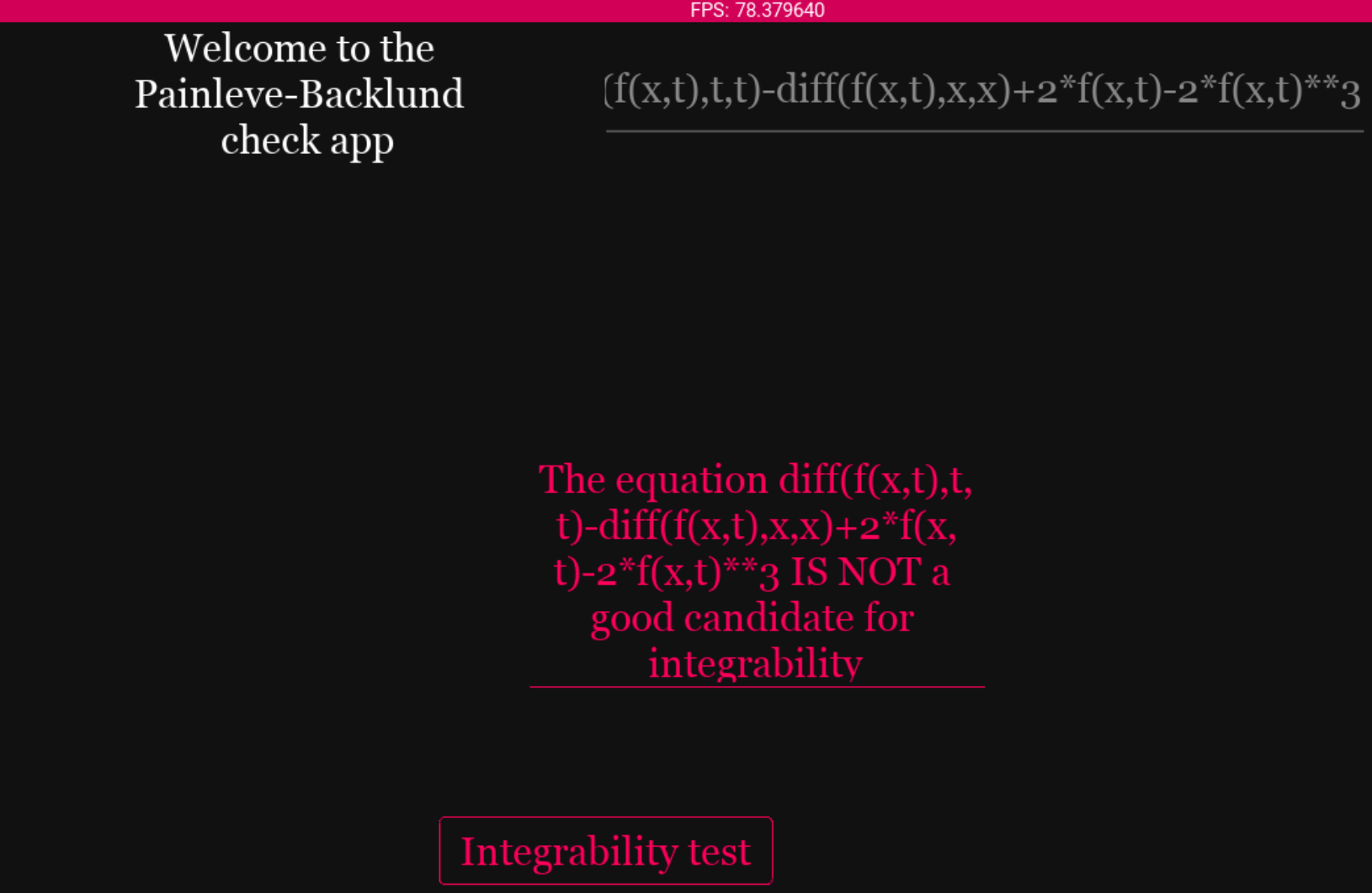}
    \caption{\label{Phi4TheoryEquationScreenshot} The various parts of the output of the PainleveBacklundCheck app with the $\phi^4$-theory based nonlinear Klein-Gordon equation \eqref{eqnRef123From12222023} as the input. The first image shows a snapshot of the execution of algorithms \ref{ContextFreeDifferentialCommutationCompatibility} and \ref{ContextSensitiveDifferentialCommutationCompatibility}, and each step of the execution therewith after pressing the `Integrability test' button can be observed to be same as seen in section \ref{backlundexamples}. Both the input and the output of the app (which might have been truncated due to the sizing of the window in the image) are sympy-compatible equations. After the failure of the test of the context sensitivity, the final result from the app comes as a statement that reads `The equation $u_{tt} - u_{xx} + 2 u^3 - 2 u = 0$ IS NOT a good candidate for integrability'.}
\end{figure*}

The algorithms in this section have been implemented in the `PainleveBacklundCheck' app with the help of two cornerstone packages in the Python programming language: Sympy (version 1.12) \cite{sympypaper}, the library for the symbolic computation, serves as the backbone of the package (the latest version of Sympy supports only integral degree polynomial rings for the series expansions, due to which we consider the implementational simplification with integer values of $\alpha$) and also the language in which the user enters the input to the GUI. The user interface is supported by Kivy (version 2.1) which envelopes the whole software into an app which can be deployed into  Android, iOS, Linux, macOS, and Windows \cite{kivy}. The output of the PainleveBacklundCheck app with the inputs of the Burgers' equation, KdV equation, the modified KdV equation, the BBM equation, KdV-Burgers' equation and the $\phi^4$ theory cases has been illustrated in the figures \ref{BurgerEquationScreenshot}, \ref{KdVEquationScreenshot}, \ref{mKdVEquationScreenshot}, \ref{BBMEquationScreenshot}, \ref{KdVBurgersEquationScreenshot} and \ref{Phi4TheoryEquationScreenshot} respectively.  These cases are all found to be in line with the theoretical analysis provided above. Additional cases with longer outputs are shown and illustrated in the different ``sandbox'' programs of test cases in \cite{painleveapprepo}.

\section{Conclusions \& Future Challenges}
In this paper we 
revisited the Painlev\'e property for nonlinear dispersive PDEs (which can be solvable ---or not--- by the inverse scattering transform) and the auto-B\"acklund transformation, which involves the precise level of truncation of the Painlev\'e expansion. An additional 
{property of the truncation explored} in this work concerns an observation regarding a pattern for the integer Painlev\'e exponent $\alpha$.
Evidence of this pattern in many different cases (in section \ref{backlundexamples})
is presented both as concerns the level of truncation at the ``just right" level, and also as concerns the contradiction with going one level lower and one level higher. Using this pattern, we present an algorithm
{that additionally addresses the existence/non-existence of the B\"{a}cklund transform by means of verifying the compatibility of the associated equations} and an implementation thereof as a Sympy-powered Kivy app 'PainleveBacklundCheck' (written in Python) \cite{painleveapprepo} is illustrated with the outputs in some of the simpler cases. 
This serves the purpose of providing the community with an open access, multi-platform,
(in our view) easy to use implementation of the notable WTC approach towards the Painlev{\'e}
test for partial differential equations.
The results for other higher-order and higher-degree PDEs are shown in \cite{painleveapprepo}. Naturally, in the present work, we have chiefly limited ourselves to the cases
of a sequence of well-established examples of integrability and have sought to ensure
the functionality of our findings in such cases. However, as new integrable systems
of different kinds are emerging, it would be particularly useful and relevant for
practitioners of integrable systems to test the methodology in other such relevant
systems and report the corresponding successes, as well as potential failures and
needs of improvement, for the formulation and algorithms presented herein. Relevant
findings and amendments, as well
as generalizations will be reported in future publications.

\begin{appendices}
\section{Painlev\'e expansions and more examples} \label{appendixA}

In this appendix section, we revisit some of the examples in section \ref{backlundexamples} in terms of the contradiction with the working with the existence and validity of the Painlev\'e expansion, going one order ahead and going one order behind the ``affine'' level of truncation of $\beta = - \alpha$. We 
examine the resulting putative
B\"{a}cklund transformation 
in terms of the context-free and context-sensitive compatibilities.
\subsection*{KdV equation} \label{appendixA_Contradiction_Of_KdV_Equation}
Continuing with the KdV equation \eqref{eqn20} extensively discussed in section \ref{backlundexamples}, if we truncate one order earlier with $u_j = 0, j \ge 2$, then {all the coefficients $P_n, n \ge 5$ vanish}, provided,
\begin{equation}
u = \frac{u_0}{\phi^2} + \frac{u_1}{\phi} = \frac{12 \sigma \phi_{xx}}{\phi} - \frac{12 \sigma \phi_x^2}{\phi^2} = 12 \sigma \frac{\partial^2}{\partial x^2}(\log{\phi}),
\end{equation}
{where $u_0 = - 12 \sigma \phi_x^2$, $u_1 = 12 \sigma \phi_{xx}$ follows from $P_0 = 0$ and $P_1 = 0$ respectively. Substituting $u_0 = - 12 \sigma \phi_x^2$ and $u_1 = 12 \sigma \phi_{xx}$ in the equation $P_2 = 0$ with $\phi_x \ne 0$ leads to
\begin{multline} \label{eqn62From12212023}
    u_1 = 12 \sigma \phi_{xx}, 24 \phi_{t} \phi_{x}^2 \sigma + 168 \phi_{xxx} \phi_{x}^2 \sigma^2 + 288 \phi_{xx}^2 \phi_{x} \sigma^2 \\
    - 18 \phi_{xx} \phi_{x} \sigma u_1 - 6 \phi_{x}^2 \sigma u_{1,x} - \phi_{x} u_1^2  = 0 \\
    \implies \phi_x \phi_t - 3 \sigma \phi_{xx}^2 + 4 \sigma \phi_x \phi_{xxx} = 0,
\end{multline}
Similarly, substituting the expressions for $u_0$ and $u_1$ in the equation $P_3 = 0$ leads to
\begin{multline} \label{eqn63From12212023}
    u_1 = 12 \sigma \phi_{xx}, -24 \phi _{,tx} \phi _{,x} \sigma - \phi _{,t} u_1 - 24 \phi _{,xxxx} \phi _{,x} \sigma^2 \\
    - 72 \phi _{,xxx} \phi _{,xx} \sigma^2 - \phi _{,xxx} \sigma u_1 - 3 \phi _{,xx} \sigma u_{1,x} \\
    - 3 \phi _{,x} \sigma u_{1,xx} + u_1 u_{1,x} = 0 \\
    \implies 2 \phi _x \phi _{xt} + \phi_t \phi_{xx} - 2 \sigma \phi_{xx} \phi_{xxx} + 5 \sigma \phi_{x} \phi_{xxxx} = 0.
\end{multline}
Simplifying $P_4 = 0$ gives,
\begin{equation} \label{eqn64From12212023}
    u_{1, t} + \sigma u_{1, xxx} = 0, u_1 = 12 \sigma \phi_{xx}.
\end{equation}
By means of systematic computation, one can show that the set $\{P_2, P_3, P_4\}$ is \textit{not} context-free compatible, as one does not encounter a zero in the sequence of the differential ideals owing to the increasing magnitude of the coefficients in the differential equations (checked up to 150 recursions). From the equations \eqref{eqn62From12212023}, \eqref{eqn63From12212023} and \eqref{eqn64From12212023}, one can systematically solve and obtain expressions for the following higher derivatives,
\begin{multline*}
\phi_t = -7 \phi_{xxx} \sigma - \frac{12 \phi_{xx}^2 \sigma}{\phi_{x}} + \frac{3 \phi_{xx} u_1}{4 \phi_{x}} + \frac{u_{1,x}}{4} \\+ \frac{u_1^2}{24 \phi_{x} \sigma}, u_{1, t} = -\sigma u_{1, xxx}, u_{1, xt} = -\sigma u_{1, xxxx}
\end{multline*}
\begin{multline*}
\phi_{xt} = \frac{\phi_{xxx} \phi_{xx} \sigma}{2 \phi_{x}} + \frac{\phi_{xxx} u_1}{6 \phi_{x}} - \frac{2 \phi_{xx}^3 \sigma}{\phi_{x}^2} + \frac{17 \phi_{xx}^2 u_1}{24 \phi_{x}^2} \\
- \frac{13 \phi_{xx} u_{1,x}}{48 \phi_{x}} - \frac{17 \phi_{xx} u_1^2}{576 \phi_{x}^2 \sigma} - \frac{3 u_{1,xx}}{16} + \frac{13 u_1 u_{1,x}}{576 \phi_{x} \sigma} - \frac{7 u_1^3}{3456 \phi_{x}^2 \sigma^2}
\end{multline*}
\begin{multline*}
\phi_{x, 4} = -\frac{7 \phi_{xxx} \phi_{xx}}{2 \phi_{x}} + \frac{\phi_{xxx} u_1}{12 \phi_{x} \sigma} + \frac{2 \phi_{xx}^3}{\phi_{x}^2} - \frac{5 \phi_{xx}^2 u_1}{24 \phi_{x}^2 \sigma} \\
+ \frac{7 \phi_{xx} u_{1,x}}{48 \phi_{x} \sigma} - \frac{\phi_{xx} u_1^2}{576 \phi_{x}^2 \sigma^2} + \frac{u_{1,xx}}{16 \sigma} + \frac{5 u_1 u_{1,x}}{576 \phi_{x} \sigma^2} + \frac{u_1^3}{3456 \phi_{x}^2 \sigma^3}
\end{multline*}
\begin{multline*}
\phi_{xxt} = \frac{\phi_{xxx}^2 \sigma}{2 \phi_{x}} - \frac{33 \phi_{xxx} \phi_{xx}^2 \sigma}{4 \phi_{x}^2} + \frac{17 \phi_{xxx} \phi_{xx} u_1}{24 \phi_{x}^2} \\
- \frac{5 \phi_{xxx} u_{1,x}}{48 \phi_{x}} - \frac{\phi_{xxx} u_1^2}{64 \phi_{x}^2 \sigma} + \frac{5 \phi_{xx}^4 \sigma}{\phi_{x}^3} \\
- \frac{19 \phi_{xx}^3 u_1}{16 \phi_{x}^3} + \frac{101 \phi_{xx}^2 u_{1,x}}{96 \phi_{x}^2} + \frac{3 \phi_{xx}^2 u_1^2}{128 \phi_{x}^3 \sigma} \\
- \frac{23 \phi_{xx} u_{1,xx}}{96 \phi_{x}} - \frac{61 \phi_{xx} u_1 u_{1,x}}{1152 \phi_{x}^2 \sigma} + \frac{\phi_{xx} u_1^3}{256 \phi_{x}^3 \sigma^2} - \frac{3 u_{1,xxx}}{16} \\
+ \frac{19 u_1 u_{1,xx}}{576 \phi_{x} \sigma} + \frac{13 u_{1,x}^2}{576 \phi_{x} \sigma} - \frac{u_1^2 u_{1,x}}{216 \phi_{x}^2 \sigma^2} + \frac{u_1^4}{20736 \phi_{x}^3 \sigma^3}
\end{multline*}
\begin{multline*}
\phi_{xxxt} = - \frac{41 \phi_{xxx}^2 \phi_{xx} \sigma}{2 \phi_{x}^2} + \frac{19 \phi_{xxx}^2 u_1}{24 \phi_{x}^2} + \frac{539 \phi_{xxx} \phi_{xx}^3 \sigma}{8 \phi_{x}^3} \\
- \frac{401 \phi_{xxx} \phi_{xx}^2 u_1}{48 \phi_{x}^3} + \frac{329 \phi_{xxx} \phi_{xx} u_{1,x}}{96 \phi_{x}^2} + \frac{73 \phi_{xxx} \phi_{xx} u_1^2}{384 \phi_{x}^3 \sigma} \\
- \frac{9 \phi_{xxx} u_{1,xx}}{32 \phi_{x}} - \frac{97 \phi_{xxx} u_1 u_{1,x}}{1152 \phi_{x}^2 \sigma} + \frac{5 \phi_{xxx} u_1^3}{1728 \phi_{x}^3 \sigma^2} - \frac{63 \phi_{xx}^5 \sigma}{2 \phi_{x}^4} \\
+ \frac{643 \phi_{xx}^4 u_1}{96 \phi_{x}^4} - \frac{301 \phi_{xx}^3 u_{1,x}}{64 \phi_{x}^3} - \frac{541 \phi_{xx}^3 u_1^2}{2304 \phi_{x}^4 \sigma} + \frac{149 \phi_{xx}^2 u_{1,xx}}{192 \phi_{x}^2} \\
+ \frac{475 \phi_{xx}^2 u_1 u_{1,x}}{2304 \phi_{x}^3 \sigma} - \frac{167 \phi_{xx}^2 u_1^3}{13824 \phi_{x}^4 \sigma^2} - \frac{23 \phi_{xx} u_{1,xxx}}{96 \phi_{x}} - \frac{\phi_{xx} u_1 u_{1,xx}}{24 \phi_{x}^2 \sigma} \\
- \frac{209 \phi_{xx} u_{1,x}^2}{2304 \phi_{x}^2 \sigma} + \frac{173 \phi_{xx} u_1^2 u_{1,x}}{6912 \phi_{x}^3 \sigma^2} + \frac{29 \phi_{xx} u_1^4}{331776 \phi_{x}^4 \sigma^3} - \frac{3 u_{1,xxxx}}{16} \\
+ \frac{19 u_1 u_{1,xxx}}{576 \phi_{x} \sigma} + \frac{55 u_{1,xx} u_{1,x}}{768 \phi_{x} \sigma} - \frac{155 u_1^2 u_{1,xx}}{27648 \phi_{x}^2 \sigma^2} - \\
\frac{281 u_1 u_{1,x}^2}{27648 \phi_{x}^2 \sigma^2} + \frac{u_1^3 u_{1,x}}{36864 \phi_{x}^3 \sigma^3} - \frac{u_1^5}{221184 \phi_{x}^4 \sigma^4}
\end{multline*}
And from there one can verify that the difference $\frac{\partial \phi_{xxxt}}{\partial x} - \frac{\partial \phi_{xxxx}}{\partial t}$ does not vanish identically (verifiable both with and without the substitution of $u_1 = 12 \sigma \phi_{xx}$), which implies that the set $\{P_2, P_3, P_4\}$ is \textit{not} context-sensitive compatible, leading to the {non-existence of the sought B\"{a}cklund transform}.
}

To complete our discussion, let us now get one order ahead of the original case, i.e., we set $u_j = 0, j \ge 4$, then {all the coefficients $P_n, n \ge 8$ vanish}, provided,
\begin{equation}
u = \frac{u_0}{\phi^2} + \frac{u_1}{\phi} + u_2 + u_3 \phi,
\end{equation}
where the relations $u_0 = -12 \sigma \phi_{x}^2$ and $u_1 = 12 \sigma \phi_{xx}$ are still valid, as before. For the coefficient $u_2$, one can substitute the coefficients $u_0$ and $u_1$ in the equation $P_2 = 0$
\begin{multline} \label{eqnRef66From12182023}
    24 \sigma \phi_x (\phi_t \phi_x + u_2 \phi_x^2 - 3 \sigma \phi_{xx}^2 + 4 \sigma \phi_x \phi_{xxx}) = 0, \\
    \implies Q_{\text{KdV, 1, +}} = 4 \sigma \phi_x \phi_{xxx} + \phi_t \phi_x  + u_2 \phi_x^2 - 3 \sigma \phi_{xx}^2 = 0
\end{multline}
which one can simplify to obtain,
\begin{equation}
    u_2 = -3\sigma \frac{\partial^2}{\partial x^2} \bigg(\log{\phi_x}\bigg) - \frac{\phi_t + \sigma \phi_{xxx}}{\phi_x^2},
\end{equation}
and a similar procedure for $P_3 = 0$ gives the coefficient $u_3$ as well,
\begin{multline} \label{eqnRef68From12182023}
    u_3 = \frac{\phi_x\phi_{xt}(2\phi_x-1)+\phi_t\phi_{xx}(\phi_x-1)}{\phi_x^4} + \sigma \\
     \bigg(\frac{3 \phi_{xx}^3 - \phi_{xx} \phi_{xxx}(2 \phi_x + 1) + \phi_x \phi_{xxxx}(2 \phi_x - 1)}{\phi_x^4}\bigg), \\
    Q_{\text{KdV, 2, +}} = u_3 \phi_x^4 - \phi_x\phi_{xt}(2\phi_x-1) - \phi_t\phi_{xx}(\phi_x-1) - \\
    \sigma (3 \phi_{xx}^3 - \phi_{xx} \phi_{xxx}(2 \phi_x + 1) \\
    + \phi_x \phi_{xxxx}(2 \phi_x - 1)) = 0
\end{multline}
The other associated compatibility conditions come from the simplification $P_k = 0, 4 \le k \le 7$,
\begin{multline} \label{neweqn43}
   P_4 = 0 \implies \phi_{txx} + \phi_{xxxxx} \sigma + \phi_{xxx} u_2 - 2 \phi_{xx} \phi_{x} u_3 \\
   + \phi_{xx} u_{2,x} - \phi_{x}^2 u_{3,x} = 0
   \implies \frac{\partial}{\partial x} Q_{\text{KdV, 3, +}} = 0, \\
   Q_{\text{KdV, 3, +}} = u_2 \phi_{xx} - u_3 \phi_x^2 + \phi_{xt} + \sigma \phi_{xxxx}
\end{multline}
\begin{multline} \label{neweqn44}
    P_5 = 0 \implies Q_{\text{KdV, 4, +}} = 12 \sigma \frac{\partial}{\partial x}\bigg(\phi_{xx} u_3\bigg) \\
    + \sigma u_{2,xxx} + u_2 u_{2,x} + u_{2,t} = 0
\end{multline}
\begin{multline} \label{neweqn45}
    P_6 = 0 \implies Q_{\text{KdV, 5, +}} = \\
    u_{3, t} + u_3^2 \phi_x + \frac{\partial}{\partial x}\bigg(u_2 u_3 + \sigma u_{3, xx}\bigg) = 0
\end{multline}
\begin{multline} \label{neweqn46}
    P_7 = 0 \implies \frac{\partial}{\partial x}\bigg(\frac{1}{2} u_3^2 \bigg) = u_3 u_{3, x} = 0 \\
    \implies Q_{\text{KdV, 6, +}} = u_{3, x} = 0
\end{multline}
{One can show that the set $\{Q_{\text{KdV, 1, +}},$ $Q_{\text{KdV, 2, +}},$ $Q_{\text{KdV, 3, +}},$ $ Q_{\text{KdV, 4, +}},$ $Q_{\text{KdV, 5, +}},$ $Q_{\text{KdV, 6, +}}\}$ is context-free compatible as $0 \in g_9$. Also, surprisingly in the similar vein, one can show that the set $\{Q_{\text{KdV, 1, +}},$ $Q_{\text{KdV, 2, +}},$ $Q_{\text{KdV, 4, +}},$ $Q_{\text{KdV, 5, +}},$ $Q_{\text{KdV, 6, +}}\}$ is also context-free compatible as $0 \in g_8$.} Consequently two reconciliation relations amongst the PDEs can be algebraically compactified in the following form.
\begin{multline} \label{eqnRef73From12212023}
{\sigma H_1 W_2 \bigg(\frac{\partial}{\partial t} Q_{\text{KdV, 1, +}} - \phi_x^2 Q_{\text{KdV, 4, +}}\bigg) - \sigma H_2 W_2} \\
{\bigg(\frac{\partial}{\partial t} Q_{\text{KdV, 2, +}} - \phi_x^4 Q_{\text{KdV, 5, +}}\bigg) - W_2 W_3 Q_{\text{KdV, 5, +}}} \\
+ W_3 W_4 Q_{\text{KdV, 4, +}} = Q_{\text{KdV, 6, +}} \bigg(u_2 W_2 W_3 + 12 \sigma \phi_{xx} W_3 W_4\bigg)
\end{multline}
\begin{multline} \label{eqnRef74From12212023}
H_3 \bigg(2 Q_{\text{KdV, 2, +}} - \phi_x^2 \frac{\partial}{\partial x} Q_{\text{KdV, 1, +}} + 2 \phi_x^2 Q_{KdV, 3, +}\bigg) \\
+ H_4 \bigg(\frac{\partial}{\partial t} Q_{\text{KdV, 2, +}} - \phi_x^4 Q_{\text{KdV, 5, +}}\bigg)= V_1 Q_{\text{KdV, 6, +}} \\
+ V_2 \frac{\partial^2}{\partial x^2}Q_{\text{KdV, 6, +}}
\end{multline}
where
\begin{multline*}
H_1 = -\phi_{ttx} \phi_{x}^2 + \phi_{tt} \phi_{xx} \phi_{x} \\
- \phi_{txxxx} \phi_{x}^2 \sigma + 4 \phi_{txxx} \phi_{xx} \phi_{x} \sigma + \phi_{txx} \phi_{t} \phi_{x} \\
+ 4 \phi_{txx} \phi_{xxx} \phi_{x} \sigma - 9 \phi_{txx} \phi_{xx}^2 \sigma - 2 \phi_{tx}^2 \phi_{x} + \phi_{tx} \phi_{t} \phi_{xx} \\
- 2 \phi_{tx} \phi_{xxxx} \phi_{x} \sigma + 4 \phi_{tx} \phi_{xxx} \phi_{xx} \sigma + 4 \phi_{tx} \phi_{x}^3 u_3 \\
- \phi_{x}^5 u_3^2 - \phi_{x}^4 u_{2,x} u_3
\end{multline*}
\begin{multline*}
H_2 = \phi_{tt} \phi_{x} + 4 \phi_{txxx} \phi_{x} \sigma \\
- 6 \phi_{txx} \phi_{xx} \sigma + \phi_{tx} \phi_{t} + 4 \phi_{tx} \phi_{xxx} \sigma + 2 \phi_{tx} \phi_{x} u_2 \\
- 12 \phi_{xxx} \phi_{x}^2 \sigma u_3 - \phi_{x}^2 \sigma u_{2,xxx} - \phi_{x}^2 u_2 u_{2,x}
\end{multline*}
\begin{multline*}
H_3 = -\phi_{ttx} \phi_{x}^2 + \phi_{tt} \phi_{xx} \phi_{x} - \phi_{txxxx} \phi_{x}^2 \sigma + 4 \phi_{txxx} \phi_{xx} \phi_{x} \sigma \\
+ \phi_{txx} \phi_{t} \phi_{x} + 4 \phi_{txx} \phi_{xxx} \phi_{x} \sigma - 9 \phi_{txx} \phi_{xx}^2 \sigma - 2 \phi_{tx}^2 \phi_{x} \\
+ \phi_{tx} \phi_{t} \phi_{xx} - 2 \phi_{tx} \phi_{xxxx} \phi_{x} \sigma + 4 \phi_{tx} \phi_{xxx} \phi_{xx} \sigma \\
+ 4 \phi_{tx} \phi_{x}^3 u_3 - \phi_{x}^5 u_3^2 - \phi_{x}^4 u_{2,x} u_3
\end{multline*}
\begin{multline*}
H_4 = -\phi_{tx} \phi_{x}^3 - \phi_{t} \phi_{xx} \phi_{x}^2 + 2 \phi_{t} \phi_{xx} \phi_{x} - 4 \phi_{xxxx} \phi_{x}^3 \sigma \\
+ 2 \phi_{xxx} \phi_{xx} \phi_{x}^2 \sigma + 8 \phi_{xxx} \phi_{xx} \phi_{x} \sigma - 6 \phi_{xx}^3 \sigma \\
- 2 \phi_{xx} \phi_{x}^3 u_2 + 2 \phi_{xx} \phi_{x}^2 u_2 - \phi_{x}^4 u_{2,x}
\end{multline*}
\begin{multline*}
W_1 = 12 \phi_{ttx} \phi_{xx} \phi_{x}^4 \sigma - 12 \phi_{tt} \phi_{xx}^2 \phi_{x}^3 \sigma + \phi_{tt} \phi_{x}^5 u_2 \\
+ 12 \phi_{txxxx} \phi_{xx} \phi_{x}^4 \sigma^2 - 48 \phi_{txxx} \phi_{xx}^2 \phi_{x}^3 \sigma^2 + 4 \phi_{txxx} \phi_{x}^5 \sigma u_2 \\
- 12 \phi_{txx} \phi_{t} \phi_{xx} \phi_{x}^3 \sigma - 48 \phi_{txx} \phi_{xxx} \phi_{xx} \phi_{x}^3 \sigma^2 \\
+ 108 \phi_{txx} \phi_{xx}^3 \phi_{x}^2 \sigma^2 - 6 \phi_{txx} \phi_{xx} \phi_{x}^4 \sigma u_2 + 24 \phi_{tx}^2 \phi_{xx} \phi_{x}^3 \sigma \\
- 12 \phi_{tx} \phi_{t} \phi_{xx}^2 \phi_{x}^2 \sigma + \phi_{tx} \phi_{t} \phi_{x}^4 u_2 + 24 \phi_{tx} \phi_{xxxx} \phi_{xx} \phi_{x}^3 \sigma^2 \\
- 48 \phi_{tx} \phi_{xxx} \phi_{xx}^2 \phi_{x}^2 \sigma^2 + 4 \phi_{tx} \phi_{xxx} \phi_{x}^4 \sigma u_2 \\
- 48 \phi_{tx} \phi_{xx} \phi_{x}^5 \sigma u_3 + 2 \phi_{tx} \phi_{x}^5 u_2^2 - 12 \phi_{xxx} \phi_{x}^6 \sigma u_2 u_3 \\
+ 12 \phi_{xx} \phi_{x}^7 \sigma u_3^2 + 12 \phi_{xx} \phi_{x}^6 \sigma u_{2,x} u_3 \\
- \phi_{x}^6 \sigma u_2 u_{2,xxx} - \phi_{x}^6 u_2^2 u_{2,x}
\end{multline*}
\begin{equation*}
{W_2 = 12 \sigma u_3 \phi_{xxx} + u_{2, t} + u_2 u_{2, x} + \sigma u_{2, xxx}}
\end{equation*}
\begin{multline*}
W_3 = \phi_{tt} \phi_{x}^5 \sigma + 4 \phi_{txxx} \phi_{x}^5 \sigma^2 - 6 \phi_{txx} \phi_{xx} \phi_{x}^4 \sigma^2 \\
+ \phi_{tx} \phi_{t} \phi_{x}^4 \sigma + 4 \phi_{tx} \phi_{xxx} \phi_{x}^4 \sigma^2 + 2 \phi_{tx} \phi_{x}^5 \sigma u_2 \\
- 12 \phi_{xxx} \phi_{x}^6 \sigma^2 u_3 - \phi_{x}^6 \sigma^2 u_{2,xxx} - \phi_{x}^6 \sigma u_2 u_{2,x}
\end{multline*}
\begin{equation*}
{W_4 = u_{3, t} + u_3^2 \phi_x + u_{2,x} u_3}
\end{equation*}
\begin{multline*}
V_1 = -\phi_{tx} \phi_{x}^7 u_2 - \phi_{t} \phi_{xx} \phi_{x}^6 u_2 + 2 \phi_{t} \phi_{xx} \phi_{x}^5 u_2 \\
- 4 \phi_{xxxx} \phi_{x}^7 \sigma u_2 + 2 \phi_{xxx} \phi_{xx} \phi_{x}^6 \sigma u_2 + 8 \phi_{xxx} \phi_{xx} \phi_{x}^5 \sigma u_2 \\
- 6 \phi_{xx}^3 \phi_{x}^4 \sigma u_2 - 2 \phi_{xx} \phi_{x}^7 u_2^2 + 2 \phi_{xx} \phi_{x}^6 u_2^2 - \phi_{x}^8 u_2 u_{2,x}
\end{multline*}
\begin{multline*}
{V_2 = -\phi_{tx} \phi_{x}^7 \sigma - \phi_{t} \phi_{xx} \phi_{x}^6 \sigma + 2 \phi_{t} \phi_{xx} \phi_{x}^5 \sigma} \\
{- 4 \phi_{xxxx} \phi_{x}^7 \sigma^2 + 2 \phi_{xxx} \phi_{xx} \phi_{x}^6 \sigma^2 + 8 \phi_{xxx} \phi_{xx} \phi_{x}^5 \sigma^2} \\
{- 6 \phi_{xx}^3 \phi_{x}^4 \sigma^2 - 2 \phi_{xx} \phi_{x}^7 \sigma u_2 + 2 \phi_{xx} \phi_{x}^6 \sigma u_2 - \phi_{x}^8 \sigma u_{2,x}}
\end{multline*}
{The equations \eqref{eqnRef73From12212023} and \eqref{eqnRef74From12212023} reveal that for non-constant functions $\phi$, $u_2$ and a non-zero function $u_3$ if $Q_{\text{KdV, 1, +}} =$ $ Q_{\text{KdV, 2, +}} =$ $ Q_{\text{KdV, 4, +}} =$ $ Q_{\text{KdV, 5, +}} = 0$ then $Q_{\text{KdV, 3, +}} = Q_{\text{KdV, 6, +}} = 0$.} {The above reconciliation relations \eqref{eqnRef73From12212023} and \eqref{eqnRef74From12212023} can be worked out upon a careful and systematic differential-algebraic recombination of the different occurences of $u_3$ and its derivatives $u_{3, x}$, $u_{3, t}$, $u_{3, xx}$ and $u_{3,xxx}$ from $Q_{\text{KdV, 1, +}}, $ $ Q_{\text{KdV, 2, +}}, $ $ Q_{\text{KdV, 4, +}}, $ and $ Q_{\text{KdV, 5, +}}$ and then comparing them with the same with corresponding higher derivatives of $Q_{\text{KdV, 3, +}}$ and $Q_{\text{KdV, 6, +}}$ to observe and eliminate the required coefficients.} 
From the equation \eqref{eqnRef66From12182023}, one can obtain $\phi_t$,
\begin{equation*}
\phi_t = \frac{3 \sigma \phi_{xx}^2 - 4 \sigma \phi_x \phi_{xxx} - u_2 \phi_x^2}{\phi_x}
\end{equation*}
and with the help of the equation \eqref{eqnRef68From12182023}, one can also obtain
\begin{multline*}
\phi_{xt} = \frac{6 \sigma \phi_{xx} \phi_{xxx}}{\phi_x} - \frac{3 \sigma \phi_{xx}^2}{\phi_x^2} + 3 u_2 \phi_{xx} - 4 u_3 \phi_{xx}^2 \\
+ 4 \frac{\partial}{\partial x} \bigg(\frac{3 \sigma \phi_{xx}^2}{\phi_{x}} - 4 \sigma \phi_{xxx} - u_2 \phi_x \bigg)
\end{multline*}
\begin{multline*}
\phi_{x, 4} = \frac{1}{\sigma} \bigg(u_3 \phi_x^2 - u_2 \phi_{xx} \\
+ \frac{\partial}{\partial x} \bigg(\frac{3 \sigma \phi_{xx}^2}{\phi_{x}} - 4 \sigma \phi_{xxx} - u_2 \phi_x \bigg)\bigg)
\end{multline*}
And from the equations \eqref{neweqn44} and \eqref{neweqn45}, one can obtain the time derivatives of $u_{2}$ and $u_3$,
\begin{equation*}
u_{2, t} = -12 \sigma u_3 \phi_{xxx} - 12 \sigma u_{3, x} \phi_{xx} - \sigma u_{2, xxx} - u_2 u_{2, x}
\end{equation*}
\begin{equation*}
u_{3, t} = - \phi_{x} u_3^2 + \sigma u_{3,xxx} + u_2 u_{3,x} + u_{2,x} u_3
\end{equation*}
The above relations can be shown to imply that the difference $\frac{\partial^4}{\partial x^4} \phi_t - \frac{\partial}{\partial t} \phi_{xxxx}$ does not vanish identically, which also goes to show that the set {$\{Q_{\text{KdV, 1, +}},$ $Q_{\text{KdV, 2, +}},$ $Q_{\text{KdV, 4, +}},$ $Q_{\text{KdV, 5, +}}\}$ is \textit{not} context-sensitive compatible, thus leading to the non-existence of the sought B\"{a}cklund transform} unless we assume $u_3=0$ in which case we fall back to the scenario studied earlier.

\subsection*{Benjamin-Bona-Mahony equation} \label{appendixA_Contradiction_Of_BBM_Equation}
Continuing with the BBM equation \eqref{NewEqnRef57} extensively discussed in section \ref{backlundexamples}, if we truncate one order earlier with $u_j = 0, j \ge 2$, one obtains,
\begin{equation} \label{eqnRef122From12102023}
u = \frac{u_0}{\phi^2} + \frac{u_1}{\phi}
\end{equation}
where $u_0 = 12 \phi_{t} \phi_{x}$ and $u_1 = \frac{-12}{5 \phi_{t} \phi_{x}}\bigg(\phi_{tt} \phi_{x}^2 + 3 \phi_{tx} \phi_{t} \phi_{x} + \phi_{t}^2 \phi_{xx}\bigg)$ follow from $P_0 = 0$ and $P_1 = 0$ respectively. The equations that follow from $P_2 = 0$, $P_3 = 0$ and $P_4 = 0$ respectively are given by,
\begin{multline} \label{eqnRef123From12102023}
g_{\text{1, BBM}, -} = 48 \phi_{ttx} \phi_{x}^2 + 72 \phi_{tt} \phi_{xx} \phi_{x} \\
+ 96 \phi_{txx} \phi_{t} \phi_{x} + 96 \phi_{tx}^2 \phi_{x} + 120 \phi_{tx} \phi_{t} \phi_{xx} \\
 + 8 \phi_{tx} \phi_{x} u_1 + 24 \phi_{t}^2 \phi_{xxx} - 24 \phi_{t}^2 \phi_{x} + 10 \phi_{t} \phi_{xx} u_1 \\
- 24 \phi_{t} \phi_{x}^2 + 8 \phi_{t} \phi_{x} u_{1,x} - 2 \phi_{x}^2 u_{1,t} - \phi_{x} u_1^2 = 0
\end{multline}
\begin{multline} \label{eqnRef124From12102023}
g_{\text{2, BBM}, -} = 12 \phi_{ttxx} \phi_{x} + 24 \phi_{ttx} \phi_{xx} + 12 \phi_{tt} \phi_{xxx} \\
- 12 \phi_{tt} \phi_{x} + 12 \phi_{txxx} \phi_{t} + 36 \phi_{txx} \phi_{tx} - \phi_{txx} u_1 - 12 \phi_{tx} \phi_{t} \\
- 12 \phi_{tx} \phi_{x} - 2 \phi_{tx} u_{1,x} - 12 \phi_{t} \phi_{xx} + \phi_{t} u_1 - \phi_{t} u_{1,xx} \\
- \phi_{xx} u_{1,t} + \phi_{x} u_1 - 2 \phi_{x} u_{1,tx} - u_1 u_{1,x} = 0
\end{multline}
\begin{equation} \label{eqnRef125From12102023}
g_{\text{3, BBM}, -} = -u_{1,txx} + u_{1,t} + u_{1,x}  = 0,
\end{equation}
and it can be shown that $P_j = 0, j \ge 5$. One can verify that the set $\{g_{\text{1, BBM}, -}, g_{\text{2, BBM}, -}, g_{\text{3, BBM}, -}\}$ is \textit{not} context-free compatible (checked up to 150 recursions with the increasing order of the differential polynomials  in the consequent differential ideal). From equations \eqref{eqnRef123From12102023}, \eqref{eqnRef124From12102023} and \eqref{eqnRef125From12102023}, one can infer that,
\begin{equation*}
u_{1,txx} = u_{1,t} + u_{1,x}
\end{equation*}
\begin{multline*}
\phi_{xtt} = -\frac{3 \phi_{tt} \phi_{xx}}{2 \phi_{x}} - \frac{2 \phi_{txx} \phi_{t}}{\phi_{x}} - \frac{2 \phi_{tx}^2}{\phi_{x}} \\
- \frac{5 \phi_{tx} \phi_{t} \phi_{xx}}{2 \phi_{x}^2} -\frac{\phi_{tx} u_1}{6 \phi_{x}} - \frac{\phi_{t}^2 \phi_{xxx}}{2 \phi_{x}^2} + \frac{\phi_{t}^2}{2 \phi_{x}} \\
- \frac{5 \phi_{t} \phi_{xx} u_1}{24 \phi_{x}^2} + \frac{\phi_{t}}{2} - \frac{\phi_{t} u_{1,x}}{6 \phi_{x}} + \frac{u_{1,t}}{24} + \frac{u_1^2}{48 \phi_{x}}
\end{multline*}
\begin{multline*}
\phi_{xxtt} = - \frac{\phi_{tt} \phi_{xxx}}{\phi_{x}} + \frac{3 \phi_{tt} \phi_{xx}^2}{\phi_{x}^2} + \phi_{tt} - \frac{\phi_{txxx} \phi_{t}}{\phi_{x}} \\
- \frac{3 \phi_{txx} \phi_{tx}}{\phi_{x}} + \frac{4 \phi_{txx} \phi_{t} \phi_{xx}}{\phi_{x}^2} + \frac{\phi_{txx} u_1}{12 \phi_{x}} + \frac{4 \phi_{tx}^2 \phi_{xx}}{\phi_{x}^2} + \\
\frac{5 \phi_{tx} \phi_{t} \phi_{xx}^2}{\phi_{x}^3} + \frac{\phi_{tx} \phi_{t}}{\phi_{x}} + \frac{\phi_{tx} \phi_{xx} u_1}{3 \phi_{x}^2} + \phi_{tx} + \frac{\phi_{tx} u_{1,x}}{6 \phi_{x}} \\
+ \frac{\phi_{t}^2 \phi_{xxx} \phi_{xx}}{\phi_{x}^3} - \frac{\phi_{t}^2 \phi_{xx}}{\phi_{x}^2} + \frac{5 \phi_{t} \phi_{xx}^2 u_1}{12 \phi_{x}^3} + \frac{\phi_{t} \phi_{xx} u_{1,x}}{3 \phi_{x}^2} \\
- \frac{\phi_{t} u_1}{12 \phi_{x}} + \frac{\phi_{t} u_{1,xx}}{12 \phi_{x}} - \frac{\phi_{xx} u_1^2}{24 \phi_{x}^2} - \frac{u_1}{12} + \frac{u_{1,tx}}{6} + \frac{u_1 u_{1,x}}{12 \phi_{x}}
\end{multline*}
From there one can verify that the difference $\frac{\partial \phi_{xtt}}{\partial x} - \phi_{xxtt}$ does not vanish identically, which implies that the set $\{g_{\text{1, BBM}, -}, g_{\text{2, BBM}, -}, g_{\text{3, BBM}, -}\}$ is \textit{not} context-sensitive compatible, leading to the {non-existence of the sought B\"{a}cklund transform}. 

Finally if we truncate one order ahead with $u_j = 0, j \ge 4$, one obtains,
\begin{equation} \label{eqnRef126From12102023}
u = \frac{u_0}{\phi^2} + \frac{u_1}{\phi} + u_2 + u_3 \phi
\end{equation}
where $u_0 = 12 \phi_{t} \phi_{x}$ and $u_1 = \frac{-12}{5 \phi_{t} \phi_{x}}\bigg(\phi_{tt} \phi_{x}^2 + 3 \phi_{tx} \phi_{t} \phi_{x} + \phi_{t}^2 \phi_{xx}\bigg)$ follow from $P_0 = 0$ and $P_1 = 0$ respectively. The equations that follow from $P_2 = 0$, $P_3 = 0$, $P_4 = 0$, $P_5 = 0$, $P_6 = 0$ and  $P_7 = 0$ respectively are given by,
\begin{multline} \label{eqnRef127From12102023}
P_2 = 0 \implies g_{\text{1, BBM}, +} = 48 \phi_{ttx} \phi_{x}^2 + 72 \phi_{tt} \phi_{xx} \phi_{x} \\
+ 96 \phi_{txx} \phi_{t} \phi_{x} + 96 \phi_{tx}^2 \phi_{x} + 120 \phi_{tx} \phi_{t} \phi_{xx} + 8 \phi_{tx} \phi_{x} u_1 \\
+ 24 \phi_{t}^2 \phi_{xxx} - 24 \phi_{t}^2 \phi_{x} + 10 \phi_{t} \phi_{xx} u_1 - 24 \phi_{t} \phi_{x}^2 u_2 \\
- 24 \phi_{t} \phi_{x}^2 + 8 \phi_{t} \phi_{x} u_{1,x} - 2 \phi_{x}^2 u_{1,t} - \phi_{x} u_1^2 = 0
\end{multline}
\begin{multline} \label{eqnRef128From12102023}
P_3 = 0 \implies g_{\text{2, BBM}, +} = 12 \phi_{ttxx} \phi_{x} + 24 \phi_{ttx} \phi_{xx} \\
+ 12 \phi_{tt} \phi_{xxx} - 12 \phi_{tt} \phi_{x} + 12 \phi_{txxx} \phi_{t} + 36 \phi_{txx} \phi_{tx} - \phi_{txx} u_1 \\
- 12 \phi_{tx} \phi_{t} - 12 \phi_{tx} \phi_{x} u_2 - 12 \phi_{tx} \phi_{x} - 2 \phi_{tx} u_{1,x} \\
- 12 \phi_{t} \phi_{xx} u_2 - 12 \phi_{t} \phi_{xx} + 12 \phi_{t} \phi_{x}^2 u_3 - 12 \phi_{t} \phi_{x} u_{2,x} + \phi_{t} u_1 \\
- \phi_{t} u_{1,xx} - \phi_{xx} u_{1,t} + \phi_{x} u_1 u_2 + \phi_{x} u_1 - 2 \phi_{x} u_{1,tx} - u_1 u_{1,x} = 0
\end{multline}
\begin{multline} \label{eqnRef129From12102023}
P_4 = 0 \implies g_{\text{3, BBM}, +} = 12 \phi_{tx} \phi_{x} u_3 + 12 \phi_{t} \phi_{xx} u_3 \\
+ 12 \phi_{t} \phi_{x} u_{3,x} + u_1 u_{2,x} - u_{1,txx} + u_{1,t} + u_{1,x} u_2 + u_{1,x}  = 0
\end{multline}
\begin{multline} \label{eqnRef130From12102023}
P_5 = 0 \implies g_{\text{4, BBM}, +} = u_1 u_{3,x} + u_{1,x} u_3 + u_2 u_{2,x}\\
- u_{2,txx} + u_{2,t} + u_{2,x}  = 0
\end{multline}
\begin{multline} \label{eqnRef131From12102023}
P_6 = 0 \implies g_{\text{5, BBM}, +} = \phi_{x} u_3^2 + u_2 u_{3,x}\\
 + u_{2,x} u_3 - u_{3,txx} + u_{3,t} + u_{3,x}  = 0
\end{multline}
\begin{equation} \label{eqnRef132From12102023}
P_7 = 0 \implies g_{\text{6, BBM}, +} = u_3 u_{3,x}  = 0
\end{equation}
{and all the other coefficients $P_k = 0, k \ge 8$.} One can verify that the set $\{g_{\text{1, BBM}, +},$ $g_{\text{2, BBM}, +},$ $ g_{\text{3, BBM}, +},$ $g_{\text{4, BBM}, +},$ $g_{\text{5, BBM}, +},$ $g_{\text{6, BBM}, +}\}$ is \textit{not} context-free compatible (checked up to 150 recursions with the increasing order of the differential polynomials  in the consequent differential ideal). A specific and interesting branch of the solution to equation \eqref{eqnRef132From12102023} is $u_3 = 0$ which takes us back to the results from the truncation $u = \frac{u_0}{\phi^2} + \frac{u_1}{\phi} + u_2$ where we have already showed the context-sensitive \textit{incompatibility} of the set $\{g_{\text{1, BBM}}, g_{\text{2, BBM}}, g_{\text{3, BBM}}, g_{\text{4, BBM}}, g_{\text{5, BBM}}\}$.

Another explicit route to a contradiction is the alternate, so-called  ``potential'' form of the BBM equation as annotated in \cite{honechapter}; for a choice of a potential function $v = \int_{-\infty}^{x} u(y, t) dy, \lim_{x \to \infty} v = C_0 \in \mathbb{R}$ (or equivalently $u = v_x$) the BBM reduces to,
\begin{equation} \label{NewEqnRef67}
    v_t + v_x + \frac{1}{2} v_x^2 - v_{xxt} = 0,
\end{equation}
with the Painlev\'e expansion $v = \frac{v_0}{\phi} + v_1 + v_2\phi + v_3 \phi^2 + \cdots$ substituted in equation \eqref{NewEqnRef67}, we get the expansion $\frac{1}{\phi^4}(P_0 + P_1 \phi + P_2 \phi^2 + P_3 \phi^3 + \cdots) = 0$,
\begin{equation}
    P_0 = \frac{1}{2} v_0 \phi_x^2 (v_0 + 12 \phi_t) \implies v_0 = -12 \phi_t
\end{equation}
and when we substitute $v_0 = -12 \phi_t$ in the equation $P_1 = 0$ we obtain,
\begin{equation} \label{NewEqnRef69}
    \phi_{tt}\phi_x^2 - 2 \phi_x \phi_t \phi_{xt} + \phi_t^2 \phi_{xx} = 0.
\end{equation}
With the choice $g = \frac{-\phi_t}{\phi_x}$, one can reduce the equation \eqref{NewEqnRef69} to the inviscid Burgers' equation $g_t + g g_x = 0$ (the Burgers' equation $u_t + u u_x - \sigma u_{xx} = 0$ with $\sigma = 0$) which can be solved using the method of characteristics \cite{sneddon} to obtain the solution to be of the form of a shock wave $g = f_0(x - g t), f_0 : \mathbb{R} \to \mathbb{R}, g = g(x, t)$, which again can be solved for $\phi$ from $\phi_{t} = g \phi_x \implies \phi_{t} = \phi_x f_1(x \phi_x - t \phi_t), f_1 : \mathbb{R} \to \mathbb{R}$ using the Charpit's method of characteristics. This appears to generically lead to the formation of shock waves in finite time. This leads to the failure of the Painlev\'e test for the potential form of the BBM equation as the coefficient at the resonant index $j = 1$ does not identically vanish. Also, apart from the application of the tests of context-free and context-sensitive compatibilities to the original PDE, one may apply the test to the potential form of the PDE (only if there exists one) to examine its integrability. 

\end{appendices}

\end{document}